\documentclass{emulateapj}
\usepackage{lscape}

\shorttitle{Physical origins of GRBs}
\shortauthors{Zhang et al.}
\begin{document}
\title{Discerning the physical origins of cosmological Gamma-ray bursts 
based on multiple observational criteria: the cases of $z=6.7$
GRB 080913, $z=8.3$ GRB 090423, and some short/hard GRBs}
\author
{Bing Zhang\altaffilmark{1}, Bin-Bin Zhang\altaffilmark{1}, 
Francisco J. Virgili\altaffilmark{1},  En-Wei Liang\altaffilmark{2}, 
D. Alexander Kann\altaffilmark{3}, Xue-Feng Wu\altaffilmark{4,5},
Daniel Proga\altaffilmark{1}, Hou-Jun Lv\altaffilmark{2}, 
Kenji Toma\altaffilmark{4}, Peter M\'esz\'aros\altaffilmark{4,6}, 
David N. Burrows\altaffilmark{4}, Peter W. A. Roming\altaffilmark{4}, 
Neil Gehrels\altaffilmark{7}}
\altaffiltext{1}{Department of Physics and Astronomy, University
of Nevada Las Vegas, Las Vegas, NV 89154, USA.}
\altaffiltext{2}{Department of Physics, Guangxi University,
Guangxi 530004, China.}
\altaffiltext{3}{Th\"uringer Landessternwarte Tautenburg, D-07778,
Tautenburg, Germany.}
\altaffiltext{4}{Department of Astronomy \& Astrophysics, Pennsylvania
  State University, University Park, PA 16802, USA.}
\altaffiltext{5}{Purple Mountain Observatory, Chinese Academy of
  Sciences, Nanjing 210008, China.}
\altaffiltext{6}{Department of Physics, Pennsylvania
  State University, University Park, PA 16801, USA.}
\altaffiltext{7}{NASA Goddard Space Flight Center, Greenbelt, MD 20771, USA.}

\begin{abstract}
The two high-redshift gamma-ray bursts, GRB 080913 at $z=6.7$
and GRB 090423 at $z=8.3$, recently detected by Swift appear as 
intrinsically short, hard GRBs. They could have been recognized by 
BATSE as short/hard GRBs should they have occurred at $z \leq 1$. 
In order to address their physical origin, 
we perform a more thorough investigation on two physically
distinct types (Type I/II) of cosmological GRBs and their 
observational characteristics. We reiterate the definitions
of Type I/II GRBs and then review the following observational
criteria and their physical motivations: supernova
association, specific star forming rate of the host galaxy, location
offset, duration, hardness, spectral lag, statistical correlations,
energetics and collimation, afterglow properties, redshift 
distribution, luminosity function, and gravitational wave signature. 
Contrary to the traditional approach of assigning the physical
category based on the gamma-ray properties (duration, hardness,
and spectral lag), we take an alternative approach to define the Type
I and Type II Gold Samples using several criteria that are more
directly related to the GRB progenitors (supernova association,
host galaxy type, and specific star forming rate). 
We then study the properties 
of the two Gold Samples and compare them with the traditional
long/soft and short/hard samples. We find that the Type II Gold
Sample reasonably tracks the long/soft population, although it
includes several intrinsically short (shorter than 1s in the rest frame)
GRBs. The Type I Gold Sample only has 5 GRBs, 4 of which are not
strictly short but have extended emission. Other short/hard GRBs
detected in the Swift era represent the BATSE short/hard sample
well, but it is unclear whether all of them belong to Type I.
We suggest that some (probably even most) 
high-luminosity short/hard GRBs instead 
belong to Type II. Based on multiple observational criteria, 
we suggest that 
GRB 080913 and GRB 090423 are more likely Type II events. In general,
we acknowledge that it is not always straightforward to discern
the physical categories of GRBs, and re-emphasize the 
importance of invoking multiple observational criteria. We 
cautiously propose an operational procedure to infer 
the physical origin of a given GRB with available multiple 
observational criteria, with various caveats laid out.
\end{abstract}

\slugcomment{2009, ApJ, in press}

\section{Introduction}

Phenomenologically, gamma-ray bursts (GRBs) have been generally 
classified into the long-duration, soft-spectrum class and the 
short-duration, hard-spectrum class in the CGRO/BATSE era based 
on the bimodal distribution of GRBs in the duration-hardness diagram
\citep{kouveliotou93}\footnote{Several analyses have suggested
the existence of an intermediate duration group
\citep{mukherjee98,horvath98,hakkila00}. However, as discussed
in the bulk of the text below, there is so far no strong indication 
of the existence of a third, physically distinct category of 
cosmological GRBs based on multiple observational data. So we will
focus on the two main phenomenological categories of GRBs in
the rest of the paper.}.
There is no clear boundary line in this diagram to
separate the two populations. Traditionally, an observer-frame
BATSE-band duration $T_{90} \sim 2$ s has been taken as the 
separation line: bursts with $T_{90} > 2$s are ``long'' and bursts
with $T_{90} < 2$s are ``short''. 

The journey was long to uncover the physical origins of these two 
phenomenologically different classes of GRBs. The discoveries and the
routine observations of the broad band afterglows of long GRBs 
reveal that their host galaxies are typically irregular (in a few
cases spiral) galaxies with intense star formation \citep{fruchter06}. 
In a handful of cases these GRBs are firmly associated with Type Ib/c
supernovae \citep[e.g.][]{hjorth03,stanek03,campana06,pian06}. This
strongly suggests that they are likely related to 
deaths of massive stars. Theoretically, the ``collapsar''
model of GRBs has been discussed over the years as the standard 
scenario for long GRBs 
\citep{woosley93,paczynski98,macfadyen99,woosley06}.

The breakthrough to understand the nature of some short GRBs
was made in 2005 after the launch of the Swift
satellite \citep{Gehrels04}. 
Prompt localizations and deep afterglow searches
for a handful of short GRBs 
\citep{gehrels05,bloom06,fox05,villasenor05,hjorth05a,barthelmy05a,berger05} 
suggest that some of them are associated with nearby early-type galaxies 
with little star formation. Deep searches of associated supernovae
from these events all led to non-detections
(e.g. Kann et al. 2008 and references therein, see also Appendix
for more references).
These are in stark contrast to the 
bursts detected in the pre-Swift era (mostly long-duration). 
On the other hand, the observations are consistent with (although
not a direct proof of) the long-sought progenitor models that
invoke mergers of two compact stellar objects, leading candidates
being NS-NS and NS-BH systems 
\citep{paczynski86,eichler89,paczynski91,narayan92}.
Although the sample with secure host galaxies is small,
a general trend in the community is to accept that the BATSE
short/hard population bursts are of this compact star
merger origin\footnote{It is widely accepted that at least
a fraction of short/hard GRBs are the giant flares
of soft gamma-ray repeaters in nearby galaxies
\citep{palmer05,tanvir05}. The observations suggest that
the contribution from such a population is not 
significant \citep{nakar06b}, but see \cite{chapman09}. 
We do not discuss these bursts in this paper.}.

The clean dichotomy of the two populations (both phenomenological
and physical) was soon muddled
by the detection of a nearby long-duration GRB without SN
association \citep{gehrels06,galyam06,fynbo06,dellavalle06}. 
GRB 060614 has $T_{90} \sim 100$s
in the Swift BAT \citep{barthelmy05b}
band, which phenomenologically definitely
belongs to the long duration category. On the other hand,
the light curve is characterized by a short/hard spike (with
a duration $\sim 5$ s) followed by a series of soft gamma-ray 
pulses. The spectral lag at the short/hard spike is 
negligibly small, a common feature of the short/hard
GRBs \citep{gehrels06}. Very stringent upper limits on the
radiation flux from an underlying SN have been established
\citep{galyam06,fynbo06,dellavalle06}. These facts are consistent 
with the compact star merger scenario. More interestingly, 
this burst looks like a more energetic version 
of GRB 050724, the ``smoking-gun'' burst of the compact
star merger population \citep{barthelmy05a,berger05}.
\cite{zhang07b} showed that if one applies the $E_p 
\propto E_{\gamma,iso}^{1/2}$ relation \citep{amati02,liang04}
to GRB 060614 and makes it as energetic as GRB 050724,
the pseudo-burst would be detected as a marginal short/hard burst
by BATSE, and would be very similar to GRB 050724 if detected
by Swift/BAT. In particular, the soft gamma-ray
tail would appear as the ``extended emission'' detected
in some ``short/hard'' GRBs including GRB 050724. 
A second, much shorter (with $T_{90} \sim 4$ s) burst without
SN association, GRB 060505, was detected around the same
time \citep{fynbo06}. However, the physical nature of this
burst is subject to intense debate 
\citep{ofek07,thone08,mcbreen08,kann09b}.

In any case, duration and hardness are not necessarily 
reliable indicators of the {\em physical} nature of a GRB any 
more. In order to determine whether or not a GRB can be 
associated with a particular physical model,
one is forced to appeal to multiple observational
criteria \citep{donaghy06}. Prompted by the 
detection of GRB 060614, we \citep{zhang07b,zhangnature06} 
suggested naming the bursts that are consistent with the
massive-star origin and the compact-star-merger origin models 
as Type II and Type I, 
respectively\footnote{The idea was to make a connection to 
the Type II and Type Ia SNe (not including Type Ib/c), which 
correspondingly have the massive star and compact star origins, 
respectively. This is however not related to the original 
definitions of Type II and Type I SNs, which are based on 
whether or not there are hydrogen lines in the spectrum.}, 
and attempted to
invoke a set of multiple observational criteria to judge
the physical category of a GRB. A more developed physical
categorization scheme was proposed by \cite{bloom08}, who
also introduced SGR giant-flare-like (non-destructive and 
likely repeating) events.
Within the destructive events, \cite{bloom08}
agreed that there are two major model types (degenerate 
and non-degenerate), which correspond to Type I and Type II
in the \cite{zhang07b}'s classification scheme. Throughout
this paper we will adopt the nomenclature of Type I/II
to denote the two physically distinct categories of 
cosmological GRB models.

The recently detected two high-$z$ GRBs, GRB 080913 at $z=6.7$ 
\citep{greiner09} and GRB 090423 at $z=8.3$ 
\citep{tanvir09,salvaterra09} introduce a further complication
to the scheme associating GRBs with particular theoretical
models. Being the two GRBs
with the highest redshifts as of the time of writing, 
these two bursts each have a 
redshift-corrected duration [$T_{90}/(1+z)$] shorter than 2 
seconds, with a hard spectrum typical for short/hard GRBs. This
naturally raises the interesting question 
regarding the progenitor system of the burst 
\citep{greiner09,perezramirez09,belczynski09,tanvir09,salvaterra09}. 
More generally, it again raises the difficult question
regarding how to use the observed properties to judge the
physical origin of a GRB. In this paper, we make
some attempts to address this difficult problem. The structure
of the paper is the following. In \S2, we present the
observational properties of GRB 080913 and GRB 090423, and show
that if the identical bursts had occurred at $z<1$,
they could have been recognized as short hard GRBs based 
on their observed properties. In \S3, we comment on the 
strengths and weaknesses of classifying GRBs based on 
physically motivated criteria.
We then reiterate the definitions of Type I/II GRBs in \S4,
and critically review a list of observational criteria
as well as their physical motivations as discussed
in the literature. In order to address
the profound questions of whether ``Type I
= short/hard/short lag'' and ``Type II = long/soft/long lag'', 
in \S5 we take an alternative approach (from the traditional one) 
to assess the problem.
Instead of associating a burst with a
particular physical model (massive star core collapses vs. 
compact star mergers) {\em a priori} 
based on its gamma-ray properties (duration, hardness, spectral
lag), we use several
observational properties that are more directly relevant to
the GRB progenitors to define the Gold Samples of Type
II and Type I GRBs. We then turn around to
evaluate the various observational properties (duration,
hardness, spectral lag, afterglow properties, empirical correlations, 
etc) of these Gold Samples, and check whether these 
properties are useful criteria to judge the physical category 
of the bursts. In \S6, we discuss the intriguing question whether
all short/hard GRBs are of the Type I origin, 
and raise the possibility
that a fraction of (probably even most) 
high-luminosity short GRBs are of the Type 
II origin. We then dedicate \S7 to discuss the possible
progenitors of GRB 080913 and GRB 090423, and argue that 
most likely they are both of the Type II origin.
We acknowledge the difficulties of discerning the physical
origins of GRBs in \S8, and cautiously propose an operational 
procedure to associate a GRB with a specific model based
on multiple observational criteria. 
Our results are summarized in \S9.

\section{GRB 080913 and GRB 090423: intrinsically short/hard GRBs 
at high-$z$}

\begin{figure}
\plotone{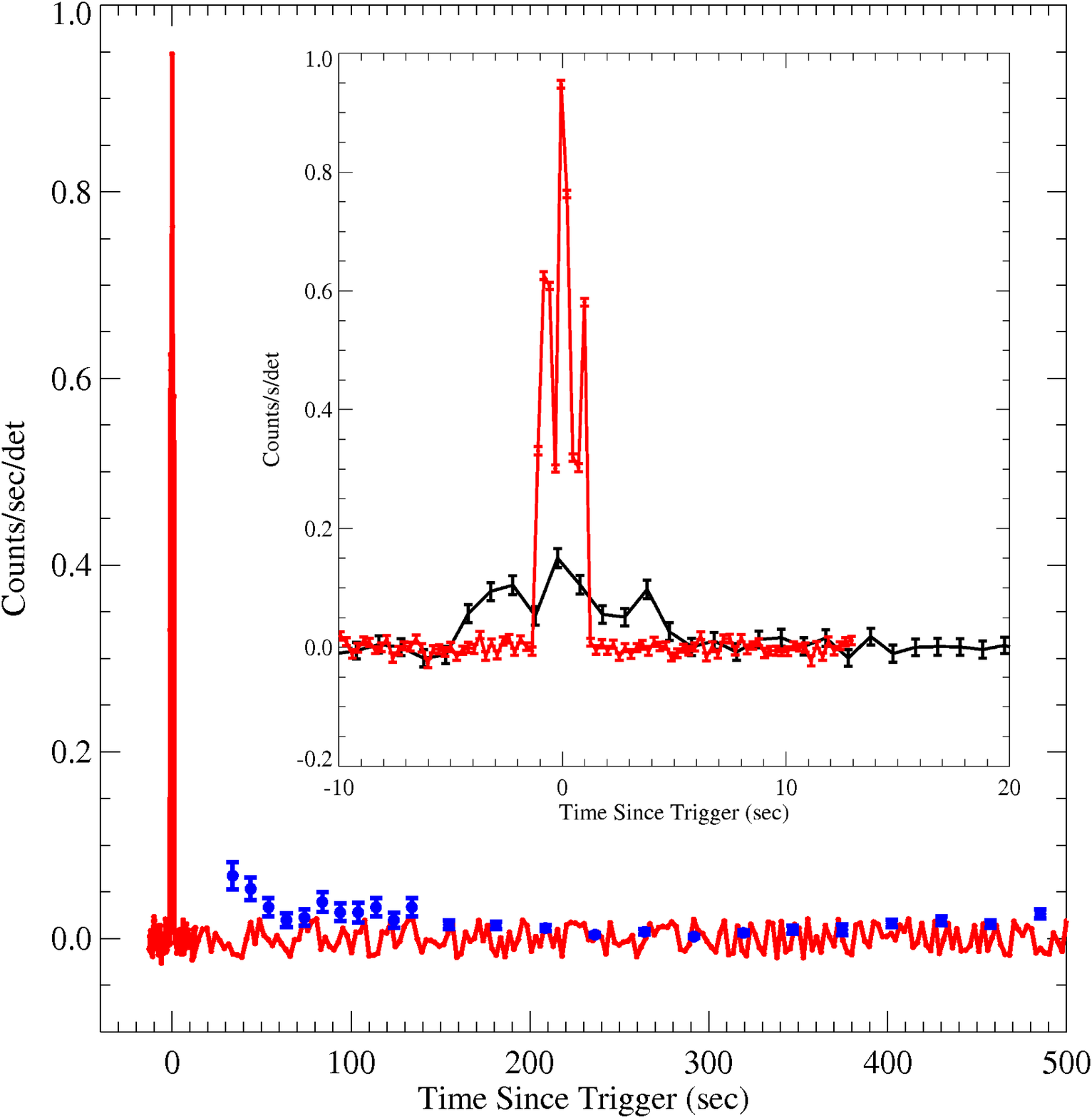}
\plotone{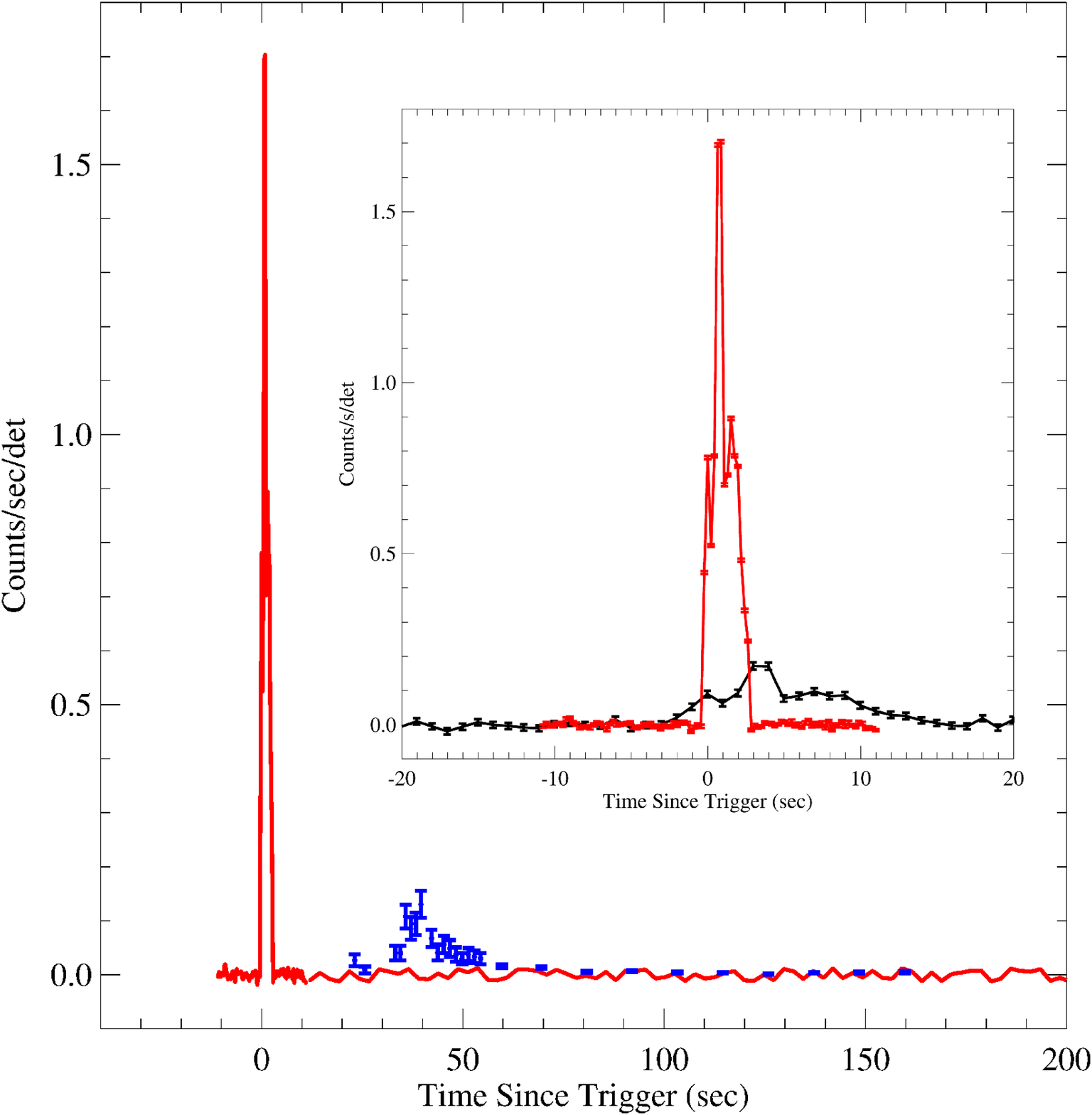}
\caption{The simulated 15-150 keV light curves of the pseudo 
GRBs obtained by placing GRB 080913 and GRB 090423 at $z=1$. The 
red curves display the extrapolated BAT data, and the blue data 
points show the extrapolated XRT data. Inset: a comparison
of the light curve of the pseudo GRBs (red) and the observed
GRBs (black). (a) GRB 080913; (b) GRB 090423.}
\label{pseudoGRB}
\end{figure}

The light curve of GRB 080913 as detected by Swift/BAT is shown
as the black solid curve in Fig.\ref{pseudoGRB}a. The burst
duration $T_{90}$ (the time interval during which 90\% of
the fluence is measured) in the BAT (15-150 keV) band is 
$8\pm 1$ s. The average BAT band spectrum can be adequately 
fit by a power law with exponential cutoff, with the peak
energy $E_p = 93 \pm 56$ keV \citep{greiner09}. A combined 
Swift/BAT and Konus/Wind (20-1300 keV) fit using the 
Band-function spectrum gives $E_p = 121^{+232}_{-39}$
keV \citep{palshin08}. Given the measured redshift $z=6.7$
\citep{greiner09}, this is translated to a rest frame 
duration of $T_{90}^{rest} \sim 1$ s, and a best-fit rest frame 
peak energy $E_p^{rest} \sim 710$ keV and $E_p^{rest} \sim 930$ 
keV for the cutoff power law and Band-function spectra, 
respectively. Although being recognized as a long duration 
burst phenomenologically, this burst has an intrinsically short
duration and an intrinsically hard spectrum.

In order to compare this burst with other phenomenologically
classified short hard GRBs,
we simulate a ``pseudo'' GRB by placing GRB 080913 at $z=1$.
We consider three factors. First, 
the specific photon flux $N(E_p)$ at $E_p$ is proportional to 
$(1+z)^2/D_L^2$, where $D_L$ is the luminosity distance. 
This can be translated to an increase of a factor of $\sim 6.8$
of $N(E_p)$ from $z=6.7$ to $z=1$. Second, we consider the BAT 
band (15-150 keV) emission of the pseudo GRB, which corresponds 
to an energy band lower by a factor of 
$(1+6.7)/(1+1) \sim 3.85$ in GRB 080913.  
We therefore extrapolate the observed BAT spectrum to lower 
energies and assume a similar light curve in that band. 
Third, we compress the time scale by a factor of $\sim 3.85$
to account for the cosmological time dilation effect.
After applying these transformations, we are able to
construct the BAT-band light curve of the pseudo
GRB at $z=1$ as shown in Fig.\ref{pseudoGRB}a. 

GRB 080913 displays a series
of early X-ray flares \citep{greiner09}.  It
is interesting to check whether they would show up in the
BAT band for the pseudo GRB to mimic the ``extended emission''
seen in a subgroup of Swift ``short/hard'' GRBs 
\citep{norris06,troja08}\footnote{Rigorously based on
the $T_{90}$ criterion, the fraction of Swift bursts that
have $T_{90} < 2$ s is much smaller than that of BATSE bursts.
Many display extended emission that extends $T_{90}$ up to
several 10s to even more than 100 seconds. The current 
approach in the community is to define a burst ``short/hard''
if it appears short in the BATSE band. A growing trend
is to also include some bursts with extended emission even in 
the BATSE band to the ``short/hard'' category.}. 
We therefore manipulate the XRT
\citep{burrows05b}
data of GRB 080913 to simulate the BAT band 
extended emission of the pseudo burst. We first extrapolate
the GRB 080913 XRT data to the BAT band according to the 
measured XRT photon spectral index. We then follow the
three steps mentioned above to shift this BAT-band ``virtual'' 
emission to the BAT band emission of the pseudo burst. 
This is shown as blue data points in 
Fig.\ref{pseudoGRB}a. By adding the appropriate noise level
for the BAT observation, we show that these extrapolated XRT
emission components stick out the background, which would
appear as the extended emission in the BAT band for the pseudo 
burst. We note that our method is based on the assumption of 
the power law extension of the X-ray flare spectrum ($0.3-10$
keV) to the BAT band of the pseudo burst ($1.3 - 39$ keV).
On the other hand, since X-ray flares are generally believed
to be due to GRB late central engine activities 
\citep{burrows05,zhang06,lazzati07,chincarini07},
they may have a Band-function or cutoff power law spectrum
\citep{falcone07}.
If the $E_p$'s of the X-ray flares are within or not far
above the XRT window, the extrapolated extended emission
would be degraded. We should therefore regard the level
of the extended emission of the pseudo burst as an upper
limit. We estimate the BAT-band duration of 
the pseudo GRB as $T_{90}({\rm pseudo}) \sim 2.0$ s without 
extended emission or $T_{90}({\rm pseudo,EE}) \sim 140$ s
with extended emission. 
In any case, the observational properties of this pseudo
burst are very similar to some ``short/hard'' GRBs detected in 
the Swift era. By comparing the flux level of the pseudo GRB
with other short/hard GRBs, we find that it belongs to
the bright end of the short/hard GRB flux distribution,
similar to, e.g. GRB 051221A \citep{burrows06,soderberg06b},
GRB 060313 \citep{roming06b}, 
GRB 060121 \citep{donaghy06},
and the recent GRB 090510 detected by Fermi LAT/GBM and Swift
\citep{hoversten09,ohno09,guiriec09,rau09}.

GRB 090423 at $z=8.3$ is amazingly 
similar to GRB 080913. It was detected by Swift/BAT with
a BAT-band $T_{90} \sim 10.3$ s \citep{tanvir09}. Given the
measured redshift $z=8.26^{+0.07}_{-0.08}$ \citep{tanvir09},  
the corresponding rest-frame duration is $\sim T_{90}/(1+z) 
\sim 1.1$ s. The peak energy measured by BAT is $E_p=48.6 
\pm 6.2$ keV, corresponding to a rest-frame value
$E_p^{rest} =451 \pm 58$ keV.
We performed a similar analysis 
on GRB 080913. The results are shown in Fig.\ref{pseudoGRB}b. 
Nearly identical conclusions can be drawn from both
bursts.

In the above analyses, the intrinsic duration of a burst is defined
as $T_{90}/(1+z)$, and the duration of the corresponding 
pseuodo GRB at $z=1$ is defined as $2T_{90}/(1+z)$. These 
calculated durations correspond to different energy bands
in the rest frame (the same observed band after redshifting). 
Strictly speaking,
in order to derive the durations of the pseudo GRBs in the observed
energy band, one needs to know the time-dependent spectral information,
which is not available for these bursts. 
Observationally, pulse widths at high energies
tend to be narrower than those at low energies 
\citep{ford95,romano06,page07}. An empirical relation 
$w \propto E^{-a}$ with $a \sim 0.3$ has been suggested
\citep{fenimore95,norris05,liang06b}. For a given observed energy band,
this suggests $w \propto (1+z)^{-a}$, which would correspond to a 
correction factor of $(1+z)^{a-1}$ rather than $(1+z)^{-1}$ to 
derive the intrinsic duration. However, GRB prompt emission is
usually composed of multiple pulses. The separations between
the pulses, which are more relevant for the $T_{90}$ definition, 
may not follow the same energy-dependence of the 
pulse widths. We therefore do not introduce this extra correction
factor of $T_{90}$ throughout the paper. For GRB 080913 and GRB
090423, if one takes the $(1+z)^{a-1}$ correction factor,
the derived intrinsic durations are in the marginal regime
between the phenomenologically-defined long and short GRBs.

\begin{figure}
\plotone{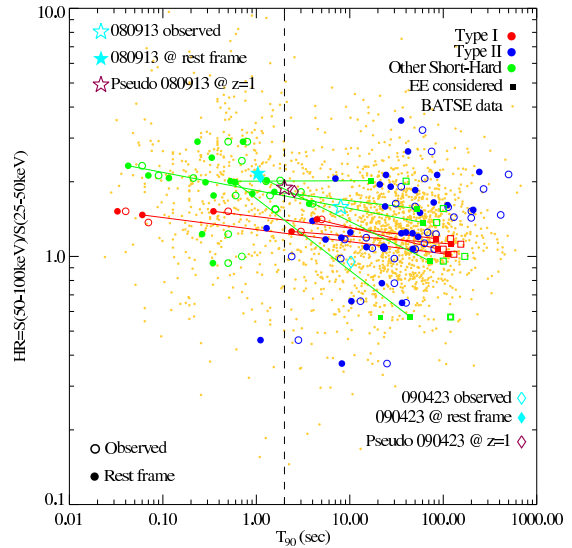}
\caption{The $T_{90} - {\rm HR}$ diagram of GRBs. The 
background orange dots are BATSE GRBs. Overplotted are
Type II Gold Sample (blue), Type I Gold Sample (red),
and other short/hard GRBs (green), mostly detected by Swift. 
Open symbols are for the observed values, while the
filled symbols are the rest-frame values. For short GRBs
with extended emission, those with the short spike only are
denoted as circles, while those including the extended emission
are denoted as squares. The same bursts (with different $T_{90}$
with or without extended emission) are connected by lines.
GRB 080913, GRB 090423, their pseudo counterparts at $z=1$, 
and their rest-frame counterparts are marked with special 
colors/symbols. }
\label{T90-HR}
\end{figure}

Figure \ref{T90-HR} displays the locations of GRB 080913, GRB 090423,
their corresponding pseudo GRBs at $z=1$, and their rest-frame 
counterparts in the traditional 
$T_{90}-$HR (hardness ratio) two-dimensional distribution plane.  
Also plotted are the BATSE GRB sample (orange), the
Gold samples of Type II (blue) and Type I (red) GRBs,
and the Other SGRB Sample (green) (see \S5 for the details
of the sample definitions). It is evident that GRB 080913 and
GRB 090423 would have been recognized as phenomenologically
short/hard GRBs should they have occurred 
at $z \leq 1$.

\section{Phenomenological vs. Physical Classification Schemes: 
Weaknesses and Strengths}

The eventual goal of GRB studies is to identify the physical
origins of every observed GRB, including its progenitor system,
central engine, energy dissipation mechanism, and radiation mechanism.
To achieve this goal, a combination of observations and theoreical
modeling is needed. The number of competive 
models and the allowed parameter space steadily
reduce as more and more observational data are accumulated.
This is evident in the history of GRB studies: while more
than 100 models were proposed before 1992 \citep{nemiroff94},
only two broad categories of progenitor models remain competitive
at the time when this paper is written. A group of GRBs are hosted 
by active star-forming dwarf galaxies \citep{fruchter06}, some of 
which have clear (Type Ic) supernova associations 
\citep{hjorth03,stanek03,campana06,pian06}. 
This points towards
a massive star origin of this group of bursts. At least a few
bursts were discovered to be associated with galaxies with a very
low star forming rate \citep{gehrels05,bloom06,barthelmy05a,berger05},
which point towards a non-massive-star origin of the bursts, likely
due to mergers of compact objects. Therefore it is now justified
to discuss at least two physically distinct categories of GRB
models as well as how to associate a particular burst with either 
category based on certain observational criteria. 

In the literature, some physical classification schemes of GRBs 
have been discussed \citep{zhang07b,bloom08}. 
Strictly speaking, these are not classifications of GRBs,
but are classifications of models that interpret GRB data.
A scientific classification scheme is based on statistical 
formalisms, which
make use of a uniform set of observational data with
instrumental biases properly corrected, and classify 
objects based on statistically significant clustering of
some measured properties. Examples include to classify supernovae
broadly into Type II/I based on whether there are/are not hydrogen
lines in the optical spectrum, and to classify GRBs into
two \citep{kouveliotou93} or three
\citep{mukherjee98,horvath98} classes based on
BATSE $T_{90}$ analyses. The classes defined by the
phenomenological data do not carry physical meanings, and 
theoretical modeling is needed to clarify whether different
phenomenological classes of objects are of different physical
origins. Compared with the SN classification schemes, which 
are based on the ``yes/no'' criteria regarding 
the existence of spectral
lines and therefore are relatively insensitive to the instrumental 
details, the GRB phenomenological classification schemes suffer
another major drawback, i.e. every parameter that one can
directly measure is strongly instrument-dependent. For example,
$T_{90}$ is strongly energy-dependent, and sensitivity-dependent,
so that a ``short'' GRB in a hard energy band would become
a ``long'' GRB in softer bands or if the detector sensitivity
is increased. The membership of a particular GRB to a particular
category (e.g. long vs. short) is not guaranteed. As a result, 
such classification schemes cannot be compared from one mission 
to another, and are of limited scientific value.

A physical classification scheme, on the other hand,
is on theoretical models that interpret the data. As a result,
it suffers the great difficulty of associating a particular
burst to a particular model category. In order to achieve 
the goal, multiple observational criteria are demanded,
but always with non-uniform instrumental selection effects.
Ideally, with infinitely sensitive detectors in all wavelengths,
it may be possible to derive a set of quantitative observational 
criteria that can be used to rigorously associate a
particular GRB to a particular model category based on
statistical properties. However, realistically this is essentially 
impossible since different criteria rely on completely different 
observational instruments with different observational bands and 
sensitivities which are quite non-uniform. Also different criteria
could carry different weights in judging the associated
model category of a particular
burst. The weighting factors of different criteria
are also difficult to quantify. Human
insights rather than pure statistical analyses are needed.
Another drawback of a physical classification scheme
is that it depends on the models, which are subject to further
development as more data are accumulated. The classification 
criteria are therefore also subject to modification based on
data. This can be diminished by invoking model-independent criteria
as much as possible. For example, the Type I/II GRB model
classification scheme discussed in this paper only appeals to 
whether the model invokes a degenerate-star or a massive-star, 
regardless of the concrete progenitor systems or energy dissipation 
and radiation mechanisms (see \S4 for full discussion).

Despite of its weaknesses, a physical classification scheme 
of models and associating a particular object to a particular
model class has
the strength to achieve a better understanding of the physical
origin of astrophysical objects. For example, in the supernova
field, there is now a consensus that only a sub-group of Type I SNe
(Type Ia) has a distinct physical origin, which is related to
explosive disruptions of white dwarfs. The other two sub-types
of Type I SNe (Type Ib/Ic) are more closely related to Type II
SNe and form together a broad physical category of SN models
that invoke massive star core collapses. 
Such a physical classification scheme of SN models 
(massive star origin vs. white dwarf origin) and the efforts
to associate the observed SNe to them reflect
a deeper understanding of the physical origins of SNe.
The same applies to GRBs. The statistical classification
of long-, short- and probably intermediate-duration GRBs has been 
established in the BATSE era. However, it took several
missions and many years of broad-band observations to reveal that 
there are at least two physically distinct types of models
that are associated with these GRBs. Although
data are not abundant enough to unambiguously associate
every individual GRB to these model categories, current data 
already revealed some perplexing observational facts (\S1) that 
demand more serious investigations of the observational criteria 
to judge the physical origin of a particular GRB
(i.e. the physical model associated with this GRB).

In the rest of the paper, we will discuss Type I/II GRBs,
which are defined as the GRBs that are associated with
two distinct physical models. This
is not a new classification scheme of GRBs to replace the
existing long/soft vs. short/hard classification scheme, but is 
a parallel classification of the models that the observed GRBs 
can be associated with based on multiple criteria data analyses.
The two approaches are complementary.
As discussed above, $T_{90}$ is energy-band-dependent and 
sensitivity-dependent, so that the membership of a particular 
GRB to a particular duration category is not always guaranteed. 
On the other hand, if adequate information is retrieved in an 
ideal observational campaign, the association membership
of a particular GRB to a particular physical model category
is almost certain regardless of the detector energy band and
sensitivity. For example, if a SN is detected to be
associated with a GRB, one can safely associate this GRB
to the Type II model category regardless of its $T_{90}$ detected
by different detectors.

\section{Type I/II GRBs, Their Observational Criteria, and Physics Behind}

We reiterate here the definitions of the Type I/II GRBs.
Improving upon the descriptions presented in \cite{zhang07b}, 
we hereby more rigorously define the following:

\begin{itemize}
\item Type I GRBs (or compact star GRBs) are those GRBs
that are associated with the theoretical models invoking
destructive explosions in old-population, degenerate, compact 
stars. The likeliest model candidate is
mergers of two compact stars. 
\item Type II GRBs (or massive star GRBs) are those GRBs
that are associated with the theoretical models invoking
destructive explosions in young-population massive stars. The 
likeliest model candidate is core collapses of massive stars.
\end{itemize}

Here we do not specify the progenitor systems of each  model
type. In reality, there could be multiple possible progenitor 
systems within each model category \citep[see also][]{bloom08}. 
Within the Type I model category, possible
progenitor systems include NS-NS mergers 
\citep{paczynski86,eichler89,narayan92,rosswog03}, NS-BH mergers
\citep{paczynski91,faber06}, and possibly BH-WD or NS-WD mergers
\citep{fryer99,king07}\citep[c.f.][]{narayan01},
see \cite{nakar07,lee07} for reviews.
On the other hand, within the Type II model
category, one may have collapses of single stars (i.e. collapsars,
Woosley 1993; MacFadyen \& Woosley 1999), or collapses of massive 
stars in binary systems \citep{fryer07}.

The definitions of Type I/II GRBs are based on the physical 
models that GRBs can be associated with
rather than their observational properties. The scheme is
therefore intended to be ``operational''. The connections between 
the physical model properties and the observational criteria 
are not straightforward, and probably very difficult for some GRBs. 

In the following, we review a list of observational criteria 
discussed in the literature that may be applied to 
differentiate the two physically distinct model categories
that GRBs can be associated with
\citep[e.g.][]{donaghy06,zhang07b,zhangnature06}, and discuss the
physical justifications of each criterion.
As justified below, some criteria (e.g. \S3.1-\S3.3)
are more directly related to the progenitor system of a GRB. 
On the other hand, the
traditional criteria invoking the observed gamma-ray 
properties (e.g. \S3.4-\S3.7) are more related to radiation
physics and have less direct connection with the progenitor system. 
Some afterglow properties (\S3.8, \S3.9) do carry information
of the progenitor, but theoretical modeling is invoked (and, hence,
less definitive because of the uncertainties inherited in the
models). The statistical properties (\S3.10, \S3.11)
can be related to the progenitor system, but again large 
uncertainties are involved in the identification of the 
explicit progenitor system and its cosmological evolutionary
scenario. The best clue may be gravitational wave signals
(\S3.12). However, they are beyond the current detector 
capability. 

\subsection{Supernova Association}

A positive detection of a supernova (SN) signature associated
with a GRB would undoubtedly establish the association
of the burst with Type II. However, the sample of the robust 
GRB-SN associations
is currently small. Non-detections of a SN signature could be
due to multiple reasons, e.g., the afterglow is too bright
so that the SN light is buried beneath the afterglow level;
the follow up observations were not ``deep'' enough or not
at the right time window; or the lack of an underlying
SN is genuine. Only the last case is helpful to judge the
physical model category of a burst, although the 
conclusion is still not 
clear cut. A genuine SN-less GRB is certainly consistent with 
the Type-I origin. However, it has been discussed in the
literature that some core collapse GRBs may not eject enough
$^{56}$Ni to power a SN 
\citep{woosley93,heger03,nagataki03,nagataki06,tominaga07},
so that the lack of a genuine SN signature may not be evidence 
completely against the Type II origin. On the other hand, we 
notice that the large uncertainties involved in
SN explosion physics prevent the models from having a definite
{\em predictive} power regarding the SN signature. Looking 
back into the history, the predictions of the SN signature 
accompanying GRBs have followed a serpentine (and ironic) path. 
The first core-collapse GRB model was dubbed ``failed supernova'' 
\citep{woosley93}, which predicts no SN signature associated with 
a GRB. Driven by the possible GRB 980425/SN 1998bw association, 
the model was developed to allow a SN associated with a GRB
within the ``collapsar'' scheme. According to \cite{macfadyen99},
the model predicts that ``collapsars will always make supernovae
similar to SN 1998bw''. Indeed the
statement that ``the data and models are consistent with, though not 
conclusive proof of, the hypothesis that ALL long-soft GRBs are
accompanied by SNe of Type Ic'' was made right before the discovery 
of GRB 060614 and GRB 060505 \citep{woosley06}. Would the discovery 
of the SN-less long GRBs (060614 and 060505) then beg for a dichotomy 
of core-collapse GRBs (one group with and another group without 
the SN association)? Although this is certainly plausible,
a simpler picture would be that all genuine SN-less GRBs 
have the Type I origin. In this paper, we take lacking a genuine
SN as a support to the Type I GRB, but do not take this criterion
alone to define a Type I GRB. On the other hand, since there is 
no observational fact that demands the existence of SN-less
Type II GRBs\footnote{In our opinion, GRB 060614 and GRB 060505
are not solid Type II candidates. As will be discussed in \S5
\citep[see also][]{gehrels06,zhang07b}, GRB 060614 is likely a
Type I GRB. GRB 060505 has a much lower overall energetics (including
gamma-ray and afterglow) than most other Type II GRBs.}, 
we do not automatically associate any genuine SN-less long
GRB with the Type II (or Type II candidate) physical model
categories unless there are other strong supports to the scenario 
(see \S4.1 and \S7 for details).

\subsection{Star Forming Rate of Host Galaxy}

Type II GRBs are 
related to massive star deaths, so they must reside in host
galaxies with active star formation. So star forming rate (SFR),
or more rigorously, specific star forming rate (SSFR, i.e. SFR
per unit mass) of the host galaxy is a critical parameter to
judge the membership of Type II GRBs. 

On the other hand, compact star mergers can occur in 
host galaxies both with and without active star formation 
\citep{belczynski06,zheng07}. If we see a GRB residing in an 
elliptical or an early 
type galaxy, we are more confident that no 
massive star is involved in the event, and that the burst 
should be associated with Type I. Those Type I GRBs 
residing in star-forming galaxies are more difficult to identify.
Due to the additional time delay required for the two compact 
objects to coalesce, a Type I GRB site is expected to be more 
aged than the site with active star formation.  As a result,
at least some Type I GRBs should preferentially reside in 
the regions with relatively low SSFR in the star forming 
host galaxy. On the other hand, there are channels of fast 
mergers \citep{belczynski06} that lead to almost ``prompt'' 
mergers of compact stars. In such a case, Type I GRBs can 
reside in high SFR regions within star-forming galaxies. 

\subsection{Position Offset with Respect to Host Galaxy}

A related criterion is the offset of the GRB location with
respect to the center of the host galaxy. The physical motivation
is that Type I GRBs invoke mergers of binaries including at least 
one NS, which likely received a ``kick'' at birth so that the 
binary system would migrate from its original birth location. 
By the time when the two compact stars coalesce, the system 
should have a large offset from the galaxy center or even 
be outside of the host galaxy \citep{bloom99}. Indeed several Gold
Sample Type I GRBs show such a property 
\citep{gehrels05,bloom06,fox05,berger05,barthelmy05a,troja08}. 
On the other hand, Type II GRBs explode right at the location where 
the progenitor stars are formed and therefore should be in the star 
forming regions inside the host galaxy \citep{bloom02b}. 
This is in general consistent with the observations of long GRBs
\citep{fruchter06}. Outliers do exist. For example, GRB 070125 
is a long GRB whose birth location is in a galactic halo 
\citep{cenko08}. 

Complications arise if a GRB is found not inside any galaxy.
It is difficult to judge whether a GRB is ``kicked'' 
out from a nearby host galaxy whose projected image is near the
location of the GRB, or it is associated with a more distant galaxy 
at high-$z$. This problem arises for a good fraction of short/hard 
GRBs detected in the Swift era. For example, GRB 060502B was suggested
by \cite{bloom07} to be associated with a nearby galaxy at $z=0.287$
(with a large angular offset),
while it is included by \cite{berger07} as one of the high-$z$ 
missing-host short/hard GRBs. 

\subsection{Duration}

Theoretically, we do not know exactly which time scale 
defines the GRB duration. In principle there are three time scales 
that are relevant. The first one is the duration of the
central engine activity $t_{engine}$. This corresponds to the
accretion time scale of an accretion-powered central engine model
(usually invoking a black hole - torus system), or the 
spindown time scale of a spindown-powered central engine model
(usually invoking a rapidly rotating millisecond magnetar or a
maximally rotating black hole whose spin energy is tapped via 
a magnetic torque through the Blandford-Znajek mechanism). 
The second time scale is the time scale $t_{jet}$ during which a 
relativistic jet is launched. 
In principle there could be epochs during
which a jet is launched, but it is not relativistic or not
relativistic enough to power the observed gamma-ray emission.
The third time scale is energy dissipation time scale $t_{dis}$.
Current Swift observations suggest that the GRB prompt emission
is ``internal'' \citep{zhang06}. This requires that the relativistic
jet dissipates energy internally before being decelerated by the
external circumburst medium. The dissipation could be via internal 
shocks \citep{rees94} or magnetic reconnection 
\citep{usov92,thompson94}. In principle, one can have an 
active central engine without launching a relativistic jet,
or have a relativistic jet without significant dissipation. 
In general, 
the observed GRB duration $T_{90}$ (which also depends on
the energy band and the sensitivity limit of the detector) should
satisfy\footnote{Here we have assumed
that $T_{90}$ records the GRB internal emission only. This is
true for most cases. Occasionally the observed prompt emission
may also include the emission from the external shocks. $T_{90}$ 
should be removed from Eq.(\ref{ts}) for these cases.}
\begin{equation}
T_{90} \leq t_{dis} \leq t_{jet} \leq t_{engine}~.
\label{ts}
\end{equation}
In most studies, however,
$T_{90} \sim t_{engine} \sim t_{jet} \sim t_{dis}$ 
has been assumed. 

If $T_{90}$ is equal to or at least is proportional to $t_{engine}$,
as is assumed by most central engine modelers, then the duration
information may be tied to the progenitor properties of GRBs.
In particular, Type II GRB progenitors have a massive envelope, 
which can power a long-duration GRB through accretion. 
According to the collapsar
scenario \citep{macfadyen99}, the duration of the burst is defined
by the envelope fallback time scale, which is typically 10s of 
seconds. The model therefore suggests that Type II GRBs should
typically have long durations.
On the other hand, NS-NS and NS-BH mergers typically 
have an accretion time scale $\sim$ 0.01-0.1 s \citep{aloy05}
if the central engine is a BH-torus system. Therefore Type I
GRBs should typically have short durations. Indeed, a 1-second
duration burst is already too long to be accommodated within
the simple merger scenarios. One needs to introduce additional 
ingredients (e.g. an intermediate neutron star phase) to increase 
the duration \citep[e.g.][]{rosswog03}.

This physically-motivated clear dichotomy was broken in the 
Swift era. Swift discovered that
X-ray flares prevail in more than half GRBs, in both long and
short duration ones \citep{burrows05,chincarini07,falcone07}. This
suggests that the GRB central engine activity is not limited to the
prompt phase, and is much longer than $T_{90}$ in both long and
short GRBs \citep{zhang06,fanwei05,lazzati07}. The progenitor
and central engine models must then be modified to invoke a much
longer accretion time scale \citep{king05,perna06,proga06},
or a non-BH-torus central engine \citep{dai06,staff07}.
More importantly, several strong Type I GRB candidates (e.g. 
GRB 050724) are not short, but have
softer, extended emission \citep{villasenor05,barthelmy05a,norris06}.
The merger models therefore must be modified to account for
this extended emission \citep{rosswog07}. Type I GRBs 
no longer must be ``short''.

The discussion above only applies to the case when the line of sight
pierces into the relativistic jet, i.e. the on-beam geometry.  In
this case, the observed time scales reflect the time scales at the central
engine \citep{kobayashi97}. In the case of an off-beam geometry, i.e.
the jet with opening angle $\theta_j$ is beaming towards an angle
$\theta_v > \theta_j$ with respect to the line of sight, the observed 
time scale no longer traces that of the central engine. For a 
discrete pulse, if the pulse duration solely reflects the duration
of the emission powered by the central engine, 
i.e. the rising and falling of
the lightcurve reflects the increase and decrease of the central
engine luminosity (in constrast to those models that interpret the
decaying wing as the high-latitude emission), the observed duration
off beam is related to the on-beam value through the ratio of the 
Doppler factor (given the same comoving value), i.e. 
\begin{equation}
\frac{t({\rm off~beam})}{t({\rm on~beam})} =
\frac{{\cal D}(\theta=0)}{{\cal D}(\theta=\theta_v-\theta_j)} =
\frac{1-\beta\cos(\theta_v-\theta_j)}{1-\beta}~.
\label{D-ratio}
\end{equation}
where the Doppler factor is defined by
\begin{equation}
{\cal D}=\frac{1}{\Gamma(1-\beta\cos\theta)}~,
\label{Doppler}
\end{equation}
and $\theta$ is the angle between the line of sight and the velocity
vector of the ejecta, which is taken as the closest approach to the jet
($\theta_v - \theta_j$). For multiple emission episodes (i.e. 
multiple pulses in the light curve), the time interval of the
quiescent episodes do not vary with the viewing direction. So
Eq.(\ref{D-ratio}) applies to the total duration of a GRB only if
the prompt emission has one single pulse. Also since the observed 
flux is lower for a lower ${\cal D}$, given a same detector
sensitivity, the off-beam $T_{90}$ tends to be shorter than that 
predicted by Eq.(2) due to the limiting flux threshold effect.

The off-beam model predicts that the
afterglow light curve should display a rising behavior initially
before the $1/\Gamma$ beam enters the line of sight 
\citep{panaitescu99,granot02}.
Broadband observations of the majority of GRB afterglows do not show 
such a signature. So the off-beam geometry, if any, is rare. 

\subsection{Hardness}

The connection between the hardness of spectrum and the GRB 
progenitor is less direct. 
It is related to the unknown internal energy 
dissipation mechanism and radiation mechanism, which in turn
depends also on the composition of the GRB ejecta. GRB spectra
are usually categorized as a smoothly-joined power-law, namely,
the Band-function \citep{band93}. The
hardness of a GRB is likely related to the location of the 
$E_p$, but the flatness of the spectral index below $E_p$ 
may also play a role. Theoretically, the spectral slope 
is more closely related to the particle acceleration 
mechanism \citep[e.g.][]{sironi09} and the ``compactness'' of the
emission region \citep[e.g.][]{peer06}. The spectral peak energy,
$E_p$, can be related to the GRB emission model parameters more 
directly, although model-dependent \citep{zhangmeszaros02c}. 
We will mainly discuss the $E_p$ models more closely in the following.

In general, $E_p$ is a function of the burst luminosity $L$, 
the Lorentz factor $\Gamma$ of the ejecta, and the radius $R$ of the 
emission site from the central engine. In order to address whether
a GRB is hard or soft, one needs to specify a particular 
emission model. In the following we discuss three internal
emission models currently discussed in the literature.

{\bf Internal shock model.} Within this model, the gamma-ray $E_p$
can be defined either by synchrotron radiation or synchrotron self-Compton
(SSC). In general one can write $E_p \sim \Gamma \hbar \gamma_e^k
(e B'/mc)$, where $k=(2,4)$ for synchrotron and SSC, respectively.
Since the comoving magnetic field strength in the ejecta flow 
satisfies $B' \propto L^{1/2}R^{-1}\Gamma^{-1}$ 
for both the ordered and the random magnetic field components 
\citep{zhangmeszaros02c}, one has 
\begin{equation}
E_p^{\rm IS}\propto \gamma_e^k L^{1/2} R^{-1} (1+z)^{-1}\propto  
\gamma_e^k L^{1/2} \Gamma^{-2} \delta t^{-1} (1+z)^{-1}~,
\label{Ep-is}
\end{equation}
where $L$ is the initial kinetic luminosity of the ejecta,
$\delta t$ is the variability time scale of the unsteady
GRB ejecta wind, and the internal
shock radius is $R \sim \Gamma^2 \delta t$. Note that $E_p$ is 
negatively correlated with the bulk Lorentz factor ($\propto 
\Gamma^{-2}$), which is contrary to the intuition that high
$\Gamma$ bursts should be hard. Here $\gamma_e$ is the 
characteristic Lorentz factor of the electrons that contribute
to the emission at $E_p$. Under the fast cooling condition, which is
generally satisfied for internal shocks, $\gamma_e$ corresponds
to the minimum ``injection'' energy of the electrons, which is
related to the ``relative'' Lorentz factor between the two
colliding shells $\Gamma_{fs}$, i.e. $\gamma_e \propto \Gamma_{fs} 
\sim (\Gamma_f /\Gamma_s + \Gamma_s/\Gamma_f) / 2$, where $\Gamma_f$ 
and $\Gamma_s$ are the Lorentz factors of the fast and slow shells,
respectively. If the $\Gamma$ variation of a flow is proportional
to the average Lorentz factor $\bar\Gamma$, i.e. $\Delta \Gamma \propto 
\bar\Gamma$ or $\Gamma_f/\Gamma_s \sim $ const, then
$\gamma_e$ essentially does not depend on $\bar\Gamma$, so that a 
higher $E_p$ should correspond to a lower $\Gamma$. On the other
hand, it is possible that high-$\bar\Gamma$ flows may be more variable,
e.g. $\Gamma_f / \Gamma_s \propto \bar\Gamma$. If this is the case,
then the negative dependence on $\Gamma$ in Eq.(\ref{Ep-is})
is canceled out (for $k=2$) or reversed (for $k=4$). In the
traditional internal shock model, 
the variability time scale $\delta t$ of the ejecta can be 
derived from the observation. Analyses of power density
spectra of GRB light curves \citep{beloborodov98} suggest
that the GRB temporal behavior may be self-similar, i.e. lacking a
characteristic time scale. In the past, the minimum variability time
scale, which can be as small as milliseconds for both short and some
long duration GRBs, has been adopted to estimate the internal shock 
radius. Alternatively, it is possible that the rapid variability
in GRB light curves may be caused by other mechanisms, such as
relativistic turbulence inside
the emission region \citep{narayan09}. Within this latter scenario, 
the outflow variability time scale relevant to internal shocks
can be much longer. Physically,
Type I GRB outflows may directly carry the variability information
from the inner central engine, i.e. the dynamic time scale of the 
innermost accretion torus around the black hole, $\delta t \sim t_{dyn}
\sim 12\sqrt{3}\pi (GM_{bh}/c^3) \sim 1 (M_{bh}/3M_\odot)$ ms (where 
$M_{bh}$ is the mass of the black hole), or the spin period of the
central magnetar or black hole, $\delta t \sim P_{engine} \sim 1$ ms. 
On the other hand, a Type II jet needs to pentrate through the 
heavy stellar envelope so that the initial temporal information
from the inner central engine may be smeared out and regulated.
The observed variability time scale may be related to that of
fluid instabilities, and therefore could be much longer.
If abundant pairs are produced, it has been argued that the pair 
photosphere would effectively screen out the
variability time scales smaller than a critical value
\citep{kobayashi02,meszaros02b}. This is more relevant to
low-$\Gamma$ events for which the internal shock radii are
below the pair photosphere. 

With all these complications in mind, one may compare the expected $E_p$
for Type I and Type II GRBs based on Eq.(\ref{Ep-is}). On average,
Type I GRBs have an isotropic gamma-ray luminosity
$L$ 2-3 orders of magnitude smaller than that of Type II GRBs
(see the theoretical argument in \S3.8, and the observational
data in \S4 and Table 1 below).
On the other hand, $\delta t$ of Type I may be smaller 
than that of Type II by 2-3 orders of magnitude. This gives
\begin{equation}
\frac{E_p^{\rm IS}({\rm I})}{E_p^{\rm IS}({\rm II})} \sim
(10 - 30) \frac{[\gamma_e^{k}\Gamma^{-2}(1+z)^{-1}]({\rm I})}
{[\gamma_e^{k}\Gamma^{-2}(1+z)^{-1}]({\rm II})}~.
\end{equation}
This suggests that in a large parameter space Type I GRBs can be 
harder than Type II GRBs. If $\gamma_e$ is similar for both types,
Type I GRBs can be harder than Type II GRBs as long as their bulk
Lorentz factors are not larger than those of Type II by a factor
more than $(3-5)$ times.  If $\gamma_e \propto \Gamma$, Type I
are generally harder than Type II for both the synchrotron model 
($k=2$, regardless of the value of $\Gamma$), and the SSC model
($k=4$, the $E_p$ ratio is positively dependent on $\Gamma$.
Theoretically, Type I GRBs should have higher $\Gamma$'s due to 
their less baryon loading as compared with Type II GRBs. This
favors a harder spectrum of Type I even more for the SSC model.
A systematically smaller redshift $z$ for Type I GRBs (due to the
merger delay with respect to star formation) also helps to increase
the hardness contrast between the two types.
In reality, there are large dispersions in $L$, $\delta t$, $\Gamma$,
$\gamma_e$ and $z$ in both types. On the other hand, the 
HR distribution of the BATSE short/hard vs. long/soft dichotomy
also shows a large dispersion (Fig.\ref{T90-HR}). In general,
the statement that Type I GRBs are harder than Type II GRBs 
can be made within the internal shock models in the statistical
sense. For individual bursts, one cannot draw a firm conclusion
regarding the hardness of a particular burst due to the large 
uncertainties involved in the parameters.

{\bf Photosphere model.} The possibility that the observed GRB 
emission has a dominant contribution from
the fireball photosphere \citep{thompson94,meszarosrees00,meszaros02b}
has gained increasing attention recently
\citep{rees05,ryde05,peer07,thompson07,ioka07,ghisellini07c,ryde09,lazzati09}.
In this model, $E_p$ is related to the observed photosphere
temperature $T_{ph}$. For a ``naked'' fireball, i.e., a fireball
expanding into a vacuum, the observed photosphere temperature
(and hence $E_p$) depends on whether the photosphere radius
$R_{ph}$ is below or above the fireball coasting radius $R_{c}$.
For a large dimensionless entropy of the fireball $\eta \geq
\eta_{c2} \sim 10^4 [L_{52} R_{0,7}^{-1}]^{1/3}$
(where $R_0$ is the initial radius of the fireball. Throughout the
text the convention $Q_n=Q/10^n$ is adopted in cgs 
units.)\footnote{This critical entropy is derived 
\citep{zhangmeszaros02c} within the discrete shell regime, 
rather than the continuous wind regime 
\citep{meszarosrees00,meszaros02b}. This is usually justified, 
since typically one has $\eta > \eta_{c1}$ in this regime.}, 
the fireball becomes transparent during the acceleration phase
(i.e. $R_{ph} < R_c$). The observed fireball temperature is
essentially the temperature at the central engine, i.e.
$T_{ph} \sim T_0$, so that
\begin{equation}
E_p^{ph,1} \sim kT_0(1+z)^{-1} \sim L^{1/4} R_0^{-1/2}(1+z)^{-1}~.
\label{Ep-ph1}
\end{equation}
This is the regime discussed in most photosphere models 
\citep{thompson94,meszarosrees00,ryde05}. On the other hand, if the fireball
becomes transparent beyond the coasting radius ($R_{ph} > R_c$), 
the photosphere temperature drops with radius due to the decrease
of residual internal energy during the expansion, so that
$T_{ph} = T_0 (R_c/R_{ph})^{2/3}$ \citep{meszarosrees00}.
The detailed parameter dependences are related to whether the opacity
is defined by a discrete shell or a continuous outflow wind
\citep{meszaros02b}. For the former ($\eta_{c1} < \eta
< \eta_{c2}$, where $\eta_{c1}\sim 250 [L_{52}
R_{0,7}^{-1}]^{1/5}$), one has \citep{zhangmeszaros02c}
\begin{equation}
E_p^{ph,2} \sim L^{-1/12} R_0^{-1/6} \Gamma (1+z)^{-1}~.
\label{Ep-ph2}
\end{equation}
For the latter ($\eta < \eta_{c1}$), one has 
\begin{equation}
E_p^{ph,3} \sim L^{-5/12} R_0^{1/6} \Gamma^{8/3} (1+z)^{-1}~.
\label{Ep-ph3}
\end{equation}
If additional energy dissipation occurs at small radii, pair
production can occur which enhances photon opacity and increases
the photosphere radius \citep{meszaros02b,rees05}.
We note that the ``naked'' fireball scenario is more relevant
to Type I GRBs.

With the presence of a stellar envelope, the photosphere emission
of a Type II GRB is likely modified.  Due to
continuous energy dissipation and heating inside the envelope, 
the jet cannot reach 
the maximum Lorentz factor but instead stores a significant energy
in heat before erupting out from the envelope. Effectively, the
GRB fireball ``central engine'' is moved from the central black hole
or magnetar to the location slightly below the stellar envelope 
\citep{thompson06,thompson07,ghisellini07c}.
The jet at this radius $R\sim R_*$ has a moderate Lorentz 
factor $\Gamma_*$ and a comoving temperature $T'_* \sim
(L/4\pi \Gamma_*^2 R^2 \sigma)^{1/4}$, and an observer frame
temperature $T_{*}=\Gamma_* T'_*$. As the jet erupts out from
the envelope, it will undergo rapid acceleration under its own
thermal pressure. If $R_*$ is greater than photosphere radius for 
a naked central engine, the fireball
would become transparent shortly after exiting the star 
due to the rapid fall of density. So the real photosphere
radius is essentially $R_{ph} \sim R_*$, and $R_{ph} < R_c$ is 
always guaranteed. The peak energy $E_p$ is defined by 
$T_{ph}=T_{*}$. This leads to a variation of Eq.(\ref{Ep-ph1}) in 
the form of
\begin{equation}
E_p^{ph,1^{'}} \sim L^{1/4} \Gamma_*^{1/2}R_*^{-1/2}(1+z)^{-1}~.
\label{Ep-ph1'}
\end{equation}

Within the photosphere models, it is not obvious why Type I
GRBs should be systematically harder than Type II GRBs. The
trend, if any, should be opposite. The logic is the following.
First, given the same parameters of $L$, $R_0$ and $z$, one
typically has $E_p^{ph,1}>E_p^{ph,2}>E_p^{ph,3}$ 
\citep[e.g. Eq.(24) of][]{zhangmeszaros02c}. Next, the stellar
envelope effectively ``raises'' the photosphere for Type II
GRBs, so Eqs.(\ref{Ep-ph2}) and (\ref{Ep-ph3}) are usually 
not relevant. One therefore may only compare the case of 
Eq.(\ref{Ep-ph1}) for the two types of GRB, 
since Eq.(\ref{Ep-ph1'}) can be related to Eq.(\ref{Ep-ph1})
through $R_0 = R_* / \Gamma_*$. Equation (\ref{Ep-ph1})
suggests that Type I GRBs, typically with smaller $L$, 
should be softer than Type II GRBs at the same redshift.
A smaller $z$ for Type I GRBs may compensate their softness,
but in general it is not straightforward to claim that Type
I GRBs should be systematically harder than Type II GRBs
for the photosphere model.
Pairs may lower the photosphere temperatures of some high-$L$
GRBs (especially for Type II), which may help to account for
the observed trend, but no handy analytical formula is 
available to perform direct comparisons.

The recent Fermi-detected GRB 080916C \citep{abdo09}
showed a series of featureless Band-function spectra. 
The expected photosphere emission component is missing,
suggesting a Poynting flux dominated flow at the base of
the central engine \citep{zhangpeer09}. At least for this
burst, the observed $E_p$ is not the thermal peak of the
photosphere emission. 

{\bf Magnetic dissipation model.} Finally, if the GRB
outflow is Poynting flux dominated 
\citep{usov92,meszarosrees97b,spruit01,lyutikov03,liangedison07},
the characteristic frequency of emission would take a different
form and have different dependences on the ejecta parameters.
The locations of the magnetic reconnection regions are unknown. 
If dissipation occurs at small radii, the effect is to
modify the photosphere emission through continuous heating
\citep{giannios08}. This is effectively a photosphere model,
which has been discussed above. 
Alternatively, a Poynting-flux-dominated outflow 
can reach a global dissipation at a large radius where
the MHD approximation is broken 
\citep{usov94,spruit01,zhangmeszaros02c,lyutikov03}.

Lacking a macroscopic reconnection model, $E_p$ in the reconnection
model is difficult to calculate. Under different assumptions, the
expression of $E_p$ may take different forms. For example, in a
random electric/magnetic field in the magnetic reconnection region, 
electron acceleration may be balanced by radiation cooling. The 
typical electron Lorentz factor is therefore $\gamma_e \propto 
B^{-1/2}$. The synchrotron peak energy may be then expressed in 
the form of 
\citep{zhangmeszaros02c}
\begin{equation}
E_p^{mag} \propto \Gamma (1+z)^{-1}~,
\label{Ep-mag}
\end{equation}
which depends on $\Gamma$ and $z$ only. Type I GRBs can then be
harder than Type II GRBs, again because Type I GRBs tend
to have a cleaner environment, and hence, less baryon
loading, than Type II GRBs. 

Similar to the duration discussion (\S4.5), the above discussion 
applies to the on-beam geometry.
For an off-beam geometry, the observed $E_p$ is smaller by
a factor of the Doppler factor ratio (Eq.[\ref{D-ratio}]).
This effect has been discussed by, e.g. \cite{yamazaki04}.

\subsection{Spectral Lag}

Soft GRB emission usually arrives later than hard emission in some 
GRBs. This ``spectral lag'' is evident for long-duration GRBs
\citep{norris00,gehrels06,liang06b}, but is typically negligible
for short-duration GRBs \citep{norris06,yi06}. Usually, the lag
may be visualized as the time differences of the peaks of the
``same'' pulse in different energy bands. Statistically it 
can be derived through a cross-correlation analysis of the
pulse profiles in different energy bands \citep{norris00}.
Technically, what is usually measured is the lag time 
$\Delta t$ between two BATSE (or Swift BAT) bands 
$E$ and $E+\Delta E$. Mathematically, this
corresponds to $|\int_E^{E+\Delta E} (dt/dE) dE|$. 
It is therefore important to study $dt/dE$ (or $dE/dt$) in
theoretical models.

Theoretically, the leading model of the spectral lag is the
``kinetic'' effect, i.e. the delay is due to the fact that the observer
is looking at the increasing latitudes with respect to the line of sight 
with time \citep[e.g.][]{salmonson00,ioka01,norris06,shen05,lu06}. One
can derive the spectral lag within this model as follows. Since
the cooling time scale in the GRB emission region is very short,
one may assume that the decay of GRB pulses is dominated by
this high-latitude ``curvature'' effect. The comoving emissivity
is assumed to be uniform everywhere across the conical jet.
Softer emission comes from higher latitudes (due to their smaller
Doppler factor) and therefore is
delayed by a time $t \sim (1+z)(R_{\rm GRB}/c) (1-\cos\theta)$
with respect to the emission from the line of sight, 
where $\theta$ is the angle from the line of sight. The observed
photon energy is related to the comoving one via $E = {\cal D}
E'$, where ${\cal D}$ is the
Doppler factor [Eq.(\ref{Doppler})]. One therefore has
\begin{eqnarray}
\frac{dE}{dt} & = & \frac{cE'}{R_{\rm GRB}(1+z)} 
\frac{d{\cal D}}{d(-\cos\theta)} \nonumber \\
& = & -\frac{cE'\beta}{(1+z)R_{\rm GRB}\Gamma 
(1-\beta\cos\theta)^2}~.
\label{lag0}
\end{eqnarray}
The negative sign denotes ``lag'', i.e. increasing $E$ (harder)
corresponds to a decreasing arrival time $t$ (earlier).
When $\theta \rightarrow 0$ (close to the line of sight),
this is simplified as
\begin{eqnarray}
\frac{dE}{dt} & = & -\frac{4cE'\Gamma^3}{(1+z)R_{\rm GRB}}
\nonumber \\
& = & -\frac{2E'}{(1+z)\tau} \Gamma = -\frac{2}{1+z}
\frac{E}{\tau}~,
\label{lag}
\end{eqnarray}
where 
\begin{equation}
\tau = \frac{R_{\rm GRB}}{2 \Gamma^2 c}
\label{tau}
\end{equation} 
is the angular spreading time, which could be related to
the observed half width (in the decaying wing) of the GRB 
pulse\footnote{In principle, $\tau$ is the upper limit
of the observed half width in the decaying wing, since part
of the tail may be buried beneath the next rising pulse.}.
Notice that $R_{\rm GRB}$ is a value one cannot directly
measure, therefore in the above expression it needs to be
combined with $\Gamma^2$ to derive $\tau$, leaving only
one power in the $\Gamma$-dependence in Eq.(\ref{lag}).
Another comment for this expression is that $dE/dt$ is
$E$-dependent, i.e. for the same GRB pulse, the lag between
$E_1$ and $(E_1+\Delta E)$ should be different from that
between $E_2$ and $(E_2+\Delta E)$. This can be understood
with Eq.(\ref{lag0}) by noticing that different $E$
corresponds to different ${\cal D}$, and hence, different
$\theta$. For $\theta > 0$, $dE/dt$ takes a different form
than Eq.(\ref{lag}), which is valid for the hardest (on
axis) pulse.

One can immediately draw the following inference from
Eq.(\ref{lag}). Since $dE/dt \sim E/\tau$,
one can get the lag $\Delta t \sim \tau$
if one takes $\Delta E \sim E$. This is to say, the lag 
time is comparable to the pulse width itself. 
So spectral lags do not carry direct information of the
progenitor. On the other hand, since pulse width can be
related to variability time scale, which may be
related to the physical types (\S3.5), one may speculate
the expected spectral lags of the two types by the 
following way. Type I GRBs have naked central engines,
so their pulse widths $\tau$ are typically much smaller than
those of Type II GRBs, whose variability time scales are
longer due to the additional modulation of the stellar
envelope. One therefore may expect that Type I GRBs
have shorter lags than Type II GRBs. This is consistent
with the fact that short duration GRBs (preferentially Type
I) have negligible lags, while long duration GRBs
(preferentially Type II) have long lags.

Another commonly discussed argument is that Type I GRBs
may have larger $\Gamma$'s than Type II GRBs (due to a
``cleaner'' environment with less baryon loading), and 
that this might be the origin of short lags 
\citep[e.g.][]{norris06}. According to Eq.(\ref{lag}),
one can argue\footnote{Notice that this is different from
\cite{norris06} who argued $\Delta t \propto \Gamma^{-1/2}$
based on the expression of the angular spreading time
rather than based on the differential property $dE/dt$
as discussed in this paper.} 
\begin{equation}
\left |\frac{dt}{dE} \right| \propto \frac{1}{\Gamma}~.
\label{lag1}
\end{equation}
if different bursts have similar $\tau$ and $E'$
\citep[see also][]{shen05}.
However, the two assumptions (same $\tau$ and $E'$) 
lack physical justifications.
In particular, $E'$ depends on the dissipation mechanism
and the properties (e.g. $B$ field strength) in the 
dissipation region, which depends on the burst parameters
such as $L$, $\delta t$, etc. Although the central engine
variability time scales ($\tau$) may be arguably similar
within the Type I or Type II category, respectively, 
they are considerably different between the two types. 
We therefore conclude that the
Lorentz factor argument is not robust. Type I GRBs can 
have larger $\Gamma$'s, but it is not the main reason
for their short spectral lags. 

A major issue of such a kinetic (high-latitude curvature effect) 
model is that the peak flux of the pulse drops rapidly with angle 
\citep{fenimore96,kumar00}, so that the flux is expected to be 
too low in softer bands to interpret the 
observed flux. None of the previous kinetic modelers
\citep{salmonson00,ioka01,shen05,lu06} have seriously
confronted the flux predictions with the data (although the
timing data have been well interpreted by the models).
One way to increase the high-latitude flux is to invoke a 
non-power-law instantaneous spectrum at the
end of internal emission (e.g. at the shock crossing time
in the internal shock model), e.g. 
a power law with exponential cutoff \citep{zhangbb09}
or a Band function \citep{qin09}. The curvature effect of these
models predicts that the spectral peak sweeps across different 
energy bands, making the flux not drop as rapidly as in the
case of a power law spectrum. Indeed, GRB 060218 can be modeled 
by an evolving cutoff power law \citep{campana06}, and GRB 
050814 can be modeled by the curvature effect model 
invoking a cutoff power law spectrum \citep{zhangbb09}. 
However, the light curve for a given band is always 
a decay function unless the spectral index before $E_p$
is much flatter than -1. This is not supported by the spectral
data of most GRBs. One is then obliged to abandon
the hypothesis of a uniform jet. A structured jet with a less
energy and/or a lower Lorentz factor at large angles from the jet 
axis and with the line of sight piercing into the wing of the 
structured jet \citep{zhangmeszaros02b,rossi02,zhang04} 
can be invoked to account for the observed spectral lag data. 

In reality,
there might be additional mechanisms that are related to
the observed spectral lags of GRBs, but the
kinetic effect must exist, and may play the dominant role
to define spectral lags in most GRBs.

\subsection{Statistical Correlations}

Observationally, some empirical correlations among several
observed quantities have been claimed (see e.g. 
Zhang 2007 for a summary). Most of these correlations
were discovered for long/soft GRBs. Here we discuss two
of them that are potentially related to Type I/II diversity.

{\bf Amati (Yonetoku) relation.} Statistically, more energetic
long GRBs are harder. This is usually expressed in terms
of $E_p (1+z) \propto E_{\gamma,iso}^{1/2}$ \citep{amati02}
and $E_p (1+z) \propto L_{\gamma,iso}^{1/2}$ \citep{yonetoku04}.
The dispersions of the correlations are large, and outliers do
exist (e.g. the nearby GRB 980425/SN 1998bw is an outlier
of the Amati-relation). It has been
argued that the observed correlations are solely due to some
selection effects \citep{nakarpiran05,band05,butler07}.
However, the fact that the correlations are valid for most 
$z$-known GRBs which cover five decades in isotropic energy 
\citep{sakamoto06} suggest that there is likely underlying
physics that drives such correlations.

Inspecting the expressions of $E_p$ in various GRB prompt
emission models discussed in \S3.5 
\citep[see also][]{zhangmeszaros02c}, one can see that 
although $E_p$ is indeed generally a function of $L$, 
it is usually also a function of other parameters, in particular,
the bulk Lorentz factor $\Gamma$
which is usually not directly measured. In order to interpret
the Amati/Yonetoku relations, one needs to introduce a rough
dependence between $\Gamma$ and $L$. We now again discuss 
the three prompt emission models and investigate how the
correlations may be interpreted within each model.
\begin{itemize}
\item For the internal shock models, the Amati/Yonetoku relations
require that $\gamma_e^k r^{-1}$ is roughly constant. If $\gamma_e$
is roughly constant, then the internal shock radius should be similar
for different bursts. 
This suggests that $\Gamma$ is essentially
independent of $L$ (since the variability time scale may be similar
among Type II GRBs), a requirement not immediately evident based
on physical arguments. Alternatively, if $\gamma_e \propto \Gamma$,
the relation can be naturally satisfied for the synchrotron model
($k=2$). This suggests that high-$\bar \Gamma$ GRBs are more variable.
In other words, while the maximum Lorentz factor of the outflows
$\Gamma_M$ can vary from burst to burst, the minimum Lorentz factors 
$\Gamma_m$ for different bursts are similar to each other, so
that $\gamma_e \propto (\Gamma_M/\Gamma_m) \propto \bar\Gamma$.
\item Within the photosphere model, the most relevant regime for
Type II GRBs is the one with a stellar envelope (Eq.[\ref{Ep-ph1'}]).
An interpretation of the Amati/Yonetoku relation can be made 
\citep{thompson06,thompson07} by
introducing another {\em assumption} that the total energy of
different GRB jets is quasi-universal, a conclusion reached in
the pre-Swift era \citep{frail01,bloom03}. Recent Swift
observations suggest that the ``achromatic'' behavior, the
demanded characteristic of a jet break, is not commonly seen
\citep{liang08}. This raises the issue of interpreting some afterglow
temporal breaks as jet breaks. Furthermore, the inferred total
energies (after beaming correction) are found to a have larger 
scatter than the pre-Swift sample \citep{liang08,kocevski08,racusin09}. 
In any case, if one  believes $E_\gamma =
E_{\gamma,iso}\theta_j^2 \sim$ const, using the argument that
baryon sheath near the breakout radius leads to
$\Gamma_* \sim \theta_j^{-1} \propto E_{\gamma,iso}^{1/2}$ 
\citep{thompson06,thompson07},
one can translate Eq.(\ref{Ep-ph1'}) into $E_p \propto
L_{\gamma,iso}^{1/2} \propto E_{\gamma,iso}^{1/2}$ by taking
the trivial proportionality 
$L \propto L_{\gamma,iso} \propto E_{\gamma,iso}$.
The first proportionality is based on the fact that the GRB 
efficiency is not a function of $L_{\gamma,iso}$ 
\citep{lloydronningzhang04};
while the second proportionality is based on the fact that the
dispersion of $T_{90}$ is not large and that $T_{90}$ is not
correlated to $L_{\gamma,iso}$. In view of the fact that GRB 080916C
disfavors the photosphere origin of $E_p$ \citep{zhangpeer09},
we regard this interpretation as no longer attractive. 
\item The magnetic dissipation model lacks a robust prediction for
$E_p$. In any case, similar to the other two models, the 
Amati/Yonetoku relation can be satisfied if one assigns a particular
$\Gamma-L$ correlation. For example, the specific model described
in Eq.(\ref{Ep-mag}) requires $\Gamma \propto L^{1/2}$. 
\end{itemize}

When the optical afterglow light curve temporal break ($t_{opt}$) is
included, a tighter correlation involving $E_p$ and $E_{\gamma,iso}$
is obtained \citep{ghirlanda04,liangzhang05} for some long
duration GRBs. However, the physical connections between the prompt 
emission properties ($E_p$ and $E_{\gamma,iso}$) and the afterglow 
properties ($t_{opt}$) or the global collimation degree of the jet 
are not straightforwardly expected. Furthermore, optical afterglow
temporal break data of Type I GRBs are still poor to draw any
conclusion. We therefore do not discuss these relations
in this paper.

{\bf Luminosity-lag relation.} \cite{norris00} discovered
a relation between the gamma-ray peak luminosity $L_{\gamma,iso}^p$
and the spectral lag $\Delta t$ between the BATSE Channel 1
(25-50 keV) and Channel 3 (100-300 keV), i.e. $L_{\gamma,iso}^p \propto
(\Delta t)^{-1.15}$. The relation was refined by \cite{gehrels06}
who corrected the observed spectral lag to that between two common bands
in the cosmic proper rest frame of GRBs. This correction takes two 
steps. First, the observed spectral lag $\Delta t$ between the bands 
from $E$ to $(E+\Delta E)$ can be expressed as $\Delta t=\Delta 
t_{rest}(E_{rest}) (1+z)$, where $\Delta t_{rest}(E_{rest})$ is 
the spectral lag between $E_{rest}$ to $(E_{rest}+\Delta E_{rest})$ 
in the cosmic proper rest frame. Second, what one cares about
is the intrinsic spectral lag between a common rest frame energy
interval, e.g. from $E$ to $(E+\Delta E)$. One needs an additional
relation between $\Delta t_{rest}(E)$ and $\Delta
t_{rest}(E_{rest})$. There is no universal relation for this,
but there is an empirical relation between pulse width and energy,
i.e. the pulses become narrower with energy with $w\propto E^{-a}$
with $a\sim (0.3-0.4)$ \citep{fenimore95,norris05,liang06b,zhangqin08}.
{\em Assuming that the spectral lag is proportional to
the pulse width} \citep[see e.g.][]{norris05}, one may derive
$\Delta t_{rest}(E) = \Delta t_{rest}(E_{rest})(E/E_{rest})^{-a}
=\Delta t_{rest}(E_{rest})(1+z)^a$. This finally gives
\begin{equation}
\Delta t_{rest}(E) = \Delta t(E) (1+z)^{a-1}~.
\end{equation}
In \cite{gehrels06}, $a\sim 1/3$ was adopted, and a correlation
\begin{equation}
L_{\gamma,iso}^p \propto [\Delta t (1+z)^{a-1}]^{-\delta}
\propto \left[\frac{\Delta t}{(1+z)^{2/3}} \right]^{-\delta}
\label{L-lag}
\end{equation} 
is found for a sample of long GRBs, with $\delta \sim 1$. Outliers 
do exist for long GRBs, and short
GRBs are noticeably off the track of the correlation.

Is there an underlying physical mechanism that justifies the observed 
$L_{\gamma,iso}^p \propto (\Delta t_{rest})^{-\delta}$ correlation? 
It is not obvious
based on the theoretical arguments above. According to Eq.(\ref{lag}),
the intrinsic lag is related to the intrinsic variability time
scale of the burst. So a $L-{\rm lag}$ negative relation may be
related to another, probably more intrinsic $L-V$ positive relation,
where $V$ is the variability parameter. Technically, there are
different definitions of the variability parameter 
\citep{fenimore00,reichart01,guidorzi06}, but in any case,
a more variable light curve (high $V$) would have shorter 
variability time scales (corresponding to $\tau$), and hence, 
shorter spectral lags ($\Delta t$). Observationally, indeed
a positive $L-V$ relation is observed, although with a large
scatter \citep{fenimore00,reichart01,guidorzi06}. The interpretation
of this correlation within the internal shock model 
invokes several assumptions: (1) The smallest variability time 
scale is defined by the collisions above the pair photosphere;
(2) The true jet energy is quasi-universal \citep{frail01};
(3) Narrower jets have higher $\Gamma$'s 
\citep{kobayashi02,meszaros02b}\footnote{The third assumption
may be in conflict with the explanation of the Amati/Yonetoku
relation within the standard internal shock model
as discussed in \S4.5. To interpret that relation,
one requires no dependence of $\Gamma$ on $L$ (and hence, on
$\theta$). }.
This is a relevant interpretation to the 
observed $L_{\gamma,iso}^p \propto (\Delta t_{rest})^{-\delta}$
relation. However, in view of the assumptions invoked in the reasoning
(the above three as well as the assumption 
that the $w-E$ correlation is similar 
to $\Delta t-E$ correlation as invoked earlier), we expect that
the correlation should not be very tight, and may not follow
the same simple power law. This is consistent with the
data (see \S5 for details). 

Would Type I GRBs satisfy a similar correlation? In principle
one can expect so if the set of assumptions discussed above
are satisfied. In reality, Type I GRBs may not effectively 
develop a pair photosphere, both because of their preferred 
higher Lorentz factors (clean environment) and because of their 
lower total energy.
This would make the observed variability time scale trace the 
central engine variability time scale, which would not vary
significantly among bursts. 

Another model to interpret the $L_{\gamma,iso}^p \propto 
(\Delta t_{rest})^{-\delta}$ correlation invokes varying
Doppler factors among different bursts \citep{salmonson00,ioka01}. 
These models assume a universal comoving properties of ALL
GRBs, and invoke the off-beam geometry to interpret longer
durations and spectral lags. As already discussed, the 
comoving properties of GRBs depend on many parameters.
The off-beam model is not supported by early afterglow 
observations. We therefore regard these early models invoking 
pure geometrical effects as no longer favorable in view 
of the recent observational progress.

\subsection{Energetics and Beaming}

Type II GRBs are generally expected to be more energetic than Type
I GRBs. In the standard BH-torus central engine model, the total
energy of the burst is positively correlated with the total available
fuel in the torus. Massive stars are much more abundant in
mass, which can reach $\sim 10 M_\odot$ of fuel in total. On the other
hand, a NS-NS merger system has a total energy budget of $\sim 2.8
M_\odot$. After the prompt collapse, the available fuel in the torus
is of order $\sim 0.1 M_\odot$. A BH-NS merger system has even less
fuel to begin with ($\sim 1.4 M_\odot$). It
would reach a similar total energy budget in 
accretion as the NS-NS system. In these models, 
Type I GRBs are expected to be
10-100 times less energetic than Type II GRBs. Alternatively, 
GRBs may be powered by the spin energy of the
central object (a rapidly spinning BH or NS). For the case of a
NS central engine, Type I GRBs may reach similar energies 
as Type II GRBs. For the case of a BH engine whose spin energy
is extracted via a magnetic torque \citep{blandford77,li02}, a Type II
GRB is again expected to be $\sim$ 10 times more energetic than
a Type I GRB if it is powered by a NS-NS merger, again because of 
the more massive BH in the Type II GRB. A BH-NS
merger Type I GRB, on the other hand, may reach the same energetics 
as Type II GRBs if the initial BH in the binary system is massive 
enough and has a large enough angular momentum.

In order to relate this theoretically motivated 
total energy budget to the observed
energy, one needs to introduce the beaming factor. The
standard GRB jet models invoke a conical jet with uniform
energy distribution \citep{meszaros98,rhoads99,sari99}. More 
complicated (maybe more realistic) jet models invoke 
distributions of jet energy with angle from the jet axis 
\citep{meszaros98,zhangmeszaros02b,rossi02,zhang04}.
Theoretically, it is difficult to model how jets are launched
from the central engine. On the other hand, one can speculate
about the collimation angle of jets from two types of GRBs
from the theoretical point of view: Type 
II GRBs should tend to have narrower jets than Type I GRBs 
due to the additional
collimation of the stellar envelope \citep{zhangw03}.
Type I GRB jets tend to be broader \citep{aloy05}. 
Observationally this prediction has not been tested
statistically. Observations
of some individual bursts seem to support this picture. 
For example, the Type I
Gold Sample burst GRB 050724 was found to have a beaming
angle wider than $\sim 25^{\rm o}$ \citep{grupe06,malesani07}.
Type II GRBs on the other hand, have a typical 
beaming angle of $\sim 5^{\rm o}$ 
\citep{frail01,bloom03,liang08,kocevski08,racusin09}.
Opposite cases are also observed in some GRBs.
For example, the short GRB 051221 (a Type I candidate) has a 
narrow jet with $\theta_j \sim 4^{\rm o}-8^{\rm o}$ 
\citep{burrows06}. This is contrary to the 
theoretical expectation if it is indeed a Type I GRB.
On the other hand, the Type II GRB 060729 \citep{grupe07,grupe09}
may have a large opening angle since its X-ray
afterglow keeps decaying without a break for hundreds 
of days.

The observed ``isotropic'' energy is the total energy 
divided by the beaming factor ($2\cdot \pi \theta_j^2
/4\pi = \theta_j^2/2$ for uniform jets under the small
angle approximation). Considering a factor of $\sim 
(20/5)^2 \sim 15$ difference in the beaming factor,
the isotropic energy of a typical Type I GRB should be 
a factor $\sim 100-1000$ times lower than that of a 
typical Type II GRB. If a Type I GRB has a high isotropic
luminosity/energy comparable to that of a typical Type II 
GRB, one must then demand a very narrow jet, with an opening 
angle even smaller than that of Type II GRBs. A BH-NS
central engine with the Blandford-Znajek mechanism as
the engine power can ease the constraint, but a
comparable beaming angle to Type II GRBs is nonetheless
needed. 

\subsection{Afterglow Properties}

Broadband afterglow emission has been interpreted as the 
external forward shock emission as the fireball is decelerated
by the circumburst medium. An ideal observational campaign can
lead to diagnostics of the circumburst medium properties
\citep{panaitescu02,yost03}, which can shed light on the 
progenitor system of the GRB \citep[e.g.][]{fan05a,greiner09,xu09}. 
In particular, if a stratified 
stellar-wind-type medium ($n \propto R^{-2}$) 
\citep{dailu98c,chevalier00} is identified, the burst can be
identified as a Type II GRB. The case of a constant density
medium is, however, less informative. Although Type I GRBs are
expected to reside in such a medium, some Gold-Sample Type II
GRBs have been found to reside in a constant medium as well
\citep{panaitescu02,yost03}. The mechanism of forming such a
medium before the death of the massive star is unknown.
In any case, one needs more information to establish the
association of a burst with a physical model category
if it goes off in a constant density
medium. For example, the afterglow luminosity of Type I GRBs
should be systematically lower than that of Type II GRBs due to the
expectation of both a lower medium density in the merger 
environment (relevant for $\nu<\nu_c$) 
and a systematically lower blastwave energy
\citep{panaitescu01b,fan05a,kann09b}. A Type I GRB can be
even ``naked'' (i.e. no detectable afterglow) if the ambient
medium density is low enough. Such GRBs are indeed observed
\citep[e.g. GRB 051210,][]{laparola06}.

A related issue is the GRB radiative efficiency. Observations
and theoretical modeling suggest that the efficiency is similar
for both Type I and Type II GRBs \citep{zhang07a,berger07c,gehrels08},
so that it cannot be regarded as a useful criterion to 
tell the model category that is associated with a particular
GRB.

\subsection{Redshift Distribution}

Statistically, redshift distributions of 
Type I and Type II GRBs should be different. Type II GRBs
generally trace the star-forming history of the 
universe\footnote{Metallicity may play additional role to
select Type II GRBs \citep[e.g.][]{wolf07,nuza07,li08}, 
but the issue is inconclusive.}.
Type I GRBs are expected to be ``delayed'' with respect to
star formation due to the long merger time scale asociated
with the shrinking of the binary orbits due to gravitational 
radiation \citep{belczynski06}. On average, it is expected
that the mean redshift of Type I GRBs is lower than that of
Type II GRBs. 

\subsection{Luminosity Function}

The luminosity function of long duration GRBs is categorized by a
broken power law with a break $\sim 10^{52}~{\rm erg~s^{-1}}$
\citep{guetta05,liang07,virgili09}. Below the break the
power law index is $>-1$, while above the break the power
law index is $<-2$. Low luminosity GRBs may
form a distinct bump at $L<10^{49}~{\rm erg~s^{-1}}$
\citep{liang07,virgili09,dai09}. As argued below, most 
long GRBs are Type II GRBs, so this luminosity function
may be regarded as that of Type II GRBs. There is however
no direct theoretical reason for such a luminosity function.
For Type I GRBs, the luminosity
function has not been studied in detail due to the limited 
sample with redshift measurements so far 
\citep[but see][]{virgili09b}. 
The luminosity function
of the BATSE short/hard GRB sample was studied \citep{guetta06},
but as discussed below, it is not justified that this 
population is identical to the Type I population. This issue 
will be further discussed in \S5.4 and \S6 below.

\subsection{Gravitation Wave Signals}

Probably the most definite criterion to differentiate Type
I GRBs from Type II GRBs is through detecting their gravitational
wave (GW) signals. Although the GW signature of a Type II GRB
is highly uncertain \citep[e.g.][]{kobayashi03}, the wave forms
of NS-NS and NS-BH mergers are well predicted 
\citep[e.g.][]{dalal06}. Detections of these signals would 
unambiguously associate some GRBs to the Type I model category.
However, this criterion can only be applied in the future
when the GW detectors reach the desired sensitivities.

\section{Type I and Type II Samples and Their Statistical Properties}

\subsection{Sample selection}

The above consideration suggests that theoretically there is no
handy, distinct criterion that can be used to 
immediately determine the physical model category 
that a burst is associated with. 
In this section, we attempt to explore the topic further from
the observational point of view. The current standard approach 
is to use three criteria, ``duration'', ``hardness'', and 
(when available) ``spectral lag'', to categorize bursts
as ``long'' (implicitly assumed to be associated
with ``Type II'') or ``short'' 
(implicitly assumed to be associated with ``Type I''), 
and use these samples to explore the 
statistical properties of other observational properties
\citep[e.g.][]{nakar07,berger09}. In other
words, it is often implicitly assumed that 
\begin{eqnarray}
{\rm Long/soft/long~lag = Type~II} \nonumber \\
{\rm Short/hard/short~lag = Type~I}. 
\label{types}
\end{eqnarray}
The problem with such an approach is that these
criteria may not be always reliable. A notable example was
GRB 060614, which is a long GRB but is very likely 
associated with Type I \citep{gehrels06,zhang07b}. 
Another complication is
related to GRB 080913, GRB 090423, and some other intrinsically 
short GRBs \citep[e.g.][]{levan07}. These GRBs can be detected
as short/hard GRBs if their redshifts were low enough, but
their physical properties are more close to those of Type II
GRBs. Some short/hard GRBs (e.g. 060121, de Ugarte Postigo
et al. 2006) are very energetic, which are not easy to be
accommodated within the Type I progenitor models.

In this paper, we adopt an alternative approach. {\em Instead of sticking
to the observed gamma-ray properties, we adopt the observational criteria that
are directly related to the progenitor systems to select the samples,
and then go back to investigate the other properties (including 
duration, hardness, spectral lag, etc) of the samples.}
The advantage of this approach is that we can start with those
GRBs whose progenitor systems are more confidently inferred. We can
then use them to verify whether the ansatz Eq.(\ref{types})
is justified.

We define the following three samples based on the criteria
detailed below.

{\bf Type II Gold Sample.} This sample is defined such that at least
one of the following two criteria are satisified. 
\begin{enumerate}
\item There is a spectrally confirmed SN
association with the GRB;
\item The specific star forming rate
(SSFR) is very high (to be specific, the SSFR satisfies 
$\log {\rm SSFR} > -0.2$ or SSFR$>0.63~{\rm
Gyr^{-1}}$ in the sample of Savaglio et al. 2009); the GRB location
does not have a large offset from the center; and there is no
stringent upper limit on the existence of a SN associated with the
GRB.
\end{enumerate}
Notice that the GRB properties (duration, hardness and lag) are not
the considerations to define the sample. Since not many GRBs have host
SSFR information published, this sample is
by no means complete, and there should be many more Type II GRBs 
that are not included. The purpose of selecting this sample is
to use the most stringent criteria to investigate how the best 
Type II GRB candidates 
look like. As a result, we do not include the GRBs
that have a claimed SN bump in the optical light curve but no
confirmed SN spectroscopic signature. The threshold of SSFR is
arbitrary. This limiting value was chosen because 
Table 11 of \cite{savaglio09} has a mix of long and short 
GRBs for $\log {\rm SSFR(Gpc^{-1})} < -0.3$,
which is the regime where confusion arises. The lower bound 
$\log {\rm SSFR} > -0.2$ can be regarded as a safe line above which
GRB hosts have very active star formation. One exception is
the short duration GRB 051221A \citep{soderberg06b,burrows06}.
The SSFR value ($\log {\rm SSFR} > 0.804$) is way above the threshold.
However, since deep searches have ruled out the association of a
1998bw-like SN \citep{soderberg06b}, we do not include it in the 
Type II sample, and will include it in the ``Other short
hard sample''. We note that many Swift long GRBs should
be associated with Type II. 
However, since no published SSFRs are available 
for most of them, we refrain from including them in the Type II Gold 
Sample\footnote{Besides those included in the 
\cite{savaglio09} sample (which covers from GRB 970228 to
GRB 061126), we only include GRB 080520 
and GRB 060602A based on the SFR criterion. They have a high
SFR (though SSFR is not measured) typical for other Type II
Gold sample GRB host galaxies. For example, GRB 080520 has 
$\sim 15M_\odot~{\rm yr^{-1}}$ \citep{malesani08}, which is 
comparable to the highest in the \cite{savaglio09} sample.}. 
This sample should be expanded significantly 
later when the host galaxy information of the Swift GRBs is released. 
Right now the Type II Gold Sample includes 33
GRBs (Table 1 Top Panel). This is already a large enough sample 
to study the statistical properties of Type II GRBs.

{\bf Type I Gold Sample.} The Gold Sample of Type I GRBs is defined
by at least one of the following two criteria. 
\begin{enumerate}
\item The host galaxy is elliptical or early type; 
\item The GRB location has a relatively low local 
SSFR, or a large offset from the center of the host galaxy; 
and deep searches reveal stringent upper limits on the
existence of an underlying SN.
\end{enumerate}
Again the GRB properties
(duration, hardness, lag) are not considered. Some
arguments \citep{belczynski06,zheng07} have suggested 
that a fraction of Type I GRBs may be located 
in star forming regions of star forming galaxies. Our criteria
do not select those, since we do not demand
completeness of sample selection. 
After systematically checking the archival data, we only
identify 5 bursts in the Type I Gold Sample: GRBs 050509B, 
050709, 050724, 060614\footnote{In the literature
GRB 060614 is usually taken as a controversial candidate for Type
I. This was mainly because of its long duration. We do not
consider duration as a criterion when selecting the Gold Sample.
This burst satisfies the criterion \#2 of the Type I Gold
Sample.}, and 061006
(Table 1 Middle Panel). The details
of individual GRBs are presented in the Appendix.

{\bf Other SGRB Sample.} Most short/hard GRBs in the Swift era 
satisfy neither of the two criteria of the Type I Gold Sample.
Some of them do not have their host galaxies convincingly identified.
Others have host galaxies with active star formation. 
These GRBs are usually regarded as Type I candidates simply because 
they are ``short/hard''. There could be a good fraction of Type I 
GRBs in this sample, but we are not sure that they can ALL
be associated with Type I. 
Since we define the Gold Samples not based 
on the GRB properties, we leave these bursts in a separate sample,
without specifying whether they are associated with
Type I or Type II. There are 20 bursts in this sample (Table 1 
Bottom Panel). The details of individual GRBs are presented in 
the Appendix.

\subsection{Duration-Hardness Distribution}

Figure \ref{T90-HR} presents the traditional $T_{90}$-hardness ratio (HR)
plot of GRBs. Superimposed on the BATSE data (orange small dots) are
the three samples defined above: Type II Gold Sample (blue), 
Type I Gold Sample (red), and other SGRB sample (green).
The HR is defined as the fluence ratio between (50-100) keV 
and (25-50) keV. For BATSE bursts, this corresponds to the fluence ratio
between channel 2 and channel 1. For other detectors 
(HETE-2, Swift/BAT, Konus/Wind, INTEGRAL) with different detector
energy bands, we perform spectral fits and use the fitted model to 
derive the HR. 
Besides the observed points (open symbols), we also
plot the corresponding ``rest-frame'' points (filled symbols) for
each burst. The HR is then defined as the flux ratio between the
rest-frame (50-100) keV band and (25-50) keV bands, which is again
derived from spectral fitting. For a power law fit, the rest frame HR
is the same as the observed one. For a curved spectrum (e.g. 
a Band function or an exponential cutoff power law), the two can be
different. 
The $T_{90}$ values are energy- and detector-dependent. We do
not make efforts to convert all $T_{90}$ to the BATSE-band, since
this requires time-dependent spectral analyses and extrapolations, 
and for many bursts the data quality is not sufficient to perform
such an analysis. Instead we simply plot $T_{90}$ measured by
different detectors (e.g. Swift and HETE). The correction to the
BATSE-band $T_{90}$ is usually not significant for most long GRBs, but
could be significant to those GRBs with soft extended emission.
Traditionally, the 
``rest frame'' $T_{90}$ are not used to defined long vs. short for a 
particular GRB. We present them here just to show how the intrinsic
distribution may differ from the observed one.
To derive the rest-frame $T_{90}^{rest}$, we simply divide
the observed value by
$(1+z)$. More rigorously one needs to again take into account the 
light curve evolution with energy. This again requires a time-dependent
spectral analysis. Since most bursts do not have such detailed
information, and since the correction would not be significant
for most bursts,
we neglect this correction for the sake of simplicity and
uniformity. For short GRBs with extended emission, we use circles
to denote the short spikes only (excluding the extended emission), 
while using squares to denote the full emission with extended 
emission included. These two locations for the 
same burst with and without extended emission are connected 
by lines. Since the mean HR is derived, the HRs including 
extended emission are usually smaller 
than those without, as the extended
emission is typically softer than the initial short spikes.

From Fig.\ref{T90-HR} one can make the following interesting
observations. First, the Type II GRBs are generally long, and
they well represent the long/soft population of the BATSE GRBs
in the $T_{90}$-HR plane. However, some Type II
GRBs have a duration close to the 2-second separation line,
and their intrinsic duration can be shorter than 2 s (e.g. 
GRB 040924 with $T_{90} = 2.39\pm 0.24$ s at $z=0.858$, and 
GRB 080520 with $T_{90}=2.82 \pm 0.67$ at $z=1.545$).
\cite{levan07} also discussed a sample of apparently-long,
intrinsically-short GRBs.
Secondly, {\em four out of five Type I Gold Sample GRBs are not
strictly ``short''}. Except GRB 050509B, all the others have
extended emission aside from the initial ``short/hard'' spike.
The spike itself is longer than 2 s for GRB 050724 and 
GRB 060614. {\em All 5 Type-I Gold Sample bursts have a moderate HR.
None has an extremely hard spectrum. }
Thirdly, the Other SGRB Sample fills in the short/hard region
in the $T_{90}-$HR diagram more uniformly, suggesting
that it represents the BATSE short/hard sample well. Some bursts
in the sample also have extended emission.

\subsection{Empirical correlations}

\begin{figure}
\plotone{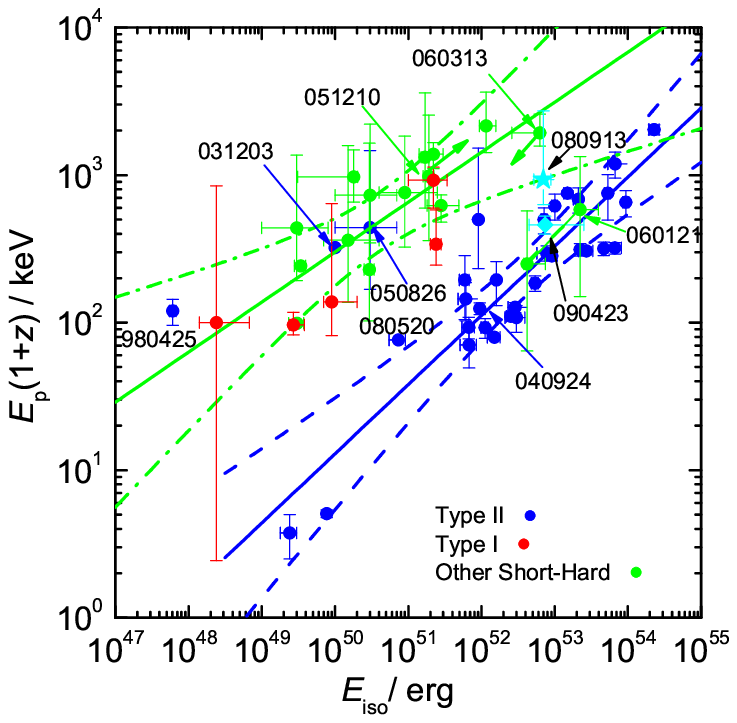}
\plotone{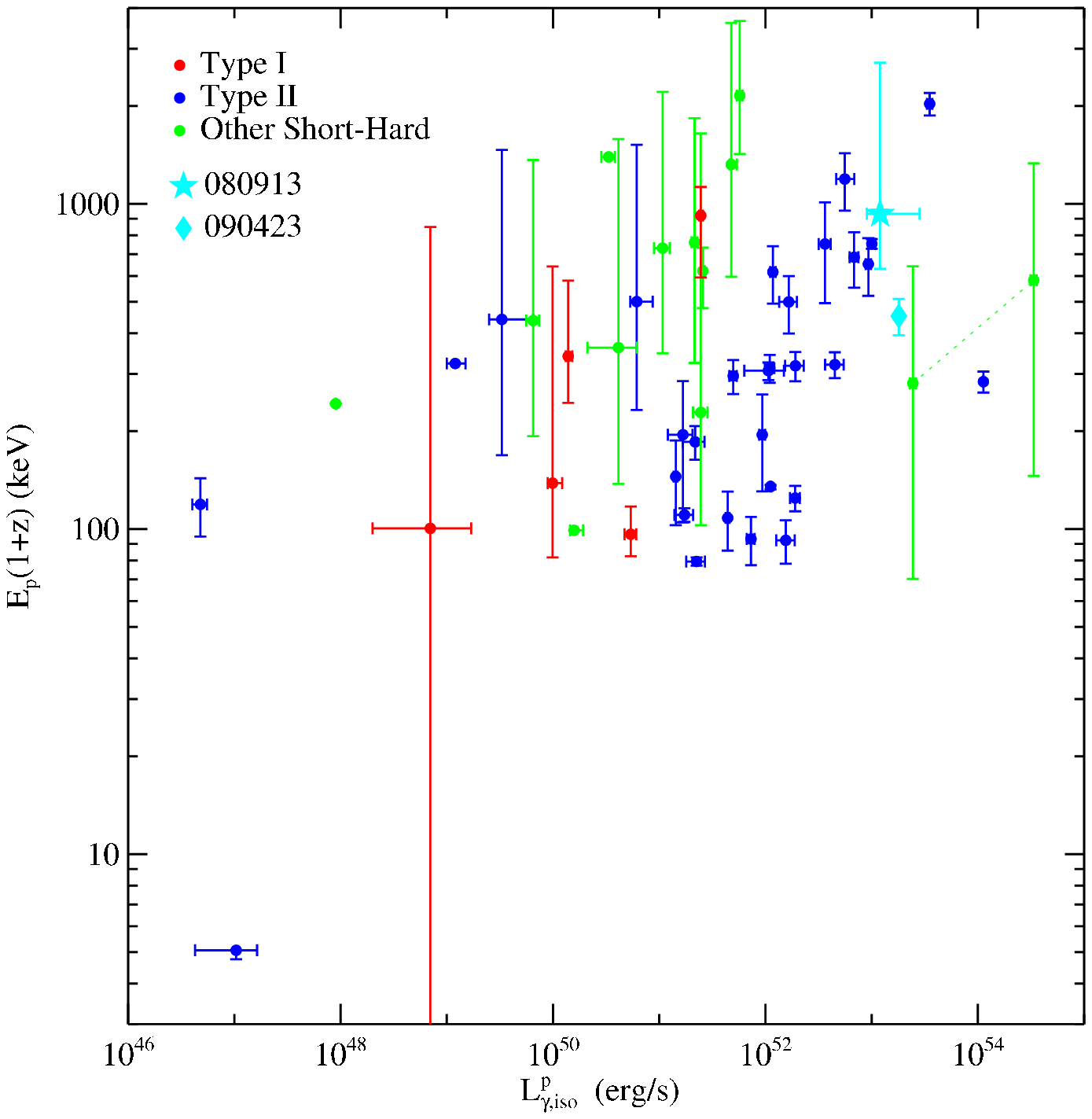}
\caption{(a) The $E_p-E_{\gamma,iso}$ diagram of the three samples
of GRB discussed in the paper: Type II Gold Sample (blue), 
Type I Gold Sample (red), and other short/hard GRBs
(green). Two possible redshifts $z=4.6,1.7$ for 
the short GRB 060121 are adopted, which satisfies the
relation well (unlike other short/hard GRBs).
GRB 080913 and GRB 090423 (cyan) are also plotted for comparison. 
The best-fit $E_p-E_{\gamma,iso}$ correlations for both Type II 
and Type I/Other SGRB samples are plotted (solid lines) with 
the $3\sigma$ boundary (dashed line) marked.
(b) The $E_p-L_{\gamma,iso}^p$ diagram. The same convention 
has been used.}
\label{amati}
\end{figure}

Figure \ref{amati}a displays the $E_p-E_{\gamma,iso}$
(Amati) relation of the three samples. The spectral parameters 
are collected from 
the published papers or GCN circular reports (see Table 1 for
references). For those GRBs with extended emission (including 
Type I Gold Sample GRBs 050724, 060614, and 061006), we only 
consider the short hard spikes. For all the bursts, the isotropic 
gamma-ray energy ($E_{\gamma,iso}$) is calculated in the GRB 
rest-frame $1-10^4$ keV band through extrapolation based on the
spectral parameters. We can see that 
most GRBs in the Type II Gold Sample indeed
follow the $E_p \propto E_{\gamma,iso}^{1/2}$ (Amati) relation.
However, there are three noticeable outliers: GRB 980425, 
GRB 031203, and GRB 050826. The first two are 
nearby low-luminosity (LL) GRBs, which have been argued
to be from a distinct population \citep[e.g.][]{liang07,virgili09,dai09}. 
Another nearby LL GRB 060218 is a soft burst \citep{campana06}
and satisfies the Amati-relation well. GRB 050826 with
$T_{90} \sim 35$ s is an
intermediate Type II GRB between the more ``classical''
Type II and the nearby LL-GRBs \citep{kann09a}, and 
deviates from the relation. We also pay special attention to 
the two intrinsically short Type II GRBs. Although GRB 040924 
is right on the Amati-relation track, GRB 080520 seems 
to be slightly off the track. The
Type I Gold Sample and the Other SGRB Sample are populated 
above the conventional Amati-relation track. Since many 
short/hard GRBs have $E_p$ outside the BAT band, their $E_p$ error 
bars are large. The values in our analyses are adopted from
\cite{butler07}. In any case, it seems that they follow a
separate track with a shallower slope than the Amati-relation.
Excluding GRBs 080913, 090423 and 060121 (which are likely Type II,
see \S6.2), a best fit to the Type I Gold and Other SGRB samples
lead to a slope 0.34, with the 3$\sigma$ limits of the slope
as (0.15-0.53) (see Fig.\ref{amati}a). GRB 080913 is marginally 
within the 3$\sigma$ regions for the Type II Amati-relation,
but is also consistent with this new track defined by Type I
and other short/hard GRBs within $3\sigma$. GRB 090423 aligns
with the Type II Amati-relation more closely \citep[see
also][]{lin09}.

A likely reason that the Type I and the Other SGRB Samples 
deviate from the Amati relation of Type II GRBs is simply 
because they have shorter durations so that they have
smaller $E_{\gamma,iso}$ values than the Type II GRBs with 
a similar $E_p$.
To test this, we plot the $E_p - L_{\gamma,iso}^p$ relation
(Yonetoku relation) in Fig.\ref{amati}b. We can see that the
distinction between Type II and Type I GRBs becomes less
significant, although the correlation now has a much larger
scatter. Noticing the large error bars of the Type I and
Other SGRB Samples, one may conclude that there is no distinct
difference among the three samples as far as the Yonetoku
relation is concerned. A similar conclusion was drawn by 
\cite{ghirlanda09} in an anaylsis of the BATSE GRBs.

\begin{figure}
\plotone{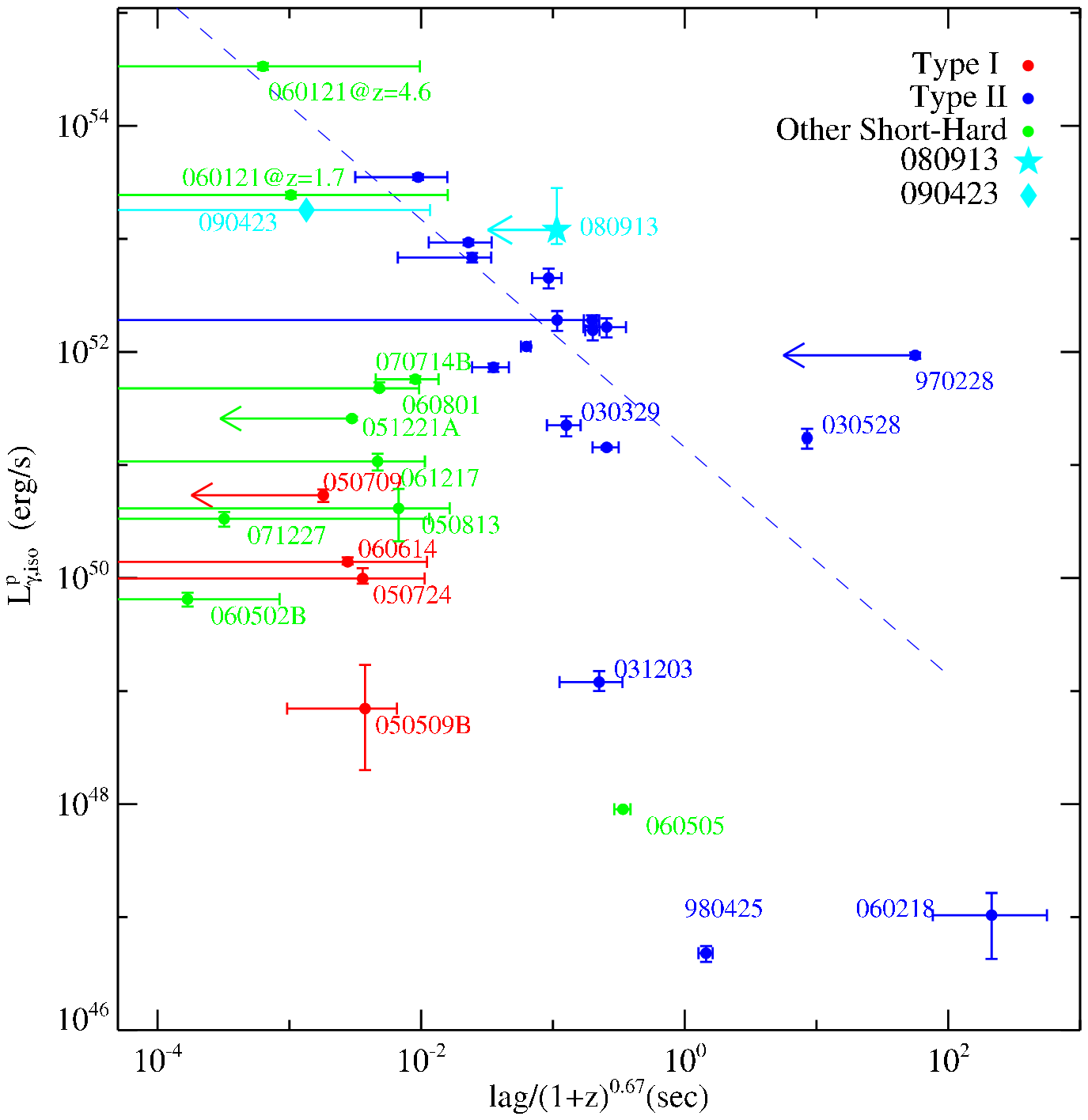}
\plotone{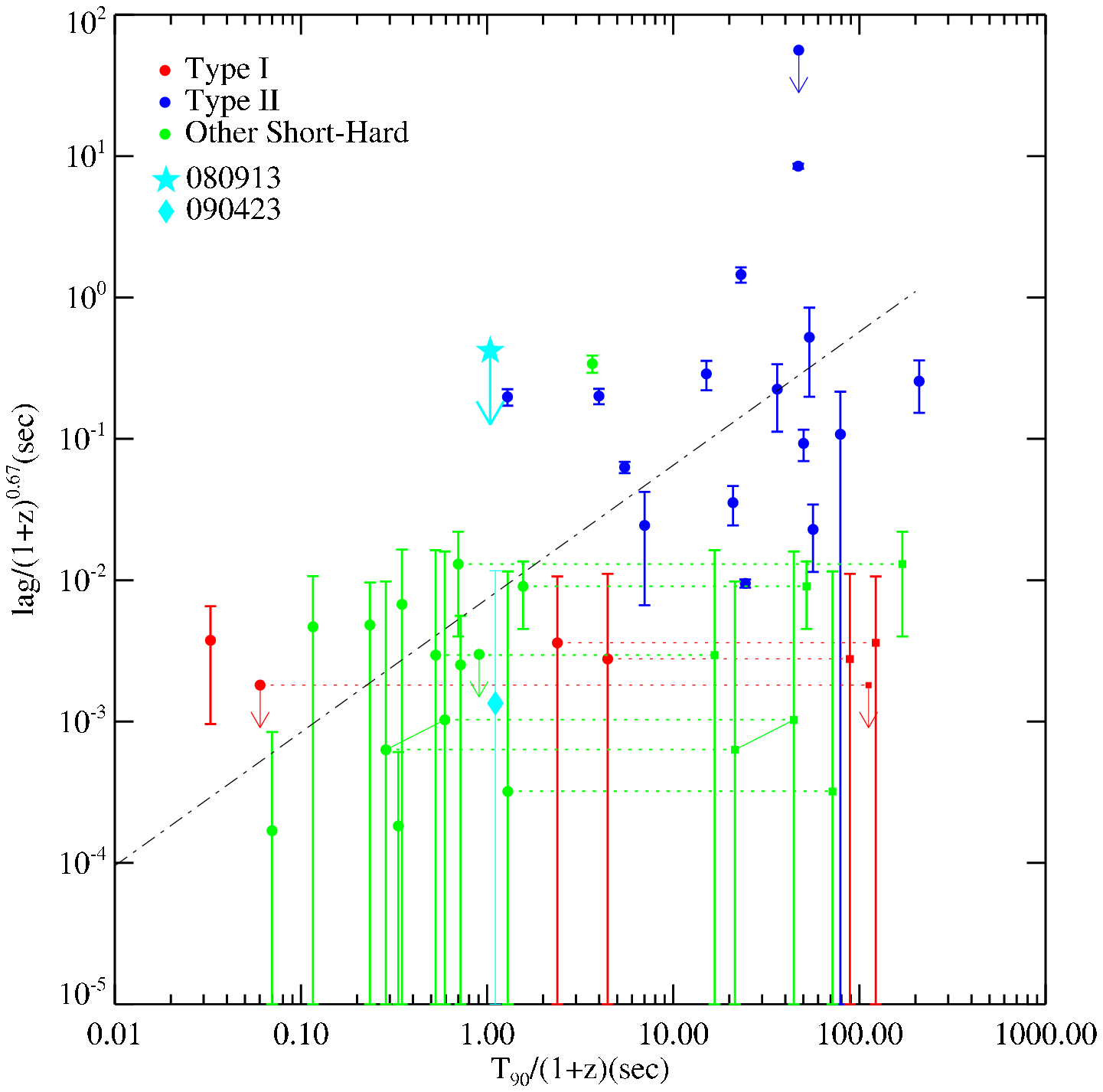}
\caption{(a) The $L_{\gamma,iso}^p-{\rm lag}$ diagram.
Same convention as Fig.3 is adopted. 
GRB 080913 and GRB 090423 satisfy both the correlation defined
by Type II GRBs and the ``zero lag'' trend defined by
Type I and Other SGRB Samples.
Two possible redshifts $z=4.6,1.7$ for 
the short GRB 060121 are adopted, which satisfies the
correlation well (unlike other short/hard GRBs).
(b) the ${\rm lag}-T_{90}$ (intrinsic) diagram
of the three samples. The same GRBs
with/without extended emission is connected by dotted lines.
The spectral lags of these GRBs are for the short/hard spikes
only. A positive correlation between duration and spectral
lag is derived (dashed line). See text for details.}
\label{Llag}
\end{figure}

Figure \ref{Llag}a displays the luminosity-spectral lag diagram
of GRBs with the three samples plotted. A group of
Gold Sample Type II GRBs indeed define a 
$L_{\gamma,iso}^p \propto (\Delta t_{rest})^{-\delta}$ correlation
track \citep{norris00,gehrels06}, although several low-luminosity,
long-lag GRBs lie below the extrapolation 
of the track \citep[see also][]{gehrels06,liang06b}.
Gold Sample Type I GRBs are clustered at the lower left corner.
This is as expected: short durations define short lags, and
smaller energy budgets define lower luminosities. 
About half of the ``Other SGRBs'' are clustered close to the Type 
I Gold Sample, suggesting that they may be associated with  
Type I as well.
Some others fill in the gap between the Type I and Type II Gold
Samples. In particular, GRB 060121 lies right on the track for 
both putative redshifts 1.7 and 4.6 \citep{deugartepostigo06}.
GRB 070714B is also close to the track. The SN-less 
GRB 060505 clusters with other nearby low-luminosity 
Type II GRBs. Finally, the two high-$z$ GRBs 080913 (notice
that only the upper limit of spectral lag is derived) and
090423 are consistent with satisfying the 
$L_{\gamma,iso}^p-{\rm lag}$ correlation of Type II,
but are also consistent with the zero-lag trend of 
Type I/Other SGRB.

As discussed in
\S4.7, the luminosity lag relation may be related to
the variability-luminosity relation, and may be more relevant
to Type II GRBs. On the other hand, the physical origin of
the relation is not clearly understood and is based on many
assumptions. Although the correlation may be taken as a 
reference, it may not be taken as the definite criterion for 
judging the physical origin of a GRB.

Based on the high-latitude-effect interpretation of spectral
lag (\S4.7), one expects that 
short spectral lags should be related to short angular spreading
times. The latter corresponds to the width of individual pulses.
If the number of pulses do not 
fluctuate significantly among bursts, one would also expect a rough
correlation between spectral lags and durations. In Fig.\ref{Llag}b
we display the $T_{90}/(1+z)-{\rm lag}/(1+z)^{2/3}$ diagram of the 
three samples of bursts. Again points of the same burst with and 
without extended emission are connected by lines. 
We investigate a possible correlation between duration and
spectral lag. Since the spectral lags are defined for the short/hard
spikes only for those GRBs with extended emission, we use $T_{90}$
excluding the extended emission for those bursts.
A positive correlation
between $T_{90}$ and lag with slope $0.94\pm 0.14$ is obtained,
with the Spearman's rank correlation coefficient $r=0.735$,
corresponding to a chance probability $P<10^{-4}$.
This is consistent with our naive expectation, suggesting
that spectral lags are closely related to durations, and
may not carry additional information in defining the
categories of GRBs.

\subsection{Luminosity and Redshift Distributions}

\begin{figure}
\plotone{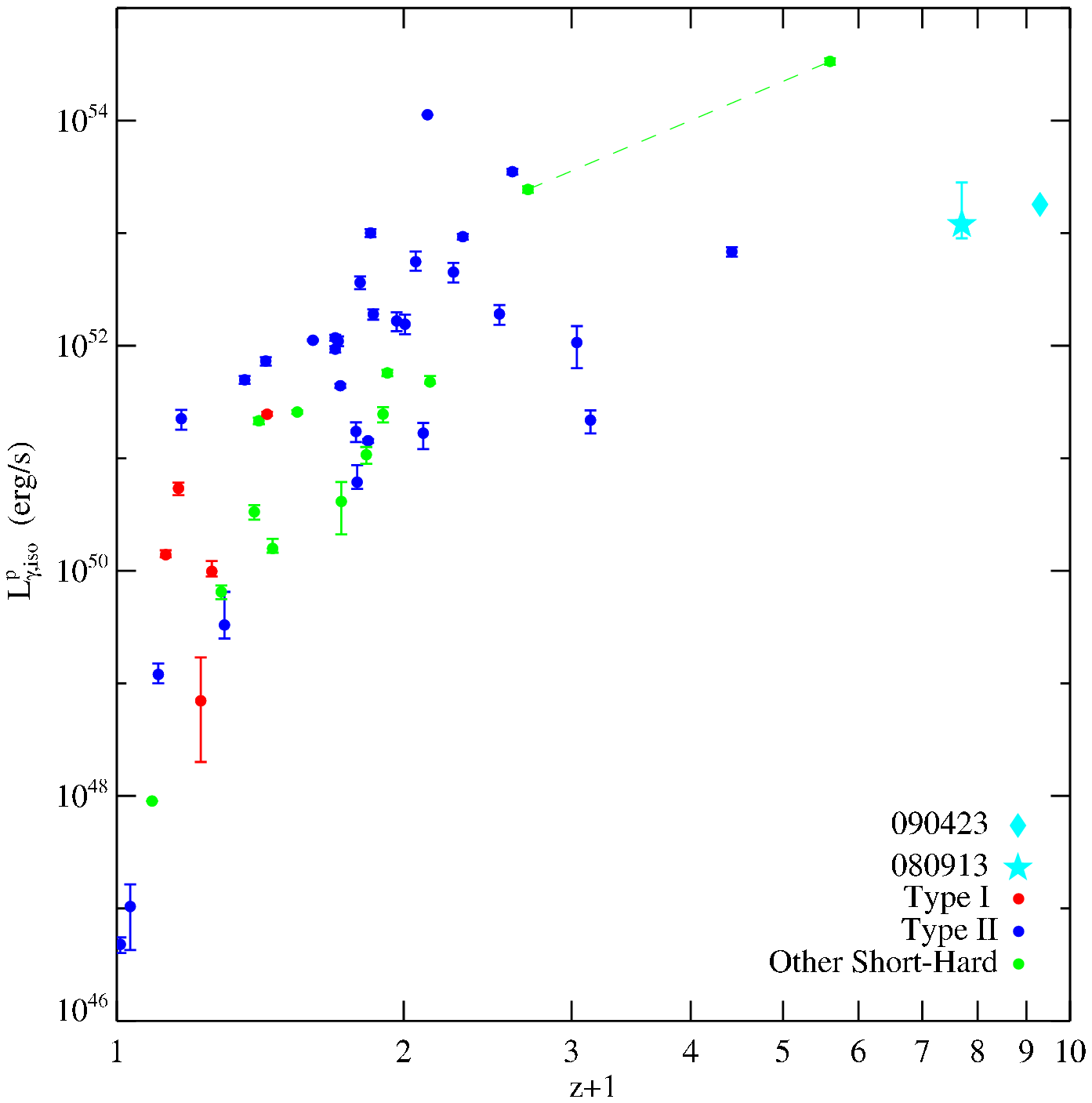}
\plotone{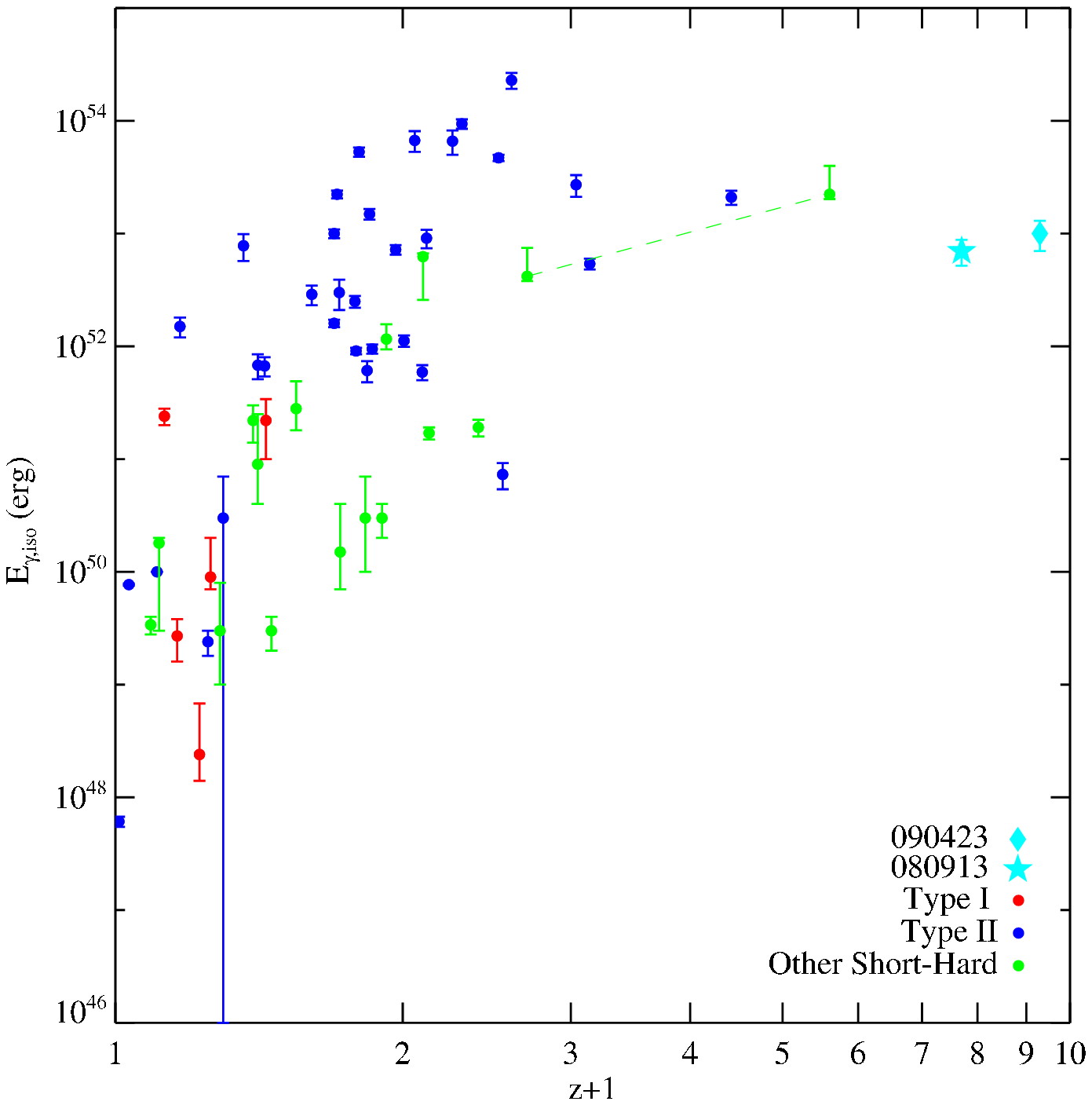}
\caption{(a) The $L_{\gamma,iso}^p-z$ diagram, and (b) the
$E_{\gamma,iso}-z$ diagram of the three samples. The same
convention as Fig.3 is adopted.}
\label{L-z}
\end{figure}

Figure \ref{L-z}(a) and (b) display the observed 2-dimensional 
luminosity-redshift ($L_{\gamma,iso}^p-z$) and energy-redshift 
($E_{\gamma,iso}-z$) distributions of the three samples. 
GRBs in the Type I Gold Sample are all at $z<0.5$. 
Including the Other SGRB Sample, the upper boundary of $z$
reaches $\sim 1$ (except GRB 060121). The Type II GRBs have
a wider span of redshift distribution, with the peak around
$z\sim 1$. In terms of luminosity distribution, the Type II
GRBs on average are $\sim$ 2 orders of magnitude more
luminous than the Type I GRBs. Type I GRBs can at least
reach a luminosity of $L_{\gamma,iso}^p \sim 2.5\times 10^{51}
~{\rm erg~s^{-1}}$ (for the Type I Gold GRB 061006). Including 
the Other SGRB Sample, several short GRBs (070714B, probably
060313, and especially the latest GRB 090510) can reach 
$L_{\gamma,iso}^p \sim 10^{52}~{\rm erg~s^{-1}}$. 
GRB 060121 even reaches $L_{\gamma,iso}^p \sim 10^{53}
-10^{54}~{\rm erg~s^{-1}}$ for the two fiducial redshifts
in discussion. This luminosity is high even for Type II GRBs.
GRB 080913 has $L_{\gamma,iso}^p \sim 1.2\times 10^{53}
~{\rm erg~s^{-1}}$. GRB 090423 has $L_{\gamma,iso}^p \sim 1.88
\times 10^{53}~{\rm erg~s^{-1}}$ \citep{nava09}.
Both are moderate to high luminosities for Type
II GRBs, and are very high when compared with the Type I and
Other SGRB Samples (except for GRB 060121). In the 
$E_{\gamma,iso}-z$ diagram, the separation between Type II
and Type I is more distinct, with most SGRB sample bursts 
lying below the Type II distribution. But GRB 080913 and 
GRB 090423 become moderate in the Type II Sample due to
their intrinsically short durations.
The clearer separation between Type II and Type I/Other SGRB
Samples is mainly due to 
the short duration of the SGRB sample, which makes them
less energetic. However, GRB 060121 is still as energetic
as the average Type II GRBs.

\subsection{Afterglow Properties}

\begin{figure}
\plotone{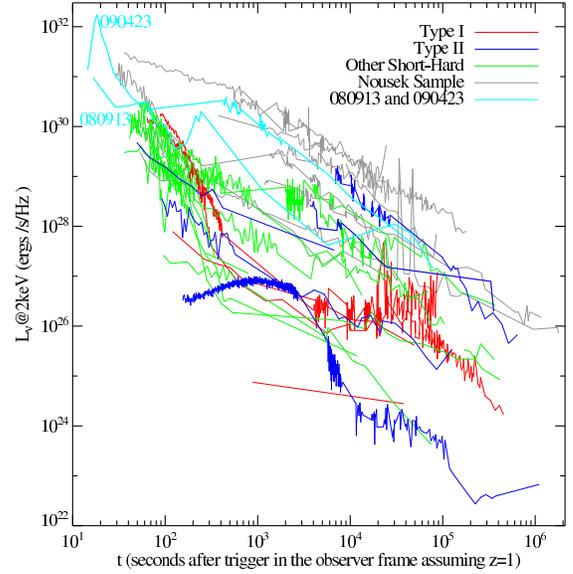}
\caption{The rest frame 2 keV X-ray afterglow luminosity light curves
of GRB 080913, GRB 090423, and the three samples. All bursts are
placed at $z=1$. The color scheme is 
the same as in the other figures.
Since most Type II Gold Sample bursts are pre-Swift ones
and have no X-ray light curves, we also add the $z$-known long
GRBs in the sample of \cite{nousek06} (grey), which are generally
believed to be Type II GRBs. GRB 080913 and GRB 090423 (cyan) 
both have bright X-ray afterglows typical of Type II GRBs.}
\label{Xray}
\end{figure}

\begin{figure}
\plotone{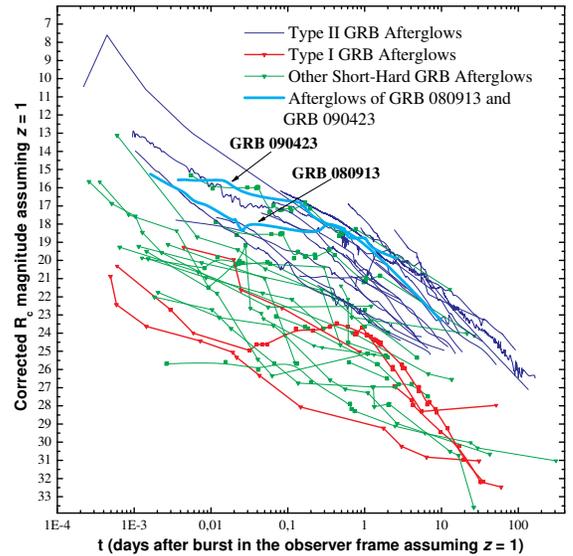}
\caption{The rest frame optical light curves of GRB 080913, GRB
090423, and the three samples. 
The color scheme is the same as in the other 
figures. Similar to \cite{kann09a, kann09b}, they are plotted 
at a common redshift of $z=1$. As with the X-ray light curves 
(Fig. \ref{Xray}), the optical afterglows of the 
Type II Gold Sample GRBs are clearly more luminous than those 
of the Type I Gold Sample and the Other Short-Hard Sample. 
The latter two populations are in good agreement with each other. 
GRB 060121 is the single short-hard GRB which is optically 
highly luminous. GRB 080913 and GRB 090423 both have bright 
optical afterglows typical of Type II GRBs.  }
\label{optical}
\end{figure}

Figures \ref{Xray} and \ref{optical} present the 
intrinsic afterglow light curves in the X-ray and optical
bands for the three samples. Figure \ref{Xray} presents the 
rest-frame 2 keV specific luminosity light curves. Since many 
Type II Gold Sample GRBs are pre-Swift, we do not have
many Type II X-ray light curves. The ones that are plotted include
two low luminosity GRBs (060218 and 050826) and two 
intermediate-to-high luminosity GRBs (080520 and 050525A). 
These do not fully represent the 
Type II GRB X-ray afterglow properties. In order to compensate 
for this weakness of sample selection, we also overplot the X-ray
light curves of a group of early Swift long GRBs in the sample
of \cite{nousek06}. Since we already demonstrated that the
Type II Gold Sample represents the BATSE long GRBs well, we 
assume that the Nousek Sample represents the Type II GRB
X-ray afterglows well. We can see that these bursts occupy the
upper portion of the light curve space in Fig.\ref{Xray}. 
By contrast, the Type I Gold Sample occupy the lower portion,
and the Other SGRB Sample populate in between with much overlap
with both Gold Samples. Low luminosity Type II GRBs have 
luminosities comparable to Type I Gold Sample GRBs.

Figure \ref{optical} presents the optical light curves with
corrected $R_c$-magnitude by moving all GRBs to $z=1$
\citep{kann09a,kann09b}. One big difference between these optical 
light curves and the X-ray light curves
(Fig. \ref{Xray}) is that most Type II GRBs are represented,
exceptions being those GRBs that had negligible optical afterglows but
strong supernovae signatures (GRBs 980425, 031203, and XRF 060218),
dark GRBs, where the optical emission was probably totally supressed
by line-of-sight extinction in the host galaxy (GRBs 990506, 000210,
020819B, 051022), and some with very sparse optical data (XRF 020903,
GRBs 030528, 050826, 060602A, 080520). Most data have
been taken from \cite{kann09a, kann09b}, where
the methods of creating the intrinsic light curves are also
presented. Similar to the
X-ray light curves, the Type II GRB afterglows form a much more
luminous group than the Type I GRB afterglows \citep{kann09b}. The
light curves of Type I Gold GRBs and those of most Other SGRBs overlap,
indicating that they are likely drawn from the same population. The
most prominent exception is again GRB 060121 with an optically 
luminous afterglow 
\citep[see][for more details]{kann09b}, which is comparable to the
afterglows of Type II GRBs. 

For both X-ray and optical afterglows, GRB 080913 and GRB 090423 
have a luminosity comparable to or higher than 
the average luminosity of the Type II
GRB afterglows \citep{greiner09,salvaterra09,tanvir09}.

\section{Nature of short/hard GRBs}

Based on the theoretical considerations and the statistical analyses 
presented above, in this section we attempt 
to address the question whether the ansatz Eq.(\ref{types}) is valid,
i.e. whether Type II GRBs are simply associated with
``long/soft/long lag'' GRBs while Type I GRBs are simply 
associated with ``short/hard/short lag'' GRBs. 
In particular, we will address the nature of the Other
SGRB Sample. It is likely that some (maybe many) GRBs in the
Other SGRB Sample are associated with Type I. The 
question is whether they are ALL associated with Type I.

\subsection{Are all short/hard GRBs associated with Type I?}

The most straightforward possibility is to accept that all short/hard
GRBs are associated with Type I GRBs. Inspecting the Other SGRB 
Sample, one may raise 
the following arguments in support of this suggestion:
(1) They indeed occupy the short/hard domain of the BATSE
$T_{90}$-HR diagram, which is in distinct contrast to Type II GRBs
that predominantly occupy the long/soft domain; (2) 
Most of them deviate from the $E_p-E_{\gamma,iso}$ relation and the 
$L_{\gamma,iso}^p-$lag relation for Type II GRBs; (3) The 
redshift and luminosity distributions of the observed sample
are different from those of Type II, although with much overlap; 
(4) The afterglow luminosities are systematically lower than those 
of the Type II majority, although with some overlap. 
However, as discussed below, there are reasons to be suspicious
of this straightforward conclusion.

\subsection{Are some short/hard GRBs associated with Type II?}

The fact that some Type II Gold Sample GRBs are intrinsically short 
naturally raises the possibility that the observed short/hard GRBs are
contaminated by Type II GRBs. A small contamination is expected given
the overlapping log-normal distributions of $T_{90}$ 
for the two populations. A
more intriguing possibility is that the contamination is not the
simple extension of the $T_{90}$ distributions, but accounts for 
a good fraction of the observed short/hard GRBs. 

Conservatively speaking none of the
4 arguments discussed in \S5.1 is conclusive. 
Comparing with the Type I Gold Sample,
one cannot straightforwardly demonstrate that the Other SGRB Sample 
and the Type I Gold Sample come from the same parent sample. 
In particular, the Type I Gold Sample GRBs are relatively ``long''
and ``soft'' within the short/hard population, with 4 out of 
5 having extended emission. The more ``classical'' short/hard
ones, such as GRBs 051221A, 060313, and 061201, do not 
satisfy the Gold Sample criteria. After the 2005 revolution
of discovering GRB 050509B
and GRB 050724, which are associated with elliptical or early
type host galaxies, it is now clear that such associations
are not very common. Most short/hard GRBs are found to be
associated with star forming galaxies, some of which have properties
close to those of Type II GRBs.  According to \cite{berger09},
the majority of short GRBs appear to occur in star forming galaxies.
Although some compact merger events can have short merger time 
scales, and therefore appear in star forming galaxies 
\citep{belczynski06,zheng07,oshaughnessy08}, 
the dominance of star-forming host galaxies
of short GRBs may raise a concern regarding whether some
short GRBs may be associated with Type II. 
Although there is an observational selection
effect that favors redshift identification for star-forming 
galaxies, the GRBs with bright host galaxies but no redshift
identifications (e.g. GRB 051210, La Parola et al. 2006; 
Berger et al. 2007) are not the predominant population of
short/hard GRBs. The fraction of Type II contamination
can be small (e.g. $<1/3$ according 
to Berger 2009), but may not be negligible. We speculate that
some high-$L$ short GRBs in star-forming galaxies may instead
be Type II GRBs. For example, GRB 
060121 \citep{deugartepostigo06} has $E_{\gamma,iso}=2.2\times
10^{53}~{\rm erg}$ and $L_{\gamma,p,iso}=3.4\times 10^{54}~
{\rm erg~s^{-1}}$ for $z=4.6$ and  $E_{\gamma,iso}=4.2\times
10^{52}~{\rm erg}$ and $L_{\gamma,p,iso}=2.4\times 10^{53}
~{\rm erg~s^{-1}}$ for $z=1.7$. These energy values are typical 
for Type II GRBs, and the luminosity values even belong to
the bright end of the Type II distribution.
Its optical afterglow luminosity is also typical for Type 
II and much brighter than those of the Type I Gold sample.
Given the possible redshifts, this burst lies right on the
Amati-relation and the luminosity-spectral lag relations 
of most Type II GRBs (see Figs.[\ref{amati}] and [\ref{Llag}]).
GRB 060313 \citep{roming06b}, whose $z\leq 1.1$, can
have $E_{\gamma,iso}\sim 3.4\times 10^{52}~{\rm erg~s^{-1}}$ 
for $z=1$. Some afterglow light curve features 
(e.g. the very shallow decay of the UVOT, Roming et al.
2005, light curve with 
flickering features) are hard to 
accommodate within the merger scenarios. 
GRB 061201 also shows a very flat early light curve
\citep{stratta07} in both X-ray and optical bands, with the
first clear X-ray ``plateau'' appearing in a short GRB.
Some models \citep[e.g.][]{kumar08} attribute X-ray plateaus 
to the signature of massive star accretion. Within such a scenario, 
GRB 061201 is then a Type II candidate.
GRB 060121 also shows strong temporal variability in the 
afterglow \citep{deugartepostigo06}, similar to GRB 060313.
Even for the not very
energetic short/hard GRB 051221A, its identity as a Type I GRB is 
not unquestionable. Its host galaxy has 
$\log$(SSFR)$=0.804$, which is greater than those of many GRBs in the
Type II Gold Sample \citep{savaglio09}. A jet break is detected 
through ToO observations with the Chandra X-ray Telescope
\citep{burrows06}, which gives a jet corrected gamma-ray energy
of $E_\gamma \sim 5\times 10^{49}~{\rm ergs}$. This is smaller than
but not far off 
from the $E_\gamma$ distribution of Type II GRBs 
\citep{frail01,bloom03,liang08,racusin09}, and it is also
much higher than some other Type I GRBs (e.g. GRB 050509B
with $E_{\gamma,iso} \sim 1.1\times 10^{48}~{\rm ergs}$,
Gehrels et al. 2005).  Although a low density
$n \sim 10^{-4}-10^{-1}~{\rm cm^{-3}}$ was inferred from afterglow
modeling \citep{soderberg06b,burrows06}, it still belongs to the
reasonable $n$-range of other Type II GRBs 
\citep{panaitescu01,panaitescu02,yost03}. 
The recent short/hard GRB 090510 detected by both
Swift and Fermi (GBM/LAT) \citep{hoversten09,ohno09,guiriec09,rau09} 
is also located in a star forming host galaxy, and has an inferred 
total (collimation-corrected) jet gamma-ray and kinetic energies 
(assuming $n\sim 1~{\rm cm^{-3}}$) $E_\gamma/E_k \sim 10^{50}$ erg. 
It is again not far-off from the $E_\gamma/E_K$ 
distributions of Type II GRBs 
\citep{frail01,bloom03,liang08,racusin09}, and a Type II origin 
is possible. Finally, \cite{nysewander09} pointed out that
the optical-to-X-ray flux ratios of short GRBs are quite similar
to those of long GRBs, suggesting a similar circumburst medium 
density for the two populations. This is consistent with most
short GRBs being associated with Type II.

Another strong argument for the Type-II association of some 
(even many) short GRBs is related to luminosity function analyses.
Following the similar methodology of modeling $L-z$ distribution
of Type II GRBs in \cite{virgili09}, \cite{virgili09b} have 
studied the required
luminosity function of Type I GRBs in order to reproduce the
observed $L-z$ distribution for both Type I and Other SGRB 
samples. The results suggest
that the underlying luminosity function (defined as $N(L) dL 
\propto L^{-q}$) must be very shallow (e.g. $q\sim 0.5$) in 
order to reproduce the $L-z$ distribution data. This shallow
luminosity function is different from Type II GRBs and other
astrophysical objects. A more severe problem is that it 
cannot reproduce the observed $\log N - \log P$ distribution of
BATSE short/hard GRBs. This apparent conflict disfavors
the hypothesis that all short/hard GRBs are associated
with Type I.

Theoretically, the duration of a GRB is defined by Eq.(\ref{ts}). 
We now discuss the three relevant time scales in turn and 
address how a short GRB can in principle be associated
with Type II. 

Firstly, recent studies of the 
collapsar model suggest that the engine time scale $t_{engine}$
may not be always long. According to the standard collapsar
model \citep{woosley93,macfadyen99,proga03}, $t_{engine}$ can last
as long as the fallback material from the collapsar envelope is 
available to fuel the accretion disk or torus.
However, one should bear in mind that the rotating torus may form 
only when the specific angular momentum of the accreting gas is 
higher than the so-called
critical specific angular momentum value, 
i.e. $l_{\rm crit}=2 R_{\rm g} c$,
where $R_{\rm g}$ is the gravitational radius. 
Note that $l_{\rm crit}$ is proportional to the mass of the BH. 
During the collapsar evolution the mass accretion rate is very high, 
therefore the BH mass and consequently
the critical angular momentum increase very fast. 
As a result, the specific angular momentum of the
rotating material, which was initially sufficient for the torus
formation (i.e., when the BH was just formed), 
may become insufficient at a later stage of the 
collapsar evolution when the BH mass increases. 
\cite{janiuk08a} showed that the simple, often cited, estimates of 
the total mass available for torus formation and consequently 
the duration of a GRB \citep{macfadyen99,proga03} are only upper 
limits. They revised these estimates 
by taking into account the long term effect so that as the BH 
accretes the minimum specific angular momentum needed for torus 
formation increases. These new estimates predict a significant 
(an order of magnitude) reduction of the total energy and overall 
duration of the central engine $t_{engine}$ because only a fraction 
of the rotating stellar envelope can form a torus. 

If a Type II GRB is powered by the black hole spin
\citep{blandford77} rather than accretion, $t_{engine}$ of 
a Type II GRB can be also short, since accretion of materials
with a very low specific angular momentum would slow down the BH and 
consequently suppress the jet production. The interplay
among the BH mass, BH spin parameter, and the critical 
specific angular momentum of accreting gas needed for the 
torus to form have been discussed by 
\cite{janiuk08b}. They studied several different cases
and reached the conclusion that depending on the 
parameter settings, $t_{engine}$ can be as short as
a second. 

Secondly, the time scale during which a relativistic jet is
launched ($t_{jet}$) may be in principle shorter than the
central engine activity time scale ($t_{engine}$). There
is no working baryon-loading model for GRBs, and it is
not clear how a clean, high entropy outflow is launched.
For example, if the engine power has several episodes and
the power in the earlier episodes is not high enough,
the earlier jet may be choked or be launched but with a 
heavy baryon loading. The ejecta may therefore become
a ``dirty'' fireball. The GRB episode is then
only related to the late ``clean'' fireball phase when baryon
loading is reduced. 

Thirdly, energy dissipation is needed to convert other forms 
(kinetic and magnetic) of energy to radiation 
\citep{rees92,meszaros93,rees94,thompson94}. For a baryonic 
fireball, a steady outflow may not generate significant 
internal, non-thermal emission without internal shocks. 
The energy dissipation time scale ($t_{dis}$) can be 
smaller than $t_{jet}$. This gives an additional room to reduce
the duration of Type II GRBs.

Finally, some other possibilities of producing short GRBs
from collapsars have been proposed in the pre-Swift era.
For example, \cite{zhangw03} proposed that a short/hard
pulse of gamma-ray emission may be associated with eruption
of the fireball from the stellar envelope. \cite{yamazaki04b}
envisioned a geometric model to unify long and short GRBs
based on a line-of-sight effect. The original
pictures proposed in these papers are no longer supported
by the current data, but some ideas may be
borrowed to associate short GRBs with Type II
model category.

\subsection{Is there a ``Type III'' model category?}

The current data do not demand the existence of a third
type of GRB models to be associated with cosmological 
GRBs\footnote{Again ``cosmological
GRBs'' do not include SGR giant flares that are believed
to account for a small fraction of short/hard GRBs.}, 
i.e. those neither associated 
with massive star deaths nor compact star mergers. 
However, the possibility is not ruled out by the
data, either. Some ``hostless'' short GRBs \citep{berger09},
some long GRBs without X-ray afterglows \citep{vetere08},
and the SN-less long-duration, low-energy GRB 060505
\citep{fynbo06,ofek07,thone08,mcbreen08,kann09b}
are oddballs that may hold the clues to identify possible 
new model categories of GRBs.

\section{Nature of GRB 080913 and GRB 090423}

We now discuss the possible origin of GRB 080913 at $z=6.7$
and GRB 090423 at $z=8.3$. We mainly focus on GRB 080913. The
case of GRB 090423 is amazingly similar to GRB 080913, and
the conclusion for GRB 080913 can be directly applied to
GRB 090423 as well.

\subsection{Prompt properties and empirical correlations}

As discussed in \S2 (Fig.\ref{T90-HR}), GRB 080913 and GRB 090423
appear as long GRBs in the observer's frame, but are intrinsically 
short/hard GRBs in the rest frame. If one applies the criterion 
for the observed $T_{90}$, both bursts are ``long'' and therefore 
may be associated with Type II according to the ansatz
Eq.(\ref{types}). However, the 
association of a particular GRB to a particular physical model
type should not have a $z$-dependence. 
The {\em identical} burst, if it have occured at $z<1$, would be 
recognized as a short/hard GRB, and hence, a Type I candidate
according to the ansatz Eq.(\ref{types}). 
So it is not straightforward 
to determine the physical model category a GRB should be
associated with based on the observed 
$T_{90}$-HR data.

We inspect the compliance of GRB 080913 and GRB 090423 with the 
empirical correlations. First, they are consistent with the 
$E_p-E_{\gamma,iso}$ Amati-relation (Fig.\ref{amati}), although 
GRB 080913 is near the 3$\sigma$ upper boundary. This has been 
regarded as one argument in support
of the Type II origin of GRB 080913 by \cite{greiner09}.
On the other hand, GRB 080913 is also consistent with the new 
track defined by Type I and short/hard GRBs within 3$\sigma$. 
This suggests that the possibility that it is a Type I GRB 
(or at least similar to other short/hard GRBs) 
is not ruled out based on this criterion. The compliance of 
GRB 090423 with the Amati-relation is more robust.
Second, inspecting
the $L_{\gamma,iso}^p-{\rm lag}$ correlation, it seems that both 
GRB 080913 and GRB 090423 are consistent with being Type II GRBs 
- this was another 
argument by \cite{greiner09}. However, both GRBs are also 
consistent with the ``zero-lag'' trend of short GRBs.

In conclusion, based on statistical properties, both GRBs
can be taken as good Type II candidates. However, the criterion
based on empirical correlations is not robust enough to claim
the case, and supports from other criteria (see below) are needed
to draw firmer conclusions.

\subsection{Afterglow properties}

The rest-frame broadband (X-ray and optical) afterglow luminosities 
of GRB 080913 are moderate, bracketed between those of Type II and 
Type I Gold Sampels (Fig.\ref{Xray} and Fig.\ref{optical}). 
Although the early luminosities are relatively low, a distinct
energy injection episode raises the afterglow luminosity level of 
this burst to those of Type II at later epochs.
\cite{greiner09} have modeled
the afterglow and suggested that the light curves are consistent
with the deceleration of a relativistic jet by a dense circumburst 
medium with constant density. The data are consistent with the
existence of an achromatic plateau in X-ray and several optical/IR
bands, which can be interpreted within the framework of a 
continuously fed forward shock.
Alternatively, the X-ray rebrightening around $10^5$ s may be
due to an X-ray flare, whose softer emission may also account
for the rebrightening in the IR/optical bands.

Improving upon \cite{greiner09}, 
we have performed a more detailed numerical modeling of the 
broad-band afterglow data. The optical emission can be well 
modeled by standard synchrotron radiation from the forward shock, 
while the X-ray emission is likely dominated by the synchrotron 
self-Compton emission (SSC). 
The following parameters can fit the optical data well (although
we do not apply a parameter search to judge whether this is the
best fit): the initial isotropic kinetic energy of the fireball 
$E_{K,iso} \sim 3.7 \times 10^{52}$ erg,  
the ambient density $n \sim 3000~{\rm cm^{-3}}$.
the electron equipartition parameter $\epsilon_e \sim 0.04$,
the magnetic field equipartition parameter $\epsilon_B \sim 
10^{-5}$, and the electron spectral index $p \sim 2.2$. 
If the late rebrightening is interpreted as an energy injection from 
the central engine with a time-dependent luminosity $L=L_0 (t/t_0)^q$,
the data are consistent with $t_0 \sim 6.5\times 10^3 ~{\rm s}$,
$L_0 \sim 2.9 \times 10^{50}~{\rm erg~s^{-1}}$, $q\sim 1$ for
$t<t_0$, and $q\leq -1$ for $t>t_0$. 
In any case, a high-density constant medium is needed. 
This is consistent with the expectation at high-$z$ \citep{gou04}
as well as the fitted density of GRB 050904 at $z=6.3$ 
\citep{frail06,gou07}. The data also demand 
a jet opening angle $\theta_j > 0.22$ rad, which corresponds 
to a geometrically-corrected total gamma-ray energy
$E_\gamma > 1.7\times 10^{51}$ ergs, and a geometrically-corrected
total kinetic energy $E_{K} > 9.0\times 10^{50}$ ergs. 
This value is consistent with those of Type II GRBs
\citep{frail01,bloom03,liang08,racusin09}.
This is probably the strongest argument in favor of 
associating the burst with Type II. 

GRB 090423 has even brighter X-ray and optical afterglow
luminosities than GRB 080913. Although we did not perform
detailed afterglow modeling, the afterglow
parameters favor those of Type II GRBs, similar to GRB 080913.

\subsection{Short Type II or high-$z$ Type I?}

Since the discovery of GRB 080913, there has been a debate
about its progenitor. As discussed above,
data analyses and theoretical modeling suggest
that GRB 080913 is very likely a Type II GRB 
\citep{greiner09}, although a Type I association
\citep{perezramirez09,belczynski09} is not ruled out.
The evidence in support of the Type II origin of GRB 080913
includes: large values 
of the geometrically-corrected gamma-ray ($E_\gamma$) and kinetic 
($E_K$) energies, moderately bright intrinsic afterglow luminosities, 
a required high density of the circumburst medium,
and the marginal compliance of the $E_p - E_{\gamma,iso}$ relation
of Type II GRBs. 

On the other hand, if GRB 080913 were a high-$z$ Type I GRB,
as suggested by its intrinsically short duration, 
it would have to be an energetic merger event, likely
due to a BH-NS merger with a rapidly rotating massive BH.
The energy tapping mechanism would have to be the \cite{blandford77}
mechanism \citep{perezramirez09}. 
For this possibility, one requires that during
the short age of the Universe at $z=6.7$, i.e. $\tau \sim 8.3
\times 10^8$ yr for the concordance universe, 
a BH-NS system is formed and merged. \cite{belczynski09} have 
modeled this possibility in detail, and claimed that the event
rates for massive star core collapses that give rise to Type
II GRBs and for compact star mergers (both NS-NS and BH-NS)
are comparable at $z=6.7$. They concluded that both scenarios 
are possible. However, there are several factors that would
change this conclusion. First, 
\cite{belczynski09} assumed that all the mergers give rise
to GRBs. In reality it may be that only a fraction of
mergers give rise to GRBs. This fraction factor may be 
calibrated through confronting the observed number ratio
of Type I and Type II GRBs with the model predictions.
In the current population synthesis models, this factor
is not taken into account (K. Belczynski, 2008, personal
communication). Secondly, GRB 080913 would be a high-luminosity
Type I GRB if it is associated with
that category. Considering the power law luminosity
function inferred for Type I GRBs \citep{virgili09b}, 
detecting one high-$L$ event would demand many more low-$L$ 
events, which would require a significant increase of 
the required event rate of compact star mergers that
is inconsistent with the results of population synthesis.
Finally, as we argued above, the large
value of the geometrically-corrected gamma-ray and 
afterglow energies do not favor a NS-NS merger model.
Only BH-NS mergers with a highly spinning BH should be
counted. This would greatly reduce the theoretically
predicted event rate that satisfies the constraint.

The detection of GRB 090423, another intrinsically short/hard,
high-$z$, high-$L$ GRB, strongly supports a Type II association
of both GRB 080913 and GRB 090423.
As argued above, the probability of detecting 
a high-$L$ Type I event at high-$z$ is much smaller than that
of detecting a moderate-$L$ Type II event. With one event,
one may still argue for a chance coincidence. With the detection
of GRB 090423, the chance probability of detecting two high-$L$ 
merger events at high-$z$ is greatly reduced, and one can 
more firmly associate both GRBs with Type II.

With two intrinsically short high-$z$ Type II GRBs detected, one must
ask why these events tend to exist at high-$z$. One possibility
would be that it is simply a threshold selection effect. Both events
have moderate gamma-ray luminosities, and were detected not far
above the threshold. It is possible that there exists other
softer pulses that are below the sensitivity threshold of Swift/BAT
(J. S. Bloom, 2009, private communication). On the other hand,
such softer emission would be easily detected by Swift/XRT if it 
was indeed there. Both GRB 080913 and GRB 090423 have X-ray
flares. However, extrapolating them into the gamma-ray band
using a simple spectral model suggests that they would appear
as low-level extended emission of short GRBs (Fig.1). One
may still argue for missing soft emission before the XRT slew.
This is not ruled out, but the XRT slew time corresponds to
a rest frame time 99.5/7.7=12.9 s for GRB 080913 and
72.5/9.3=7.8 s for GRB 090423. The intrinsic $T_{90}$'s
have to be in any case smaller than these values for these
bursts. 

A more intriguing possibility would be that this 
is due to a physical origin and reflects the intrinsic
property of the high-$z$ massive stars. These high-$z$
stars may not be spinning as rapidly as their low-$z$
sisters, so that only a smaller mass is left after the 
prompt collapse. More high-$z$ GRB data are needed to
test whether such a scenario is demanded by the data.

\section{How to Associate a Burst with a Physical Model Category?}

The extensive discussion presented above suggests that 
it is not always easy to associate a particular GRB
to a particular physical model category
based on observational 
criteria. The multiple observational criteria discussed 
in this paper are summarized
in Table 2. This is an extension of Figure 2 of 
\cite{zhangnature06}. New criteria are added based on
the discussion in this paper. A new column lays out
the issues of each criterion. The criteria are sorted
by relevant observations. The first six rows (duration,
spectrum, spectral lag, $E_{\gamma,iso}$, $E_p-E_{\gamma,iso}$
relation, and $L_{\gamma,iso}^p$-lag relation, are based on
the gamma-ray properties only. The next five rows 
(supernova association, circumburst medium type,
$E_{K,iso}$, jet opening angle, and the geometrically
corrected energies $E_\gamma$ and $E_K$), are 
based on follow-up broadband observations and afterglow
modeling. The next three rows (host galaxy type, specific
star forming rate of the host galaxy, and offset of the
GRB from the host galaxy) are based on observations of the
host galaxies. The next two rows (redshift distribution
and luminosity function) are statistical properties.
The final row is the gravitational wave criterion.
In general, most of
these criteria are not ``conclusive'', i.e., one cannot
draw a firm conclusion based on a single criterion.
Nonetheless, there are several criteria which, if satisfied,
would unambiguously associate a GRB to a certain
physical model category.
These are marked in bold in Table 2. In particular, if
a GRB is found in an elliptical or an early type galaxy, 
or if the SSFR of its host galaxy is very low, one would be
able to associate it with Type I. 
On the other hand, a SN association or the identification 
of a wind-type medium in a GRB would establish its
association with Type II.

Unfortunately, the above four criteria are usually not satisfied
for most GRBs. One is then obliged to use multiple criteria 
since there are overlapping predicted properties between the 
two physical model types 
for each individual criterion. In Fig.\ref{flowchart} we cautiously
propose an operational procedure to discern the physical
origin of a GRB based on the available data. 

Several features are worth commenting on in Fig.\ref{flowchart}.
(1) The criteria to define the physical model category a
burst is associated with are less stringent compared with those 
used to define the
Gold Samples in \S5.1. This is because the purpose of the
Gold Samples was to allow us to perform statistical analyses.
After reviewing the statistical properties 
in \S5, we have gained confidence on additional criteria
so that more bursts can be analyzed. (2) There are five
outcomes in the flowchart. Besides the solid Type I/II
identifications, we also define Type I/II ``candidates''
and the ``unknown'' category. The Type I/II
candidates refer to those with evidence of associating a
burst to a particular physical model category, 
but the evidence is not strong enough to
make a firm claim. The unknown category includes the oddball
GRBs that do not obviously fit into any criteria discussed
in this paper, or the observational data are not adequate 
for us to make the judgement. They may be associated with 
Type I, Type II or a completely new type of models. 
(3) Some qualitative rather than
quantitative criteria have been used (e.g. high/low SSFR,
large offset, large/small $E_\gamma$, $E_K$). The reason is 
that it is very difficult to adopt quantitative criteria at
the current stage, since the distributions of these quantities 
predicted by both physical model types and displayed
in the statistical analyses of the Type I/II Gold Samples
are continuous, without sharp transitions. 
The ``high/low'' and ``large/small'' definitions are based
on the statistical properties, and therefore
in the relative sense. If confusion occurs (e.g. the quantity
is near the boundary and not easy to judge whether it is
high/low, large/small, one can follow the ``?'' sign to 
go down the flowchart. The flowchart is reasonably 
operational, i.e. essentially every GRB with reasonable 
afterglow follow up observations can find a destiny in 
the chart. For example, the SN-less long-duration GRB 060614 
\citep{gehrels06,galyam06,fynbo06,dellavalle06} is 
associated with
Type I (based on low SSFR), and the other SN-less
GRB 060505 \citep{fynbo06,ofek07,thone08,mcbreen08} can be
associated with a Type I candidate based on its small
energetics, or an ``unknown'' burst if one
argues that the $L_{\gamma,iso}^p-$lag relation is satisfied
for this burst \citep{mcbreen08}. 
GRB 080913 and GRB 090423 (the main topic of this paper) find
their homes as Type II candidates based on the $E_p-E_{\gamma,iso}$
correlation. GRB 060121 (a high-$z$ short GRB) satisfying
the $E_p-E_{\gamma,iso}$ is also found to be associated
with the ``Type II candidate'' outcome in the flowchart.
(4) It is possible 
that the procedure and the criteria may be further revised as 
more data are accumulated. The current procedure 
only reflects the best knowledge for the time being.

In the flowchart, there are five thick arrows that bridge
the short-duration and long-duration GRBs. This suggests that
the duration information sometimes is misleading. 
Some long duration GRBs can be associated with Type I 
(e.g. GRB 060614 and probably GRB 080503, Peyley et al. 2008), 
and some short duration GRBs can be associated with
Type II (e.g. GRB 060121, GRB 080913 and GRB 090423). 
We also present 
two dashed arrows in the flowchart. These
two tracks (a short GRB associated with a SN and a long GRB
with an elliptical/early type host galaxy) are in principle 
possible, but such bursts have never been observed so 
far\footnote{The GRB field is full of surprises. If
some short/hard GRBs are indeed associated with
Type II as argued in this paper, one may someday discover a SN 
associated with a short/hard GRB. We encourage continuous SN 
searches for all nearby GRBs, both long and short.}. 
The order of the criteria in Fig.8 is based on the 
``definiteness'' of the criteria,
with the higher-level ones carrying more
weight than the lower-level ones. Notice
that ``hardness'' is generally not regarded as a definitive 
criterion in the flowchart (except for the relative hardness 
of the short spike and the extended emission). 

\section{Summary}

\begin{figure}
\plotone{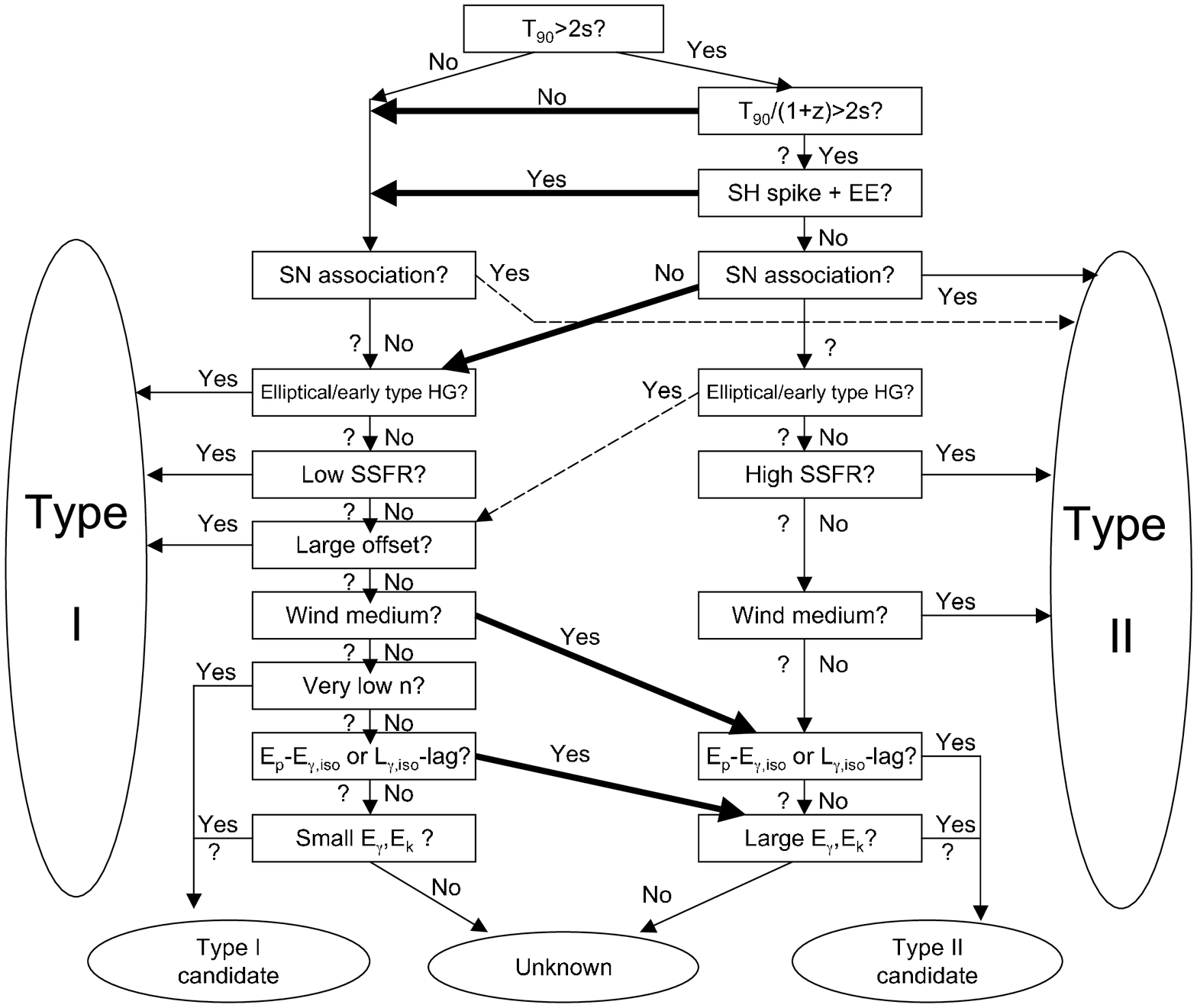}
\caption{A recommended procedure to judge the association
of a particular GRB to a particular physical model category.
Multiple observational criteria have been applied.
Question marks stand for no information being available to judge
the validity of the criterion. The two dotted arrows stand
for the possibilities that are in principle possible but have
never been observed. Five thick arrows bridge the long-duration
and short-duration GRBs, suggesting that the there can be
long duration Type I and short duration Type II GRBs.}
\label{flowchart}
\end{figure}

Prompted by the interesting question whether the $z=6.7$ GRB 
080913 and $z=8.3$ GRB 090423 are intrinsically short GRBs 
associated with the Type II physical model category
\citep{greiner09} or high-$z$ GRBs 
associated with the Type I physical model category
\citep{perezramirez09}, 
we performed a more thorough investigation on the two physically
distinct categories of GRB models and their predicted
observational characteristics. We further developed the ``Type I/II''
concept proposed in \cite{zhang07b} in the following directions.
(1) We have reviewed and expanded the possible multiple 
observational criteria, and discussed their physical origins
from the theoretical point of view. By doing so, we are able
to differentiate those criteria that are more closely related
to the progenitor types and those that are more directly
related to radiation physics. In particular, we argue that
SN association, host galaxy properties (type and SSFR),
and the offset of the GRB location in the host galaxy are
more directly related to the progenitor types. The gamma-ray
properties, such as duration, hardness, spectral lag, 
empirical correlations, are more related to jet dissipation 
and radiation processes in the emission region, and can only
be related to progenitors indirectly. Afterglow and statistical 
properties can be used to diagnose GRB progenitor, but 
theoretical modeling is needed. Gravitational wave
signals may be the best criterion to directly probe the
progenitor system, but they are too faint for the current 
detectors to detect. 
(2) We use several key observational
criteria that are directly related to GRB progenitors to
define the Gold Samples for Type I and Type II, respectively.
These criteria do not involve GRB gamma-ray emission properties
such as duration, hardness, spectral lag, etc.
We then use these samples to investigate their statistical
properties, especially their distribution in the 
duration-hardness space. We found that the Type II Gold Sample
represent the BATSE long/soft population well. The Type I
Gold Sample, on the other hand, is not very representative 
of the short/hard population. The Type I Gold Sample GRBs
are typically ``long'' and not particularly ``hard''. 
(3) Although some short/hard GRBs detected
in the Swift era may share a similar origin as the Type I
Gold Sample, we suggest that {\em some (maybe most) high-$L$ 
short GRBs may be instead associated with Type II,
namely, of a massive star origin}. 
(3) We summarized the multiple 
observational criteria
needed to discern the physical origin of a GRB
in Table 2, with various issues laid out. We emphasize that 
it is not always straightforward to judge the physical model 
category a particular GRB is associated with, and we cautiously 
proposed an operational procedure 
to discern the physical origin of GRBs (Fig.\ref{flowchart}). 
(4) According to this procedure,
GRB 080913 and GRB 090423 are Type II candidates. 
Although a specific Type I scenario invoking the Blandford-Znajek 
mechanism of a BH-NS merger system is not completely ruled out,
the fact that two such GRBs
are detected at high-$z$ indeed suggest that a Type I 
association of these bursts is essentially impossible. 

The proposed procedure to associate a particular GRB to
a particular physical model category
is subject to further test with new observational data. More detailed
analyses may allow more quantative criteria to discern the physical
origin of GRBs. Based on past experience, the chances are high that 
new observations will bring surprises that continuously call for 
modifications of the criteria, which would further our understanding
of the physical origins of cosmological GRBs.

\acknowledgments
We thank the referee, Jon Hakkila, for insightful comments
that significantly improved the presentation of the paper.
This work is partially supported by NASA (through grants NAG05GB67G,
NNX08AN24G, and NNX08AE57A at UNLV, and NNX08AL40G at PSU), 
the National Natural Science Foundation (Award ID 0908362 for BZ),
the National Natural Science Foundation of China (grant 10873002
for EWL, and grants 10503012, 10621303, 10633040 for XFW),
and the National Basic Research Program of China (''973" Program
2009CB824800 for both EWL and XFW), and the research foundation 
of Guangxi University (Grant M30520 for EWL). BBZ acknowledges
the President's Fellowship and GPSA Awards from UNLV.  FJV
acknowledges the Nevada Space Grant Consortium Fellowship.
We acknowledge helpful discussion/comments from K. Belczynski, 
J. S. Bloom, S. Covino, D. B. Fox, A. Fruchter, J. Fynbo, J. Greiner, 
D. Malesani, P. O'Brien, and S. Savaglio.

\bibliography{ms}

 \clearpage
        \LongTables 
        \begin{landscape}

\begin{deluxetable}{lllllllllllll}
\tablecolumns{11}
\tabletypesize{\tiny}
\tablewidth{-10pt}
\tablecaption{Samples}
\tablehead{
\colhead{GRB} &
\colhead{ $z$} &
\colhead{ log\ SSFR} &
\colhead{SN?} &
\colhead{$T_{90}$ } &
\colhead{$T_{90}$ w/ EE } &
\colhead{HR$^a$} &
\colhead{$lag^b$} &
\colhead{$E_{p}$ } &
\colhead{$E_{\gamma,iso}$} &
\colhead{$L_{p,iso}$ } &\\
\colhead{name} &
\colhead{redshift} &
\colhead{Gyr$^{-1}$} &
\colhead{} &
\colhead{sec} &
\colhead{sec} &
\colhead{$\frac{S(50-100keV)}{S(25-50keV)}$} &
\colhead{sec} &
\colhead{keV} &
\colhead{$10^{52}$ erg } &
\colhead{$10^{50}$ erg/s} 
}
\startdata
\multicolumn{11}{c}{Type II Gold}\tabularnewline
\hline\noalign{\smallskip}

970228& 0.695 & 0.082           &  ?  &       $\sim80$  &      n/a          & 1.07 &    $0^c$        &   $115\pm 38$         &     $1.6\pm 0.1$  &   $93.3_{-6.1}^{+5.7}$    \\

970508& 0.835 & 0.534           &  ?  &     $\sim 23.1$  &      n/a          & 1.09 &    $0.384^{+0.090,b}_{-0.026}$        &   79$\pm 23$ &$0.61\pm 0.13$ & $14.3^{+0.5}_{-0.6}$   \\

971214& 3.418 & 0.467           &  ?  &     $31.0\pm 1.2$   &      n/a          & 1.63 &    $0.066^{+0.026}_{-0.048}$        &   155$\pm 30$ &$21\pm3$ &$684\pm65$  \\

980425& 0.0085 & -0.883        &  Y  &     $23.3\pm 1.4$   &      n/a          & 1.08&    
        1.46$\pm$0.18 &   $119\pm 24$             &    $(6.1\pm0.62)\times 10^{-5}$ &  $4.8_{-7.8}^{+7.5}\times 10^{-4}$ \\

980613& 1.0964 & 1.184        &  ?  &     50   &      n/a          & 1.59&    
  ...$^d$                  &  $93\pm 43$      &                         $0.59\pm 0.09$ &                           $16.7_{-4.7}^{+3.9}$\\

980703& 0.966 & 0.885        &  ?  &     411.6$\pm$9.3   &      n/a          & 1.47&    
$0.402^{+0.162}_{-0.134}$ &  $254\pm 51$            &                         $7.2\pm 0.7$  &                          $166_{-31}^{+32}$ \\

990123& 1.6 & 0.340        &  ?  &     63.4$\pm$0.3   &      n/a          & 2.06&    
$0.018^{+0.012}_{-0.012}$ &  $781\pm 62$            &                         $229\pm 37$  &                         $3517_{-198}^{+210}$ \\

990506& 1.30658 & -0.081        &  ?  &     130.0$\pm$0.1   &      n/a          & 1.44&    
$0.04\pm 0.02$ & $283\pm 57$             &                         $94\pm 9$ &                         $930_{-52}^{+54}$ \\

990712& 0.4331 & 0.093        &  ?  &     $\sim 30$   &      n/a          & 0.98&    
0.045$\pm$0.014 &  $65\pm 11$           &                         $0.67\pm 0.13$  &                         $73.1_{-6.4}^{+5.9}$ \\

991208& 0.707 & 1.121       &  ?  &     $\sim 68$   &      n/a          & 1.25&    
... &  $183\pm 18$             &                         $22.3 \pm 1.8$  &                         $110\pm 11$\\

000210& 0.846 & 0.049       &  ?  &     $\sim 15$   &      n/a          & 1.19& ...
 &  $408\pm 14$            &                         $14.9\pm 1.6$  &       $1003_{-79}^{+80}$       \\

000418& 1.1181 & 0.757       &  ?  &     $\sim 30$   &      n/a          & ?&    
... &  $134\pm 10$           &                         $9.1\pm 1.7$  &           $11.3_{-4.1}^{+4.0}$\\

000911& 1.0585 & -0.124       &  ?  &     $\sim 500$  &      n/a          &  2.14 &    
... &  $579\pm 116$             &                         $67\pm 14$  &                         $558_{-95}^{+128}$\\

000926& 2.0379 & -0.165       &  ?  &     $\sim 25$  &      n/a          &  0.37 &    ... &  $101\pm 6.5$             &                         $27.1\pm 5.9$  &                        $107\pm 43$\\

011121& 0.362 & -0.464      &  Y  &     $\sim 30$  &      n/a          &  0.78 &    
... &  $217\pm 26$            &                         $7.8\pm 2.1$ &                       $49.8\pm 4.0$\\

011211& 2.14 & -0.084      &  ?  &     $\sim 270$  &      n/a          &  1.87 &    
...&  $59\pm7$            &                        $5.4\pm 0.6$  &                       $21.8_{-5.2}^{4.8}$\\

020405& 0.695 & -0.174      &  Y  &     $\sim 60$  &      n/a          &  3.23 &    
... &$364\pm 73$           &                         $10\pm 0.9$  &                        $117_{-6.7}^{+7.2}$\\

020813& 1.255 & 1.167     &  ?  &     113.0$\pm$1.1  &      n/a          & 1.58 &    
$0.16\pm 0.04$ &  $142\pm 13$             &                         $66\pm 16$  &                          $450_{-86}^{+94}$\\

020819B& 0.41 & -0.664     &  ?  &     $\sim 50.2$  &      n/a          & 1.07 & ... &  $50\pm 15$            &                         $0.68\pm 0.17$  &                         ...\\

020903& 0.25 & 0.555     &  Y  &     $\sim 13$  &      n/a          & 0.66 &  ... &  $3\pm 1$            &                         $(24\pm6)\times 10^{-4}$ &                         ...\\

021211& 1.006 & -0.841     &  Y  &    $ \sim 8$  &      n/a          & 0.98 &  $0.32\pm 0.04$
 &  $46\pm7$        &                        $1.12\pm 0.13$  &                        $155_{-29}^{+33}$\\

030328& 1.52 & 0.680     &  ?  &     $\sim 199.2$  &      n/a          & 1.43 &    
 $0.2\pm 0.2 $ &  $126\pm 13$           &         $47\pm 3$             &             $191\pm 38$\\

030329& 0.1685 & 0.304     &  Y  &     $\sim 62.9$  &      n/a          & 1.13 &    
$0.58_{-0.36}^{+0.60}$ &  $68\pm 2$          &                        $1.5\pm 0.3$ &                         $22.5\pm 4.5$\\

030528& 0.782 & 1.355     &  ?  &     $\sim 83.6$  &      n/a          & 1.23 &    
$12.5\pm 0.5$ &  $62\pm 3$           &                       $2.5\pm 0.3$  &                         $17.3_{-3.4}^{+3.6}$\\

031203& 0.1055 & 1.287     &  Y  &     $\sim 40$  &      n/a          & 0.65 &    
0.24$\pm$0.12 &  $\sim 292$        &                      $\sim 0.01$ &                         $0.12_{-0.02} ^{+0.03}$\\

040924& 0.858 & 0.071     &  ?  &     2.39$\pm$0.24  &      n/a          & 1.00 &    
$0.3\pm 0.04$ &  $67\pm 6$           &                       $  0.95\pm 0.09$  &    $191\pm 20$\\

041006& 0.716 & -0.131     &  ?  &     17.40$\pm$0.25  &      n/a          & 1.08 &    
... &  $63\pm 13$            &                         $3\pm 0.9$ &                 $44_{-1.8}^{+1.7}$ \\   

050525A& 0.606 & ?    &  Y  &     8.830$\pm$0.004  &      n/a          & 1.17 &    
$0.0865^{+0.0065}_{-0.008}$ &  $84.1\pm 1.7$      &                       $ 2.89\pm 0.57$ &                         $111.8 \pm 2.1$\\

050826& 0.297 & 0.172     &  ?  &    35.5$\pm$1.2  &      n/a          & 1.91 &    
... &  $340^{+790}_{-210}$       &                    0.03$\pm$0.04    &      $0.33_{-0.08}^{+0.32}$\\

051022& 0.8 & 0.142     &  ?  &    $\sim 200$  &      n/a          & 1.52 &
... &  $418\pm 143$        &                       $53\pm 5$ &                        $364_{-47}^{+48}$\\

060218& 0.033 & -0.061     &  Y &    $\sim 2000$  &      n/a          & 0.76&    
218$_{-140}^{+356}$ & $4.9\pm 0.3$          &                         $(77\pm1.4) \times 10^{-4}$ &                        $1.0\pm0.6\times 10^{-3}$\\

060602A& 0.787 & ?   &  ? &    75.0$\pm$0.2  &      n/a          & 2.65&    
... & $280^{+570}_{-150}$   &             $0.91\pm 0.06$         &   $6.14_{-0.80}^{+2.54}$                 ...\\

080520& 1.545 & ?     &  ? &    2.82$\pm$0.67  &      n/a          & 0.46&    
... & $\sim 30$         &          $0.073\pm 0.019$   &                         ...\\
\hline\noalign{\smallskip}
\multicolumn{11}{c}{Type I Gold}\tabularnewline
\hline\noalign{\smallskip}

050509B& 0.2248 & -0.853     &  N &    0.040$\pm$0.004  &      n/a          & 1.52&    
0.0043$\pm 0.0032$ & $82_{-80}^{+611}$           &                       $ 2.4_{-1}^{+4.4} \times 10^{-4}$ &                        $0.07_{-0.05}^{+0.10}$ \\

050709& 0.1606 & -0.512    &  N &    0.07$\pm$0.01  &      $130\pm 7$          & 1.37/1.02$^j$&    
$0\pm 0.002$ &  $83_{-12}^{+18}$      &                        $(2.7\pm 1.1)\times 10^{-3}$  &                         $5.4_{-0.69}^{+0.67}$\\

050724& 0.2576 & -0.367    &  N  &    3$\pm$1  &      $154.20\pm 1.12$          & 1.26/1.12&    
$-0.0042\pm 0.0082$ &  $110_{-45}^{+400}$          &         $9_{-2}^{+11} \times 10^{-3}$                  &                         $0.99_{-0.10}^{+0.23}$\\

060614& 0.1254 & -0.863    &  N &    $\sim 5 $  &      $106.0\pm 3.3$          & 1.41/1.07&    
$0.003 \pm 0.009$ &  $302_{-85}^{+214}$            &                    $0.24\pm 0.04$ &                         $1.39_{-0.07}^{+0.13}$\\

061006& 0.4377 & -2.189    &  N &    $\sim 0.5$  &      $120.00\pm 0.04$          & 1.52/1.18&    
... & $640_{-227}^{+144}$          &                      $0.22\pm 0.12$  &                         $24.60_{-0.77}^{+1.22}$\\

\hline\noalign{\smallskip}
\multicolumn{11}{c}{Other Short-Hard Bursts}\tabularnewline
\hline\noalign{\smallskip}

000607& 0.14 & ?    &  ? &    $\sim 0.008$  &      n/a          & 2.18&    
... &  ...         &                         ...  &                         ...\\

050813& $\sim$0.72 & ?   &  N &    0.6$\pm$0.1  &   n/a          & 1.76&    
$-0.0097 \pm 0.014$ & $210_{-130}^{+710}$           &                  $(1.5_{-0.8}^{+2.5})\times 10^{-2}$  &                         $4.13\pm 2.02$\\

051210$^g$& $>$1.4    & ? &  ? &   1.27$\pm$0.05  &   40        & 2.01&    
$-0.0053\pm 0.024$& $410_{-260}^{+650}$            &                      $>0.191\pm 0.032$ &                         ...\\

051221A& 0.5464 & 0.804    &  ? &   1.4$\pm$0.2  &   n/a          & 1.74&    
$0\pm 0.004$ & $402_{-93}^{+72}$          &                      $0.28_{-0.1}^{+0.21}$     &                       $ 25.8\pm 0.9$\\

060121& 1.7/4.6    & ? &  ? &   1.60$\pm$0.07  &   $\sim 120$        & 1.55/0.57$^h$&    
0.017$\pm$0.009$^i$ &  $104_{-78}^{+134}$           &                      $4.18_{-0.39}^{+3.29}$/$22.3_{-2.07}^{+17.5}$   &                         $2445\pm 162 /33574\pm 2226$\\

060313& $\le 1.1$  & ?&  ? &   0.7$\pm$0.1  &   n/a       &  2.43&  $(3 \pm 7)\times 10^{-4}$  
 & $922_{-177}^{+306}$      &           $\leq 6.24_{-3.66}^{+0.43}$            &                         ...\\

060502B& $0.287$  & ? &  ? &   0.09$\pm$0.02  &   n/a       &  2.12& $(-2 \pm 8)\times 10^{-4}$  
 &  $340_{-190}^{+720}$       &                        $3_{-2}^{+5}\times 10^{-3}$  &                        $0.65\pm 0.09$\\

060505& $0.0889$  & -0.777 &  ? &   4$\pm$1  &   n/a       &  1.63&  0.36$\pm$0.05  
 &$\sim 223$         &                         $(3.39\pm 0.60)\times 10^{-3}$  &                      $\sim 0.009^k$\\

060801& $1.131$  & ?&  ?&   0.5$\pm$0.1  &   n/a       &  2.89&  $0.008\pm0.008$  
 &      $620_{-340}^{+1070}$   &                         $0.17\pm 0.021$ &                   $47.6_{-1.6}^{+6.2}$\\

061201& $0.111?$  & ?& ? &   0.8$\pm$0.1  &   n/a       &  2.90&  $2.7^{+3.3}_{-2.4}\times 10^{-3}$  
 &       $873_{-284}^{+458}$      &                        $0.018_{-0.015}^{+0.002}$ &                         ...\\

061210& $0.4095$  &? &  ? &   $\simeq$0.06  &  85$\pm$5      &  2.32/1.37&  ... 
 & $540_{-310}^{+760}$       &                      $0.09_{-0.05}^{+0.16}$  &                         $21.5\pm 1.4$\\

061217& $0.8270$  & ?&  ? &   0.212$\pm$0.041 &  n/a     &  2.07 &  -0.007$\pm$0.009$^j$ 
 & $400_{-210}^{+810}$        &                          $0.03_{-0.02}^{+0.04}$   &                        $10.8\pm 1.8$\\

070429B& 0.9023    & ?&  ? &   0.5$\pm$0.1  &   n/a          & 1.23&    
... &  $120_{-66}^{+746}$            &                       $0.03\pm 0.01$  &                      $ 24.6\pm 3.8$\\

070714B& 0.9225    & ?&  ? &   $\sim 3$  &  $ \sim 100 $         & 1.82/1.56&    
$0.014\pm 0.007$ & $1120_{-380}^{+780}$           &                      $1.16_{-0.22}^{+0.41}$ &                         $57.3\pm 3.6$\\

070724A& 0.457    & ?&  ? &   0.50$\pm$0.04  &   n/a          & 0.94&... 
&  $\sim 68$       &                       $0.003\pm 0.001$ &              $1.58_{-0.14}^{+0.34}$\\

071227& 0.3940    & ? &  ? &   1.8$\pm$0.4  &   $\sim 100$        & 2.02/0.96&    
$(0.4\pm 14)\times 10^{-4,l}$ & $\sim 1000$          &                 $0.22 \pm 0.08$&                    $3.34\pm 0.49$ \\

080503& ...  & ? &  N &   $\sim $0.7  &  170$\pm$ 40     &  1.0 &  -0.013$\pm$0.009$^m$
 & ...      &  ...                        &                         ...\\



  \hline\noalign{\smallskip}
080913& 6.7 &  ?& ? & $8\pm 1$ & n/a & 1.58& $0\pm 0.42$ & $121_{-39}^{+232}$ & $7\pm 1.81$& $ 1200_{-300}^{+1622}$\\

090423& 8.3 & ? & ? & $10.3\pm 1.1$ & n/a & 1.50 & $0.046^{+0.085}_{-0.058}$ & $48^{+6}_{-5}$ & $10 \pm 3$ & $\sim 1880$ \\
  \hline\noalign{\smallskip}

\enddata
\tablecomments{Values of $E_p$ and $E_{\gamma,iso}$ are taken from \cite{amati08} and $L_{p,iso}$ are caculated in this work unless otherwise stated below. Futher references are:
 \textbf{GRB970228}- $z$:\cite{1998IAUC.6896....3W}; SSFR:\cite{savaglio09}; $T_{90}$,spectrum$^c$: \cite{1998ApJ...493L..67F}; lag:\cite{2007AandA...474L..13B}; \textbf{GRB970508}-$z$:\cite{1997Natur.387..878M};SSFR:\cite{savaglio09};$T_{90}$: \cite{1999ApJS..122..465P}; spectrum: \cite{1997IAUC.6660....1D}; lag: \cite{norris00}. \textbf{GRB971214}-$z$:\cite{1998Natur.393...35K};SSFR:\cite{savaglio09};$T_{90}$: \cite{1999ApJS..122..465P}; spectrum: \cite{2000AandA...355..454D}; lag: \cite{norris00} 
\textbf{GRB980425}-$z$:\cite{1998IAUC.6896....3W};SSFR:\cite{savaglio09};$T_{90}$: \cite{1998Natur.395..670G}; spectrum: \cite{2004AIPC..727..416Y}; lag: \cite{2008ApJ...685.1052Z}. \textbf{GRB980613}-$z$:\cite{1999GCN...189....1D};SSFR:\cite{savaglio09};$T_{90}$: \cite{1998IAUC.6938....1S}; spectrum: \cite{2001grba.conf..201S}; ;lag: \cite{norris00}. \textbf{GRB980703}-$z$:\cite{1998ApJ...508L..17D};SSFR:\cite{savaglio09};$T_{90}$: \cite{1999ApJS..122..465P}; spectrum: \cite{ghirlanda04}; lag: \cite{norris00}. \textbf{GRB990123}-$z$:\cite{1999Sci...283.2075A};SSFR:\cite{savaglio09};$T_{90}$: \cite{1999ApJS..122..465P}; spectrum: \cite{ghirlanda04}; lag: \cite{norris00}. \textbf{GRB990506}-$z$:\cite{2003AJ....125..999B};SSFR:\cite{savaglio09};$T_{90}$: \cite{1999ApJS..122..465P}; spectrum: \cite{ghirlanda04}; lag:\cite{2007ApJ...660...16S}. \textbf{GRB990712}-$z$:\cite{2001ApJ...546..672V};SSFR:\cite{savaglio09};$T_{90}$: \cite{1999IAUC.7221....1H}; spectrum: \cite{2001ApJ...550L..47F}; lag:\cite{2006ApJ...646.1086H}. \textbf{GRB991208}-$z$:\cite{1999GCN...475....1D};SSFR:\cite{savaglio09};$T_{90}$: \cite{2000ApJ...534L..23H}; spectrum: \cite{2000ApJ...534L..23H}. \textbf{GRB000210}-$z$:\cite{2002ApJ...577..680P};SSFR:\cite{savaglio09};$T_{90}$: \cite{2002ApJ...577..680P}; spectrum: \cite{2002ApJ...577..680P}. \textbf{GRB000418}-$z$:\cite{2003AJ....125..999B};SSFR:\cite{savaglio09};$T_{90}$: \cite{2000GCN...642....1H}. \textbf{GRB000911}-$z$:\cite{2002ApJ...573...85P};SSFR:\cite{savaglio09};$T_{90}$: \cite{2002ApJ...573...85P}; spectrum: \cite{2002ApJ...573...85P}. \textbf{GRB000926}-$z$:\cite{2000GCN...851....1C};SSFR:\cite{savaglio09};$T_{90}$: \cite{2000GCN...801....1H}; spectrum: \cite{2005NCimC..28..351U}. \textbf{GRB011121}-$z$:\cite{2003ApJ...599.1223G};SSFR:\cite{savaglio09};$T_{90}$: \cite{2003ApJ...599.1223G}; spectrum: \cite{2003ApJ...599.1223G}. \textbf{GRB011211}-$z$:\cite{2002AJ....124..639H};SSFR:\cite{savaglio09};$T_{90}$: \cite{2002AJ....124..639H}; spectrum: \cite{2005ApJ...623..314P}. \textbf{GRB020405}-$z$:\cite{2002GCN..1330....1M};SSFR:\cite{savaglio09};$T_{90}$: \cite{2003ApJ...589..838P}; spectrum: \cite{2003ApJ...589..838P}. \textbf{GRB020813}-$z$:\cite{2003ApJ...584L..47B};SSFR:\cite{savaglio09};$T_{90}$,spectrum,: \cite{2005ApJ...629..311S}; lag:\cite{2007ApJ...660...16S}. \textbf{GRB0208019B}-$z$:\cite{2005ApJ...629...45J};SSFR:\cite{savaglio09};$T_{90}$,spectrum: \cite{2005ApJ...629..311S}. \textbf{GRB020903}-$z$:\cite{2004ApJ...606..994S};SSFR:\cite{savaglio09};$T_{90}$,spectrum: \cite{2005ApJ...629..311S}. \textbf{GRB021211}-$z$:\cite{2003GCN..1785....1V};SSFR:\cite{savaglio09};$T_{90}$,spectrum: \cite{2005ApJ...629..311S};lag:\cite{2007ApJ...660...16S}. \textbf{GRB030328}-$z$:\cite{2003GCN..1980....1M};SSFR:\cite{savaglio09};$T_{90}$,spectrum: \cite{2005ApJ...629..311S};lag:\cite{2007ApJ...660...16S}. \textbf{GRB030329}-$z$:\cite{2003GCN..2020....1G};SSFR:\cite{savaglio09};$T_{90}$,spectrum: \cite{2005ApJ...629..311S}; lag:\cite{gehrels06}. \textbf{GRB030528}-$z$:\cite{2005A&A...444..425R};SSFR:\cite{savaglio09};$T_{90}$,spectrum: \cite{2005ApJ...629..311S}; lag:\cite{2007ApJ...660...16S}. \textbf{GRB031203}-$z$:\cite{2004ApJ...611..200P};SSFR:\cite{savaglio09};$T_{90}$,spectrum: \cite{2004Natur.430..646S}; lag:\cite{2004Natur.430..646S};$E_{p}$,$E_{iso}$:\cite{2006MNRAS.372.1699G}. \textbf{GRB040924}-$z$:\cite{2004ApJ...611..200P};SSFR:\cite{savaglio09};$T_{90}$,spectrum: \cite{2004Natur.430..646S}; lag:\cite{2007ApJ...660...16S}. \textbf{GRB041006}-$z$:\cite{2005ApJ...626L...5S};SSFR:\cite{savaglio09};$T_{90}$:\cite{2008PASJ...60..919S},spectrum: HETE2 website$^e$. \textbf{GRB050525A}-$z$:\cite{2006IAUC.8696....1D};SSFR:\cite{savaglio09};$T_{90}$:\cite{2005GCN..3479....1C};spectrum:this work$^f$; Lag:\cite{gehrels06};$E_{p}$,$E_{iso}$:\cite{2005GCN..3474....1G} \textbf{GRB050826}-$z$:\cite{2006GCN..5982....1H};SSFR:\cite{savaglio09};$T_{90}$,spectrum:this work$^f$.  \textbf{GRB051022}-$z$:\cite{2005GCN..4145....1D};SSFR:\cite{savaglio09};$T_{90}$:\cite{2005GCN..4139....1H};spectrum:\cite{2005GCN..4156....1G}. \textbf{GRB060218}-$z$:\cite{2006GCN..4804....1S};SSFR:\cite{savaglio09};$T_{90}$:\cite{liang06b};spectrum: this work$^f$; lag:\cite{gehrels06}. \textbf{GRB060602A}-$z$:\cite{2007GCN..6997....1J};SSFR:n/a;$T_{90}$,spectrum: this work$^f$. \textbf{GRB050820}-$z$:\cite{2008GCN..7757....1J};SSFR:n/a;$T_{90}$,spectrum: this work$^f$; $E_{p}$,$E_{iso}$:\cite{2008GCN..7761....1S}. \textbf{GRB050509B}-$z$:\cite{gehrels05};SSFR:\cite{savaglio09};$T_{90}$:\cite{gehrels05},spectrum:this work$^f$; lag:\cite{gehrels06};$E_{p}$,$E_{iso}$:\cite{butler07}. \textbf{GRB050709}-$z$:\cite{fox05};SSFR:\cite{savaglio09};$T_{90}$:\cite{villasenor05},spectrum:\cite{villasenor05}; lag:\cite{gehrels06};$E_{p}$,$E_{iso}$:\cite{butler07}. \textbf{GRB050724}-$z$:\cite{barthelmy05a};SSFR:\cite{savaglio09};$T_{90}$:\cite{barthelmy05a};spectrum:this work$^f$;lag:\cite{gehrels06};$E_{p}$,$E_{iso}$:\cite{butler07}. \textbf{GRB060614}-$z$:\cite{dellavalle06};SSFR:\cite{savaglio09};$T_{90}$:\cite{kann09b};spectrum: \cite{zhang07b}; lag:\cite{gehrels06}; $E_{p}$,$E_{iso}$:\cite{2006GCN..5264....1G}. \textbf{GRB061006}-$z$:\cite{berger07};SSFR:\cite{savaglio09};$T_{90}$,spectrum:this work$^f$;$E_{p}$,$E_{iso}$:\cite{butler07}. \textbf{GRB000607}-$z$:\cite{nakar06b};$T_{90}$,spectrum:\cite{2002ApJ...567..447H}. \textbf{GRB050813}-$z$:\cite{prochaska06};$T_{90}$:\cite{2005GCN..3793....1S},$HR$:\cite{2005GCN..3793....1S}; lag:\cite{ferrero07},\cite{gehrels06};$E_{p}$,$E_{iso}$:\cite{butler07}. \textbf{GRB051210}-$z$:\cite{berger07};$T_{90}$,spectrum:\cite{laparola06}; lag:\cite{gehrels06};$E_{p}$,$E_{iso}$:\cite{butler07}.
\textbf{GRB051221A}-$z$:\cite{2006ApJ...650..261S};SSFR:\cite{savaglio09};$T_{90}$:\cite{2005GCN..4365....1C},spectrum:\cite{2005GCN..4394....1G}; lag:\cite{gehrels06};$E_{p}$,$E_{iso}$:\cite{2005GCN..4394....1G}. \textbf{GRB060121}-$z$:\cite{2006ApJ...648L...9L},\cite{deugartepostigo06};$T_{90}$,lag,spectrum:\cite{donaghy06};$E_{p}$,$E_{iso}$:\cite{butler07}. \textbf{GRB060313}-$z$,$T_{90}$,spectrum:\cite{roming06b},Lag:\cite{roming06b}; $E_p,E_{iso}$:\cite{2006GCN..4881....1G}. \textbf{GRB060502B}-$z$:\cite{bloom07};,$T_{90}$,spectrum:\cite{2006GCN..5064....1S}; lag:\cite{gehrels06}; $Ep,E_{iso}$:\cite{butler07}. \textbf{GRB060505}-$z$:\cite{ofek07};SSFR:\cite{savaglio09};$T_{90}$:\cite{2006GCN..5076....1P};spectrum:\cite{2006GCN..5142....1H}; lag:\cite{mcbreen08};$E_{p}$,$E_{iso}$:\cite{2006GCN..5142....1H};$L_p$:\cite{mcbreen08}. \textbf{GRB060801}-$z$:\cite{2006GCN..5470....1C};$T_{90}$,spectrum:\cite{2006GCN..5381....1S}; lag:\cite{gehrels06};$E_{p}$,$E_{iso}$:\cite{butler07}. \textbf{GRB061201}-$z$:\cite{stratta07};$T_{90}$:\cite{2006GCN..5882....1M};spectrum:\cite{2006GCN..5890....1G}; lag:\cite{stratta07};$E_{p}$,$E_{iso}$:\cite{butler07}. \textbf{GRB061210}-$z$:\cite{berger07c};$T_{90}$,spectrum:\cite{2006GCNR...20....1C};$E_{p}$,$E_{iso}$:\cite{butler07}. \textbf{GRB061217}-$z$:\cite{berger07c};$T_{90}$,spectrum,lag:\cite{2006GCNR...21....2Z};$E_{p}$,$E_{iso}$:\cite{butler07}. \textbf{GRB070429B}-$z$:\cite{cenko08b};$T_{90}$,spectrum:\cite{2007GCNR...51....1M};$E_{p}$, $E_{iso}$:\cite{butler07} \textbf{GRB070714B}-$z$:\cite{cenko08b};$T_{90}$:\cite{kann09b};spectrum:this work$^f$; lag:\cite{cenko08b};$E_{p}$,$E_{iso}$:\cite{butler07}. \textbf{GRB070724A}-$z$:\cite{cucchiara07};;$T_{90}$:\cite{kann09b};sepctrum:this work$^f$; $E_{p}$,$E_{iso}$: estimated with $\Gamma - E_p$ relation. \textbf{GRB071227}-$z$:\cite{berger09};$T_{90}$:\cite{2007GCN..7148....1S},\cite{2007GCN..7156....1S};spectrum:this work$^f$,\cite{2007GCN..7148....1S}; lag:\cite{2007GCN..7156....1S};$E_{p}$,$E_{iso}$:\cite{2007GCN..7155....1G}. \textbf{GRB080503}-$T_{90}$,spectrum,Lag:\cite{2008GCNR..138....1M}. \textbf{GRB080913}-$z$,$T_{90}$,spectrum:\cite{2008arXiv0810.2314G}; $E_{p}$:\cite{palshin08},$E_{iso}$:this work. \textbf{GRB090423}-$z$,$T_{90}$,$E_{iso}$:\cite{tanvir09};spectrum:this work;lag:\cite{krimm09};$L_p$:\cite{nava09};$E_p$:\cite{salvaterra09}.}

\tablenotetext{a}{{HR=S(50-100keV)/S(25-50keV)}}
\tablenotetext{b}{{Lag between 25-50keV and 50-100 keV.}}
\tablenotetext{c}{{absence of lag between 2-26keV and 40-700 keV }}
 \tablenotetext{d}{{BATSE data are not completed or recoreded (\cite{norris00})}}
 \tablenotetext{e}{{\url{http://space.mit.edu/HETE/Bursts/GRB041006/}}}
\tablenotetext{f}{{\url{http://grb.physics.unlv.edu}}}
\tablenotetext{g}{{We adopted z=1.4 for this burst. }}
\tablenotetext{h}{{only EE }}
\tablenotetext{i}{{lag between 6-40keV and 80-400 keV }}
\tablenotetext{j}{{lag between 25-50keV and 100-350 keV}}
\tablenotetext{k}{{got from $\frac{E_{\gamma,iso}}{t_{90}}$ }}
\tablenotetext{l}{{Lag between 25-50keV and 100-350 keV}} 
\tablenotetext{l}{{Lag between 15-25keV and 50-100 keV}}


\label{tabTypeISample}
\end{deluxetable}
\clearpage
       \end{landscape}

\begin{table}
\begin{center}
\caption{Observational criteria for physically classifying GRBs.\label{tbl-2}}
\begin{tiny}
\begin{tabular}{llll}
\tableline
Criterion & Type I & Type II & Issues \\
\tableline
Duration & Usually short, but can & Long without short/hard spike, & No clear separation line. \\
& have extended emission. & can be shorter than 1s in rest frame. & \\ 
Spectrum & Usually hard (soft tail) & Usually soft & Large dispersion, overlapping \\
Spectral Lag & Usually short & Usually long, can be short. &
Related to variability time scale \\
$E_{\gamma,iso}$ & Low (on average) & High (on average) & Wide distribution
in both, overlapping \\
$E_p-E_{\gamma,iso}$ & Usually off the track. & Usually on the track. & 
Some Type II off the track.\\
$L_{\gamma,iso}^p-$lag & Usually off the track. & Usually on the
track. &
Some Type II off the track.\\
\tableline
SN association & No. & {\bf Yes.} & Some Type II may be genuinely SNless.\\
Medium type & Low-$n$ ISM. & {\bf Wind} or High-$n$ ISM. & Large
scatter of $n$ distribution. \\
$E_{K,iso}$ & Low (on average) & High (on average) & Large dispersion, overlapping\\
Jet angle & Wide (on average) & Narrow (on average) & Difficult to
identify jet breaks \\
$E_{\gamma}$ and $E_K$ & Low (on average) & High (on average) & Type I
BH-NS BZ model $\sim$ Type II.\\
\tableline
Host galaxy type & {\bf Elliptical, early} and late & Late & 
Deep spectroscopy needed.\\
SSFR & {\bf Low} or high & High (exception GRB 070125) & overlapping \\
Offset & Outskirt or outside & Well inside & How to claim association 
if outside?\\
\tableline
$z$-distribution & Low average $z$ & High average $z$ & overlapping \\
$L$-function & Unknown & Broken power law, 2-component & overlapping \\
\tableline
GW signals & Precisely modeled & Unknown & No data yet \\
\tableline
\end{tabular}
\end{tiny}
\end{center}
\end{table}

\appendix
\section{Details of the Type I Gold Sample}

\begin{itemize}
\item GRB 050509B: No optical afterglow is detected.
The host galaxy is very likely a bright cD
elliptical galaxy in a nearby galaxy cluster at $z=0.2248\pm 0.0002$. 
The Swift XRT error circle is
offset from the elliptical galaxy. Deep upper limit on
SN association achieved \citep{gehrels05,bloom06,hjorth05b}. 
Satisfies both criteria \# 1 and 2 of the Gold Sample.
\item GRB 050709: The host galaxy is identified with the optical
afterglow, and is a star forming galaxy at $z=0.1606\pm 0.0001$.
The location of the GRB is at the outskirt of the host. Deep 
upper limit on SN association
achieved \citep{fox05}. Satisfies criterion \# 2 for the 
Gold Sample.
\item GRB 050724: The host is an early type galaxy at
$z=0.2576\pm 0.0004$. The afterglow
is at the outskirt of the host \citep{barthelmy05a,berger05}. 
Deep upper limit on SN association achieved \citep{malesani07}. 
Satisfies both criteria \#1 and 2 for Gold Sample.
\item GRB 060614: The host galaxy has a low SSFR 
\citep{galyam06,savaglio09}. Very stringent upper limits for any
associated SN fainter than any known SN achieved 
\citep{galyam06,fynbo06,dellavalle06}. Satisfies criterion \# 2 
for the Gold Sample.
\item GRB 061006: A faint host galaxy is detected at $z=0.4377\pm
0.0002$ \citep{berger07}. The SSFR is very low
\citep{berger07,savaglio09}. Satisfies criterion \# 1 for the Gold 
Sample.
\end{itemize}

\section{Details of Other Short/Hard GRB Sample}

\begin{itemize}
\item GRB 000607: This was a IPN-localized GRB. A putative host galaxy
at $z=0.14$ was proposed \citep{galyam08}. 
\item GRB 050813: No optical afterglow detected. The X-ray error
circle is associated with a galaxy cluster at $z \simeq 0.72$
\citep{prochaska06}. Moderately deep SN limit (relevant to this
low-$z$ interpretation) was reported by \citep{ferrero07}.
\item GRB 051210: A Swift GRB with X-ray afterglow \citep{laparola06}. 
A galaxy appeared outside the error box, but is likely the host.
No lines are observed. It is argued that $z>1.4$ \citep{berger07}.
\item GRB 051221A: Host galaxy is a star forming galaxy 
at $z=0.5464$ with a very
high SSFR \citep{soderberg06b,savaglio09}. A bright SN such as
1998bw is ruled out, but the limit is still consistent with the 
existence of faint SN 2002ap-like event \citep{soderberg06b}.
So it does not satisfy the citeria of the Type I Gold Sample.
\item GRB 060121: HETE-2 GRB with a faint optical afterglow,
leading to the discovery of an extremely faint host galaxy. The
redshift of the afterglow can be estimated as either 4.6 or 
1.7 \citep{deugartepostigo06,berger07}.
\item GRB 060313: Bright short/hard GRB with bright afterglow
\citep{roming06b}. A faint host galaxy is identified whose
redshift is unknwon \citep{berger07}. Spectral analysis of the
UVOT data suggests $z\leq 1.1$ \citep{roming06b}.
\item GRB 060502B: No optical afterglow is detected. The XRT 
position is close to a nearby early type galaxy at $z=0.287$
\citep{bloom07}. The chance probability for the association is 
0.03. There is another faint object in the field of view,
which could be the host galaxy at a high redshift \citep{berger07}.
\item GRB 060801: No optical afterglow is detected. Two possible
sources may be considered as the host. One has a
redshift $z=1.131$ (which is slightly outside the XRT error box
using the UVOT-aligned XRT position). Another source is within
the error box, but is likely even farther 
away \citep{berger07}.
\item GRB 061201: Optical afterglow was detected by UVOT. No host
galaxy was identified. Candidates include galaxy cluster Abell 995
($z=0.0865$), a star-forming galaxy at $z=0.111$, or a missing host
at an even higher $z$ \citep{stratta07}.
\item GRB 061210: A Swift GRB with delayed X-ray afterglow
detection. No optical afterglow detected. The host galaxy is 
likely a star forming galaxy at
$z=0.4095\pm 0.001$ \citep{berger07}.
\item GRB 061217: A faint Swift burst without optical afterglow
detection. Within the XRT error circle, there is a star forming
galaxy at $z=0.8270$ \citep{berger07}.
\item GRB 070429B: A Swift GRB with delayed X-ray afterglow
detection. The host galaxy is likely a faint galaxy at
$z=0.9023\pm 0.0002$ \citep{perley07,cenko08b}.
\item GRB 070707: Detected by INTEGRAL and have X-ray and optical
afterglow detected. A very faint host galaxy candidate was reported
with no redshift information \citep{piranomonte08}.
\item GRB 070714B: A Swift GRB with optical afterglow.
A secure host galaxy at $z=0.9225\pm0.0001$ is 
identified \citep{graham08,cenko08b}.
\item GRB 070724A: A Swift GRB with X-ray afterglow. A potential
host galaxy is detected, which is a star forming galaxy at
$z=0.457$ \citep{cucchiara07}.
\item GRB 070729: A Swift GRB with faint X-ray afterglow. A putative
red host galaxy is identified \citep{bergermurray07}. No redshift is known.
\item GRB 070809: A Swift GRB with X-ray and optical afterglow.
A nearby, edge-on spiral galaxy may be the host, with $z=0.2187$ 
\citep{perley07b}.
\item GRB 071227: A Swift GRB with X-ray and optical afterglows.
A host galaxy is identified as an edge on spiral galaxy. The
redshift is $z=0.3940$ \citep{berger09,davanzo09}. 
\item GRB 080123: A Swift GRB with X-ray afterglow. No host galaxy
detection is reported.
\item GRB 080503: A Swift GRB with short initial spike and very bright
extended emission. X-ray and optical afterglows are detected. There is
no galaxy directly at the GRB position. There are faint galaxies 
nearby, but one cannot make firm statements regarding their 
association with the GRB \citep{perley09}. 
\end{itemize}

After this work is finished, two more interesting short/hard GRBs
were detected, whose observational properties strengthen the main
theme of this paper. We include them here as follows.
\begin{itemize}
\item GRB 090426: This is rest-frame 0.35 s short/hard GRB at $z=2.6$
\citep{levesque09}. It has a blue, very luminous, star-forming
putative host galaxy with a small angular offset of the afterglow
location from the center. It is very likely associated to
Type II as argued in this paper \citep{levesque09}.
\item GRB 090510: This is bright short/hard GRB detected by both
Swift and Fermi (GBM/LAT) \citep{hoversten09,ohno09,guiriec09} with 
bright X-ray and optical afterglows
\citep{grupe09b,kuin09,olivares09}. With a redshift $z\sim 0.9$
\citep{rau09}, the isotropic energy and luminosity of this burst
all belong to the high end of the distribution
for Other Short/Hard GRBs presented in Fig.5. The
afterglow is consistent with a uniform density medium. The X-ray
lightcurve shows an early break at $t \sim 1500$ s since the trigger 
with a post-break decay index $\sim -2.16$. If it is interpreted 
as a jet break, then the total (collimation-corrected) jet 
gamma-ray and kinetic energies (assuming $n\sim 1~{\rm cm^{-3}}$)
are $E_\gamma/E_k \sim 10^{50}$ erg. This is relatively small
as compared with (but not far-off from) the $E_\gamma/E_K$ 
distributions of Type II GRBs 
\citep{frail01,bloom03,liang08,racusin09}. Although one may argue 
for a Type I association based on this, a Type II 
assoication is not strongly disfavored.
\end{itemize}


\end{document}